\pdfoutput=1
\documentclass[11]{article} 

\usepackage{natbib} 
\usepackage{graphicx}
\usepackage{amsmath}
\usepackage[titletoc,title]{appendix}
\usepackage[top=1in,bottom=1in,left=0.75in, right=0.75in]{geometry}
\usepackage{authblk}
\usepackage{indentfirst}
\usepackage{tikz}
\usepackage{mathtools}
\usepackage{subfig}
\usepackage{tabularx}
\usepackage{float}
\usepackage{setspace}
\usepackage{caption}
\usepackage[font=footnotesize]{caption}
\usetikzlibrary{shapes.geometric, arrows}
\date{}

\newtheorem{myDef}{Definition} 
\newtheorem{theo}{Theorem}
  
\newtheorem{pr}{Proof} 
\newtheorem{rem}{Remark}

\title{\bf Stochastic factors and string stability of traffic flow: Analytical investigation and numerical study based on car-following models}

\author{\textsc{\small Marouane Bouadi}}
\author{\textsc{\small Bin Jia}\thanks{Corresponding authors: bjia@bjtu.edu.cn (Bin Jia), jiangrui@bjtu.edu.cn (Rui Jiang)}}
\author{\textsc{\small Rui Jiang*}}
\author{\textsc{\small Xingang Li}}
\author{\textsc{\small Zi-You Gao}}

{\affil{\small  Institute of Traffic System Science, Beijing Jiaotong University, Beijing 100044, China \\
Key Laboratory of Transport Industry of Big Data Application Technologies for Comprehensive Transport, Ministry of Transport, Beijing Jiaotong University, Beijing 100044, China 
}

\begin{document}

\maketitle

\abstract{

The emergence dynamics of traffic instability has always attracted particular attention. For several decades, researchers have studied the stability of traffic flow using deterministic traffic models, with less emphasis on the presence of stochastic factors. However, recent empirical and theoretical findings have demonstrated that the stochastic factors tend to destabilize traffic flow and stimulate the concave growth pattern of traffic oscillations. In this paper, we derive a string stability condition of a general stochastic continuous car-following model by the mean of the generalized Lyapunov equation. We have found, indeed, that the presence of stochasticity destabilizes the traffic flow. The impact of stochasticity depends on both the sensitivity to the gap and the sensitivity to the velocity difference. Numerical simulations of three typical car-following models have been carried out to validate our theoretical analysis. Finally, we have calibrated and validated the stochastic car-following models against empirical data. It is found that the stochastic car-following models reproduce the observed traffic instability and capture the concave growth pattern of traffic oscillations. Our results further highlight theoretically and numerically that the stochastic factors have a significant impact on traffic dynamics.

}\hspace{10pt}


\section{Introduction}

Traffic congestion has attracted the interest of researchers in various fields of knowledge. The traffic congestion causes long time delays and high fuel consumption, which ultimately bring enormous economic loss all around the world. In this context, \cite{fc} estimated that the annual loss due to traffic congestion had attained 100 billion dollars in the USA alone. The transition from free flow to congestion and the emergence dynamics of traffic instability have also been widely studied, see, \cite{bando,hel2,schu,ward,lec1,ward2,tr11,chen1,sqe,chen3}. Some important findings include: (i) Usually the capacity drop occurs when traffic transits from free flow into congestion, (ii) Traffic flow is unstable, and traffic jams will emerge spontaneously when the density is high. These findings have constituted a solid basis for understanding the traffic flow dynamics.

In the deterministic framework, stability analysis has offered insights into the traffic stability condition, which indicates that traffic flow could be unstable  under certain circumstances. \cite{herm2} and \cite{her} are known to be the first researchers to explain the emergence of traffic oscillations using the concept of traffic instability. Depending on the number of vehicles involved, there are two well-known types of instability: local instability (or follower instability \citep{mar}) and string instability. Local instability occurs when a small perturbation grows with time for a single follower. In string instability, which is the ubiquitous type of instability \citep{trbb}, the magnitude of a small perturbation grows when transferred along a platoon. In this respect, different elegant stability methods have been exploited to extract local and string stability conditions of different traffic flow models, including the indirect Lyapunov method based on eigenvalues analysis \citep{ward,trbb,luigi,zh} and frequency domain analysis \citep{cui,ploe,mar}, See also the review made by \cite{sun}. Moreover, the analytical investigation and numerical simulation have demonstrated that the deterministic traffic models yield an initial convex growth pattern of traffic oscillations \citep{lit}.

Recently, a series of experimental investigations have been performed to study the growth pattern of traffic oscillations. The experiments demonstrated that traffic oscillations grow in a concave way. Several research studies have demonstrated that the stochastic factors are responsible for the concavity of the growth pattern of traffic oscillations \citep{jiang2,tian1,jiang3,wang1,trb3}. In this context, \cite{jiang1,jiang2} and \cite{tian1} have suggested that the traffic instability is due to the cumulative effect of stochastic factors. The results suggest that the presence of stochastic factors should be explicitly considered for a better prediction of traffic instability.

The stochastic factors have various causes, e.g., additional stimuli, the vehicle power train, temporal variations of the driver’s behavior, or the true randomness of some unconscious decisions of the drivers. Thus, they can be formulated in different ways. 

From one perspective, recent research studies have opted to develop traffic models that can better capture the behavior of the vehicle/driver system. For instance, (i) \cite{ciu}  argued that the gear-shifting strategy has a non-negligible effect on acceleration dynamics, (ii) \cite{mak1} proposed a Microsimulation Free-flow aCceleration model (MFC) in which the nature of drivers (timid and aggressive) and the power-train characteristics have been considered, (iii) \cite{mak2} extended MFC to model congested traffic. It is found that considering both human driver factors and vehicle power-train characteristics can reproduce the observed period and the growth pattern of traffic oscillations. 

Nevertheless, exhaustive modeling of all factors that influence the car-following maneuvers might be a cumbersome task. In an attempt to improve the traffic models, researchers have introduced stochastic elements in traffic models reflecting our limited knowledge \citep{pz1} of all the factors that might affect the car-following process. From this perspective, many traffic flow models have been proposed to consider stochastic elements. A usual way is to inject the noise into the vehicle’s acceleration, see, e.g., \cite{laval,trb3,ngod,tian1,wang1}. Some other examples include, (i) \cite{jiang2} characterized the stochastic factors by introducing the temporal variations of the driver's desired time gap. (ii) \cite{tian4} demonstrated that the changing rate of the wave traveling time varies over time stochastically. (iii) \cite{trb3} modeled the stochastic elements based on the concept of action point.

To our knowledge, NaSch cellular automaton model developed by \cite{ns} is the earliest model showing that the presence of randomness is crucially important for the spontaneous formation of traffic jams at high density. \cite{trb2} incorporated noise in the HDM (Human Driven Model) to model the drivers' imperfection and pointed out the destabilizing effect of stochasticity. \cite{laval} developed a parsimonious traffic model considering the presence of a traffic state independent noise through the Wiener process. It is found that stochasticity due to human uncertainties alone may trigger the emergence of traffic instability. \cite{trb3} carried out a spectrum analysis of a stochastic IDM and demonstrated that the presence of white noise induces traffic oscillations even in the sub-critical regime (below the string stability threshold of the deterministic model). Recently, \cite{ngod} carried out a string stability analysis of a stochastic optimal velocity model. It is found that the presence of stochasticity deteriorates traffic performance. The stochastic model was calibrated and validated against empirical data and showed good performance. \cite{wang1} performed a rigorous analytical analysis of the growth pattern (the speed standard deviation) of traffic oscillations. It is found that the presence of noise induces concave traffic oscillations even in the asymptotically stable regime. \cite{yuan} have shown that human acceleration imperfection contributes significantly to the capacity drop. \cite{xu} have studied a stochastic and simple traffic model in which stochasticity was proportional to the velocity. The authors demonstrated that the presence of noise triggers traffic instability. In their context, the noise impacted not only the period and amplitude of traffic oscillations but also the average velocity at the bottleneck. \cite{zheng2} studied the effect of a stochastic fundamental diagram in the macroscopic model and observed a qualitative change in the traffic oscillation growth feature. 

Apart from the modeling efforts that dealt specifically with the role of stochasticity in generating traffic instability, the stochastic traffic models \citep{ker1,jab1,jab2,ker2,lv,ngod2,tian5} have also been used to investigate the effect of randomness in the observed traffic phenomena, such as the scatted fundamental diagram \citep{ngod2}, the empirically observed time series of traffic speed \citep{tian5} and the intervention of randomness in the probabilistic aspect of traffic breakdown \citep{ker1,ker2}.

Among the above mentioned efforts, the linear stability analysis carried out by \cite{ngod} is the only known stability analysis of a stochastic car-following model including Brownian motion dW. It is found that the presence of stochasticity deteriorates traffic performance. \cite{ngod} performed a local and string stability analysis of a stochastic optimal velocity model (SOVM) where Cox-Ingersoll-Ross (CIR) process has been considered. According to this method, the string stability of a linearized SOVM in the form $dx=Axdt + RxdW$ and in the mean-square sense is reduced to searching a condition where the eigenvalues of the quantity $\Re{(A)}+0.5R^2$ are negative, i.e. $\Re{(A)}+0.5R^2\leq0$ while almost sure linear stochastic stability corresponds to $\Re{(A-0.5R^2)}\leq0$. To our knowledge, this method yields accurate results only for one-dimensional linear stochastic differential equations where the solution of the stochastic differential equation can be derived explicitly. On the other hand, the eigenvalue-based stability analysis has been proven feasible only when the matrices A and R commute for almost sure exponential stability. Hence, it is of utmost importance to conduct more studies on the string stability of the SOVM and the stochastic traffic models in general.

To fill the research gap, the present work will first extend the work of \cite{ngod} by presenting a general stochastic continuous car-following model that considers not only the inter-vehicular gap but also the velocity difference. Subsequently, we will use the direct generalized Lyapunov method \citep{zhang1,maob,zhang2,tob} to extract a string stability condition of the general stochastic car-following model. More importantly, we will demonstrate analytically and by numerical simulations that our methodology significantly improves the stability condition obtained by \cite{ngod}. As it will be clear in the following section, we will show that the present study is needed to address many methodological issues in deriving the stability condition of an important class of stochastic traffic models. Finally, we will carry out an extensive calibration and validation of different deterministic and stochastic traffic models. Our objective is to assess the potential of the studied stochastic models in reproducing the spontaneous emergence of traffic oscillations, the concave growth pattern of traffic oscillations, and more importantly predicting the emergence of traffic instability. After the extensive calibration of traffic models and numerical simulations, we will show that the stochastic models outperform the deterministic models in predicting the emergence of traffic instability.

Our paper will be organized as follows. In section 2, we will present a general formulation of stochastic continuous car-following models. In section 3, we will present a string stability analysis based on the generalized Lyapunov method and extract a general string stability condition. Numerical simulations will be conducted in section 4 to validate our theoretical analysis. In section 5, we will calibrate and validate different stochastic and deterministic traffic models against empirical data. Finally, section 6 will be devoted to a conclusion.

\section{The stochastic continuous car-following model }

In this work, we will deal with the following general stochastic continuous car-following equation taking the form:

\begin{equation} 
\frac{dv_{n}(t)}{dt}=f_{1}(s_{n},v_{n},v_{n-1})+f_{2}(v_{n})
\end{equation}

\noindent where $f_{1}(s_{n},v_{n},v_{l})$ denotes the deterministic component and $f_{2}(v_{n})$ denotes the stochastic one. The symbols $s_{n},v_{n},v_{n-1}$ denote respectively the spacing, the velocity of the vehicle $n$ and the leader's velocity $n-1$. 

In equation (1), different from the work of \cite{laval} in which the Ornstein Uhlenbeck-like process has been studied, we have considered a noise strength that is dependent on the traffic state. In this context, without loss of generality, we assume that the traffic state is only dependent on the velocity \citep{hel2,wu,trbb,trb6}. In this context, \cite{xu} have proposed an expression in which the standard deviation tends to zero when approaching a given high velocity. \cite{ngod} have studied an expression in which the standard deviation tends to zero when the velocity tends to zero. In this work, we will study the function $f_{2}(v_{n})$ proposed by \cite{ngod} which is given by: 


\begin{equation} 
f_{2}(v_{n})=\sigma \sqrt{v_{n}}\frac{d{W}_{n}}{dt}
\end{equation}

\noindent where $\sigma$ denotes the noise's strength or dissipation term and $d{W}_{n}$ denotes the Wiener process. The dependency on velocity in equation (2) and the normal distribution (generated by the Wiener process) are in agreement with empirical observations. When the velocity tends to zero, the term $\sqrt{v_{n}}$ will also tend to zero. Conversely, the standard deviation term will be maximal when the velocity tends to the maximal velocity (See Figure 13 in \cite{hel2}, the normal distribution of the acceleration in Table 3 in the paper of \cite{wu}, \cite{trbb} and \cite{trb6}). These features have motivated us to adopt equation (2) in this study. Note that the following methodology is valid for many possible forms of equation (2). 

To analyze the traffic instability, we approximate equations (1) and (2) by the following linearized stochastic continuous car-following equation:

\begin{equation} 
d\delta v_{n}(s_{n},v_{n},v_{n-1})= (\alpha_{1}\delta s_{n} + \alpha_{2}\delta v_{n} + \alpha_{3}\delta v_{n-1})dt + \frac{\sigma }{2\sqrt{v_{e}}} \delta v_{n} d{W}_{n}  
\end{equation}




\noindent where $s_{e}$ and $v_{e}$ are respectively the equilibrium spacing and velocity. $\delta s_{n}$, $\delta v_{n}$ and $\delta v_{n-1}$ are small perturbations around the equilibrium spacing and velocity. 


The coefficients $\alpha_{1}$, $\alpha_{2}$ and $\alpha_{3}$ represent the sensitivities with respect to the variation of the spacing, the velocity, and the leader's velocity which are given by: 

\begin{equation} 
\alpha_{1} = \frac{\partial a_{n}}{\partial s_{n}}
\end{equation}

\begin{equation} 
\alpha_{2} = \frac{\partial a_{n}}{\partial v_{n}}
\end{equation}

\begin{equation} 
\alpha_{3} = \frac{\partial a_{n}}{\partial v_{n-1}}
\end{equation}

\noindent where $a_{n}=\frac{dv_{n}(t)}{dt}$ is the acceleration function, and the derivatives are all valuated at the point $(s_{e},v_{e})$. 

For conciseness, we adopt the notation $\mu=\frac{\sigma}{2\sqrt{v_{e}}}$. Then, we consider small perturbations around the equilibrium in the form of wave equations $\delta s_{n}=\hat{s}(t)\exp{(i\omega n)}$ and $\delta v_{n}=\hat{v}(t) \exp{(i \omega n)}$. Considering periodic boundary conditions, equation (3) and also the relation $d\delta s_{n}= (\delta v_{n-1} - \delta v_{n})dt$, we get finally the following linear stochastic equation: 

\begin{equation} 
dx=Axdt+RxdW
\end{equation}

\noindent where 

\begin{equation} 
x= {[\delta s_{n},\delta v_{n}]}^T
\end{equation}

\begin{equation}
A= {\begin{bmatrix}
    0      & \exp{(-i\omega)}-1  \\
  \alpha_{1}    &  \alpha_{2}+\alpha_{3} \exp{(-i\omega)}\\
\end{bmatrix}} 
\end{equation}

\noindent and 

\begin{equation}
R= {\begin{bmatrix}
    0      & 0 \\
  0    &  \mu \\
\end{bmatrix}} 
\end{equation}

We will see in the following section that equation (7) has useful properties which we will exploit to extract a string stability condition of the car-following model given by equations (1) and (2). For an illustration of the string instability and local instability in the stochastic car-following models, one may refer to Appendix A.

\section{Linear string stability analysis of the stochastic continuous car-following model}

In this section, we will first explain the stability of stochastic differential equations based on the direct Lyapunov method and the mean-square stability. Then, we will derive a stability condition of the stochastic car-following model given by equations (1) and (2).

\subsection{Stability of linear stochastic differential equations}

In this subsection, we will recall the direct Lyapunov method to extract a stability condition of the stochastic car-following model given by equations (1) and (2). The generalized Lyapunov method is suitable for studying stochastic differential equations. Next, we will recall the mean-square stability of stochastic differential equations, which is the stricter stability condition (if we exclude higher moments). 


\subsubsection{Generalized Lyapunov equation}

Following the recent findings on the stability of linear stochastic differential equations, we recall the basic formalism of stochastic stability and how to deal with the stability of complex-valued linear stochastic differential equations.  

\begin{myDef}

The equilibrium position of the general linear stochastic differential equation of the form:

\begin{equation} 
dx=f(x,t)dt+g(x,t)dW
\end{equation}
		
\noindent	is called stochastically stable or stable in probability if a positive definite function $V(x)$ exists such that $LV(x)<0$.  $V(x)$ is the Lyapunov function which should be twice differentiable in x and once in t. $LV(x)$ is related to the differential of $V(x)$ as the following Ito formula \citep{maob}: 

\begin{equation} 
dV(x,t)=LV(x,t)+V_{x}(x,t)g(x,t)dW 
\end{equation}

\noindent More precisely: 

\begin{equation} 
LV(x,t)=V_{t} (x,t)+V_{x} (x,t)f(x,t)+\frac{1}{2} trace[g^{t}(x,t) V_{xx}(x,t)g(x,t)]
\end{equation}

\end{myDef}

\noindent In control theory, it is well known that there is no proof of the uniqueness of the Lyapunov function. In this context, the following typical quadratic Lyapunov function is commonly used: 

\begin{equation} 
V(x)=x^{T}Px
\end{equation}

\noindent where P should be a symmetric and positive definite matrix. 
For real valued autonomous stochastic differential equation (7), we get the following expression of $LV(x)$ after considering equation (14) and calculating the corresponding derivatives: 

\begin{equation} 
LV(x,t)=x^{T}(PA+A^{T} P+R^{T} PR)x
\end{equation} 

\noindent From equation (15), we define the following equation as the generalized Lyapunov equation \citep{maob,tob}:

\begin{equation} 
L_{A,R}= PA+A^{T} P+R^{T} PR
\end{equation} 

Equation (7) is called stable in probability if a symmetric and positive definite matrix P exists (thus a Lyapunov function (14)) such that the real parts of eigenvalues of $L_{A,R}$ in equation (16) are negative. 

However, in our context, equation (7) has complex-valued matrix A and the standard Lyapunov formalism applies only to real systems. Fortunately, \cite{zhang1} and \cite{zhang2} have demonstrated that the stability of equation (7) (considering that the matrices A and R are complex-valued) is equivalent to the stability of the following system:

\begin{equation} 
dx_{s}=A_{s} x_{s} dt+R_{s} x_{s} dW
\end{equation} 

\noindent where 

\begin{equation} 
x_{s}= {[x_{r},x_{i}]}^T
\end{equation}

\begin{equation}
A_{s}= {\begin{bmatrix}
    A_{r}      & -A_{i} \\
  A_{i}    &  A_{r} \\
\end{bmatrix}} 
\end{equation}

\noindent and 

\begin{equation}
R_{s}= {\begin{bmatrix}
    R_{r}      & -R_{i} \\
  R_{i}    &  R_{r} \\
\end{bmatrix}} 
\end{equation}

The equation (17) can be deduced by separating the real part and the imaginary part of equation (7). $x_{r}$ and $x_{i}$ in equation (18) are respectively the real part and the imaginary part of $x$ in equation (7) , that is $x=x_{r}+i x_{i}$. The $2\times2$ matrices $A_r$ and $A_i$ in equation (19) are respectively the real part and the imaginary part of matrix $A$ in equation (9). The same reasoning holds for matrix $R$.   
Therefore, we can use equation (17) to study the stability of equation (7), where the $4\times4$ matrices $A_{s}$ and  $R_{s}$ will be used instead of the $2\times2$ matrices $A$ and $R$. 

\subsubsection{Mean-square stability}

\begin{myDef}
     A stochastic system of the form (17) is said to be asymptotically mean-square stable if: 
		
	\begin{equation}
\lim_{t\to \infty} E{||x(t)||}^{2}=0  \nonumber
\end{equation}

\noindent For any initial state $x(0) \in \Re^n$. This means that the second moment of the solution $x(t)$ will tend to zero. 

\end{myDef}

Given a Lyapunov function (14) (where $x_{s}= {[x_{r},x_{i}]}^T$), \cite{zhang1} and \cite{zhang2} have demonstrated that the linear stochastic system (17) is asymptotically mean-square stable if the real parts of the eigenvalues of equation (16) (by considering $A_{s}$ and $R_{s}$ instead of $A$ and $R$) are negative.

Note that for equation (17), mean-square stability implies the almost sure asymptotic stability $\lim_{t\to \infty} E{||x(t)||}=0  \nonumber$ \citep{maob}. 

\subsection{Stability analysis of the stochastic traffic model}

In literature, the Lyapunov method has been used only for studying the local stability of deterministic traffic models \citep{ch, sun} and a stochastic OVM model \citep{ngod}. In this subsection, we use the generalized Lyapunov equation to extract the string stability condition of the stochastic car-following model given by equations (1) and (2). In this context, the most challenging task is to find a suitable symmetric and positive matrix P for equation (14) leading to a plausible stability condition where the wave number $\omega$ is eliminated. Hence, based on the previously mentioned analysis, we have established the following theorem:

\begin{theo}
 
The single class stochastic continuous car-following model in equations (1) and (2) is mean square stable if 

\begin{equation} 
4\alpha_{1}<2(\alpha_{2}^{2}-\alpha_{3}^{2})+\mu^{2}(\alpha_{2}-\alpha_{3})
\end{equation}

\end{theo}

\begin{pr}

We consider the corresponding linearized car-following equation (7). \cite{trbb} have demonstrated that traffic flow instability is usually of long-wave lengths. Accordingly, we perform a Taylor approximation to the exponential in the matrix (9) as the following:

\begin{equation} 
\exp{(-i\omega)}-1\approx -i\omega - \omega^{2}/2
\end{equation}

\noindent and 

\begin{equation} 
\alpha_{3} \exp{(-i\omega)} \approx \alpha_{3} - i\omega\alpha_{3}
\end{equation}

\noindent As explained in the previous section, we decompose matrix A in equation (9) into a real part and an imaginary part. This process leads to equation (17), where the matrices $A_{r}$ and $A_{i}$ in equation (19) take the following forms:

\begin{equation}
A_{r}= {\begin{bmatrix}
    0      & -\frac{\omega^{2}}{2}\\
  \alpha_{1}    &  \alpha_{2}+\alpha_{3} \\
\end{bmatrix}} 
\end{equation}

\begin{equation}
A_{i}= {\begin{bmatrix}
    0      & -\omega \\
  0    &  -\alpha_{3}\omega \\
\end{bmatrix}} 
\end{equation}

\noindent The dissipation matrix R is entirely real. We get finally the following expressions for the $4 \times 4$ matrices $A_{s}$ and $R_{s}$: 

\begin{equation}
A_{s}= {\begin{bmatrix}
    0      & -\frac{\omega^{2}}{2} & 0 & \omega \\
  \alpha_{1}    &  \alpha_{2}+\alpha_{3} & 0 & \alpha_{3} \omega \\
	 0   &  -\omega & 0 & -\frac{\omega^{2}}{2}\\
	 0    &  -\alpha_{3} \omega & \alpha_{1} & \alpha_{2}+\alpha_{3}\\
\end{bmatrix}} 
\end{equation}

\begin{equation}
R_{s}= {\begin{bmatrix}
    0      & 0 & 0 & 0 \\
  0    &  \mu & 0 & 0 \\
	 0   &  0 & 0 & 0\\
	 0    & 0 & 0 & \mu\\
\end{bmatrix}} 
\end{equation}

\noindent Then, we search for a suitable Lyapunov function (14) and calculate the corresponding generalized Lyapunov equation:

\begin{equation} 
L_{A_{s},R_{s}}= PA_{s}+A_{s}^{T} P+R_{s}^{T} PR_{s}
\end{equation}

\noindent We define a matrix P having the following form:

\begin{equation}
P= {\begin{bmatrix}
    U      & V\\
  -V    &  U \\
\end{bmatrix}} 
\end{equation}

\noindent where $U$ is a diagonal block matrix given by: 

\begin{equation}
U= {\begin{bmatrix}
    \frac{2\alpha_{1}(\alpha_{2}+\alpha_{3})}{\omega^2(\alpha_{2}-\alpha_{3})}      & 0\\
  0    &  1 \\
\end{bmatrix}} 
\end{equation} 

\noindent and $V$ is a skew-symmetric block matrix ($V^{T}=-V$) given by:

\begin{equation}
V= {\begin{bmatrix}
 0       & -\frac{2\alpha_{1}}{\omega(\alpha_{2}-\alpha_{3})} \\
  \frac{2\alpha_{1}}{\omega(\alpha_{2}-\alpha_{3})}   &  0\\
\end{bmatrix}} 
\end{equation}

\noindent The matrix (29) is symmetric. Finally, after replacing in equation (28), we get a diagonal matrix where two entries are zero while the two others are given by:

\begin{equation}
\lambda \normalfont= \frac{-(-2\alpha_{2}^{2}-\alpha_{2}\mu^{2} +2\alpha_{3}^{2}+\alpha_{3}\mu^{2}+4\alpha_{1})}{\alpha_{2}-\alpha_{3}} 
\end{equation}

\noindent From equation (28) and equation (15), the Lyapunov function derivative $LV(x)$ has the following form:

\begin{equation}
LV(x)= \frac{-(-2\alpha_{2}^{2}-\alpha_{2}\mu^{2} +2\alpha_{3}^{2}+\alpha_{3}\mu^{2}+4\alpha_{1})}{\alpha_{2}-\alpha_{3}}(\delta v_{r}^{2}+\delta v_{i}^{2})
\end{equation}

\noindent where $\delta v_{r}$ and $\delta v_{i}$ are respectively the real and imaginary components of $\delta v_{n}$ in equation (8).  

\noindent Moreover, the matrix P in equation (29) has two eigenvalues which are positive if the following condition is fulfilled:  

\begin{equation}
-8\alpha_{1}\omega^{2}(-\alpha^{2}_{2}+\alpha^{2}_{3}+2\alpha_{1})>0
\end{equation}

\noindent Assuming that every driver tends to accelerate when: (i) The gap increases; (ii) The ego-velocity decreases; (iii) The leader's velocity increases; we have $\alpha_{1}> 0$ and $\alpha_{2}-\alpha_{3}<0$. Subsequently, the negativity of equation (33) implies that the matrix (29) is positive definite. Hence, the stability condition is given by equation (21).

\end{pr}

\begin{rem}

Mathematically, the Lyapunov method does not ensure the uniqueness of the Lyapunov function for a given stochastic differential equation. Hence, there might exist many Lyapunov functions yielding to different stability conditions. Consequently, the theorem 1 is sufficient but not sufficient and necessary condition. 

\end{rem}

\begin{rem}
 
In some traffic models, it is more convenient to express the stability condition in term of the sensitivity to the velocity difference $\Delta v=v_{n}-v_{n-1}$ instead in term of the sensitivity to the leader's velocity.  In this case, we can perform the replacement: $\alpha_{1} \rightarrow \tilde{\alpha}_{1}$,  $\alpha_{2} \rightarrow \tilde{\alpha}_{2}+\tilde{\alpha}_{3}$ and $\alpha_{3}\rightarrow -\tilde{\alpha}_{3}$. 

\noindent $\tilde{\alpha}_{1}$, $\tilde{\alpha}_{2}$ and $\tilde{\alpha}_{3}$ denote respectively the derivative with respect to the space, velocity, and velocity difference. Thus, the condition (21) will be: 

\begin{equation}
4\tilde{\alpha}_{1}<2\tilde{\alpha}_{2}^{2}+4\tilde{\alpha}_{2}\tilde{\alpha}_{3}+ \mu^{2}(\tilde{\alpha}_{2}+2\tilde{\alpha}_{3})
\end{equation}

\noindent The sign before $\tilde{\alpha}_{3}$ will be negative if one defines $\Delta v=v_{n-1}-v_{n}$.
\end{rem}

\begin{rem}

For the deterministic case $\mu=0$, we recover the following stability condition of deterministic traffic models \citep{trbb,ward,zh,cui,mar}:  

\begin{equation} 
\alpha_{1}<\frac{1}{2}(\alpha_{2}^{2}-\alpha_{3}^{2}) 
\end{equation}

\end{rem}

Equations (21) or (35) are the string stability conditions of the stochastic car-following model given by equations (1) and (2). From equations (21) and (35), we can deduce that the presence of noise can destabilize traffic flow, which will be clear in the following section. This is in accordance with empirical observations, e.g. \cite{jiang2} and previous theoretical investigations \citep{wang1,trb3,ngod}. Moreover, equation (35) suggests that the contribution of noise is not only proportional to the sensitivity to the gap \citep{ngod} but also the sensitivity to the velocity difference.

\section{	Numerical simulations }

In this section,  we compare our theoretical results with numerical simulations and give a visual representation of the impact of stochasticity on the stability phase diagram. To perform these tasks, we will consider three typical car-following models: The stochastic optimal velocity model (SOVM), the stochastic full velocity difference model (SFVDM), and the stochastic intelligent driver model (SIDM). In the following, we will consider a circular road of length $L=1000 \ \mathrm{m}$ and a time step $\Delta t=0.02\ \mathrm{s}$. Initially, vehicles are equally spaced with equilibrium spacing $s_{e}$ and velocity $v_{e}$. Next, we monitor the traffic evolution. For the deterministic case, we perturb the position of one vehicle by 1 m around the equilibrium position. We will consider a vehicle length of $l=5 \ \mathrm{m}$ and a maximal velocity of $V_{max}=20\ \mathrm{m/s}$.

\subsection{The stochastic optimal velocity model}

In the OVM, each vehicle aims to attain the desired velocity $V_{op}$ within a given relaxation time \citep{bando}. The stochastic version of the OVM (SOVM) reads \citep{ngod}: 

\begin{equation} 
dv_{n}(t)=\beta(V_{op}(s_{n}(t))-v_{n}(t))dt + \sigma \sqrt{v_{n}(t)} {dW}_{n}
\end{equation}

\noindent $V_{op}$ is given by:

\begin{equation} 
V_{op}(s) = \frac{V_{max}}{2}(\tanh{(\frac{s_{n}}{s_{c}}-k)}+\tanh{(k)}) 
\end{equation}

\noindent where $s_{c}$ is the critical headway, $\beta$ is the reaction coefficient (the inverse of the relaxation time), and $k$ is a constant. In the following, we will denote the derivative of $V_{op}$ simply as $V'$.
From equations (21), (35) and (37), we can deduce the stability condition of the SOVM:

\begin{equation} 
V'<\frac{\beta}{2}- \frac{\mu^2}{4}
\end{equation}

\noindent In the deterministic case, we find $V'<\frac{\beta}{2}$ \citep{bando}.

Note that \cite{ngod} have found the following asymptotic mean-square stability condition: 

\begin{equation} 
\mu^2 \leq \frac{V'}{\beta}(\beta-2V')
\end{equation}

\begin{figure}[H]
\centering
\subfloat[]{\includegraphics[width=.7\textwidth]{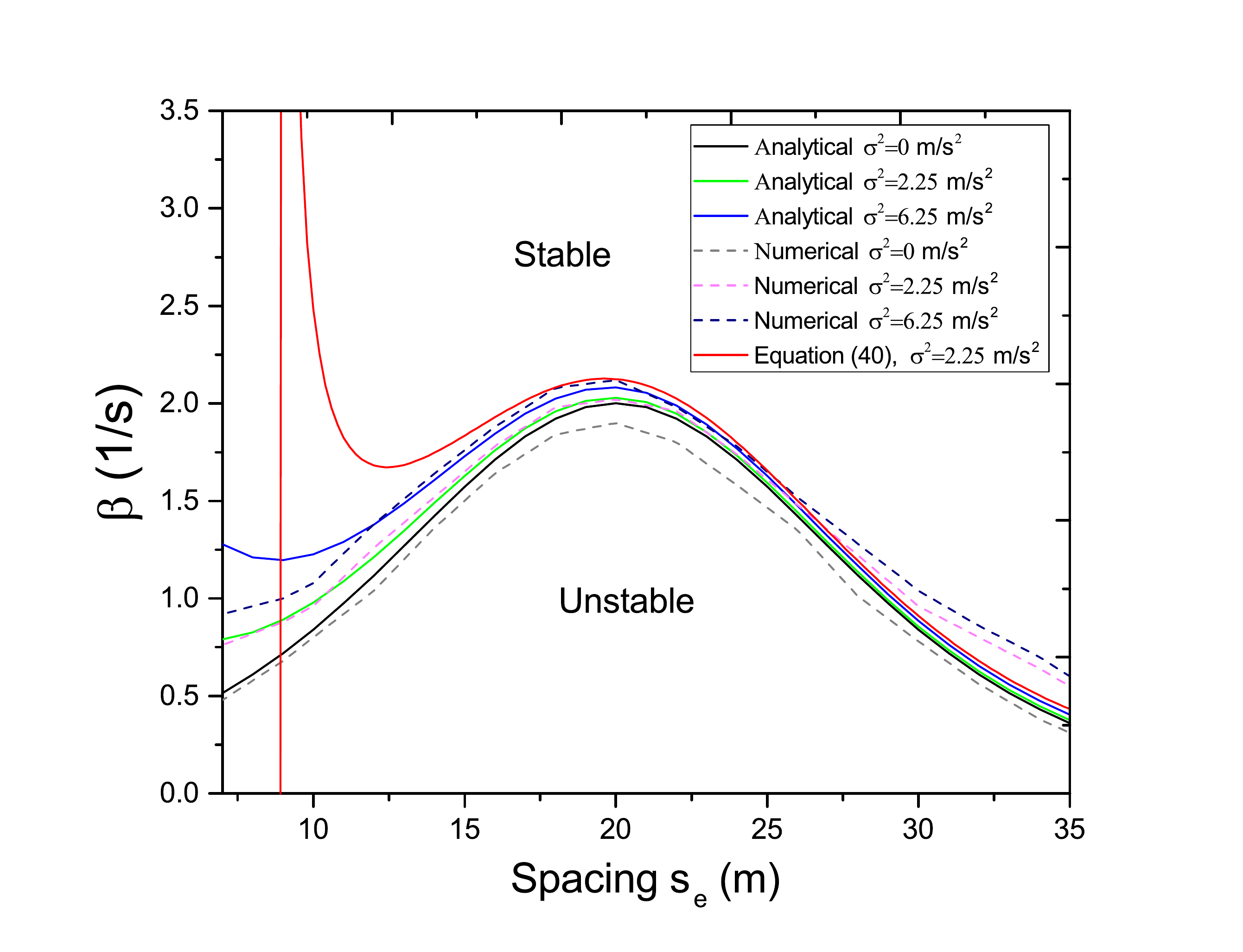}} \\
\subfloat[]{\includegraphics[width=.7\textwidth]{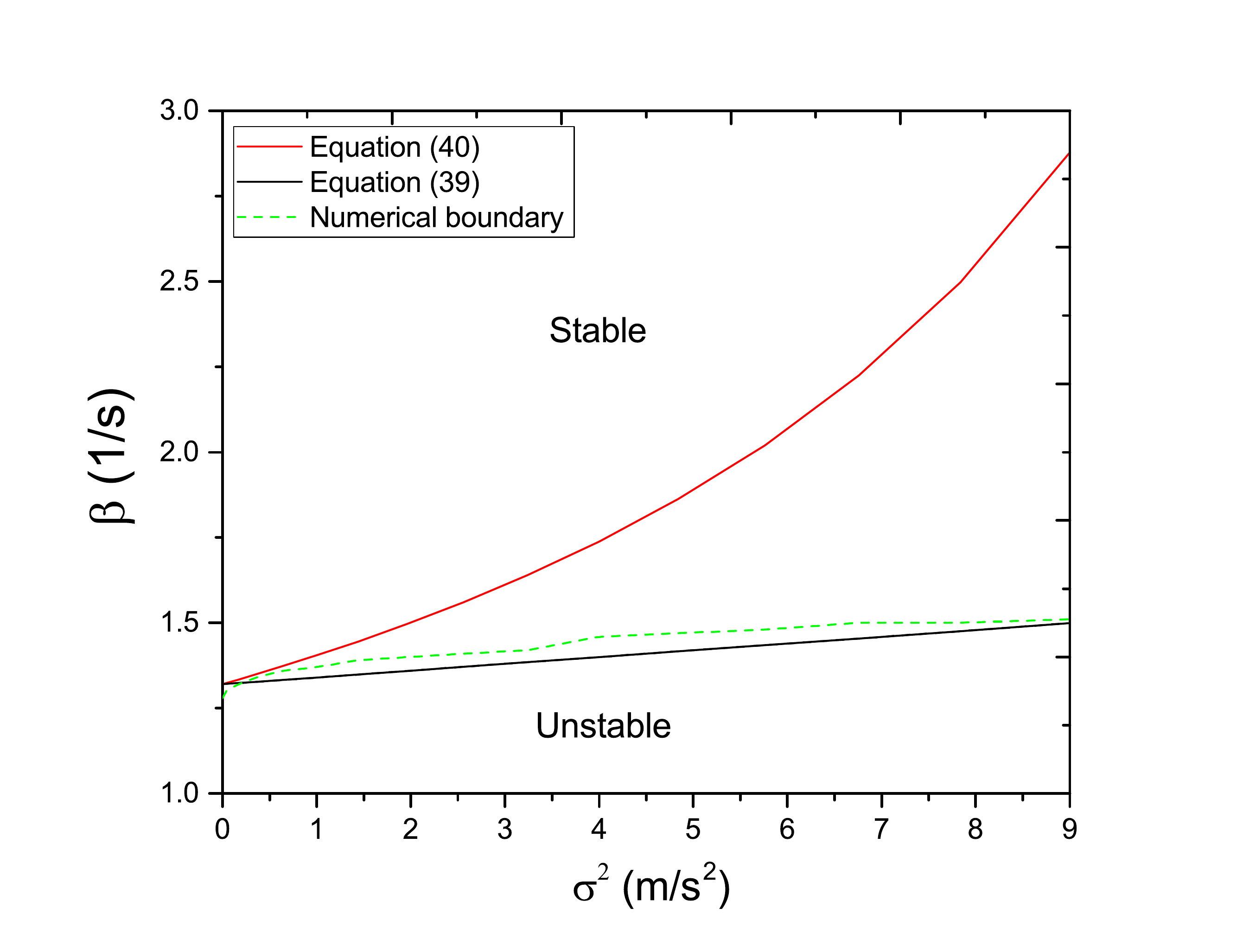}} 
\caption{Stability phase diagrams of the SOVM in the $(\beta, s_{e})$ plane and the $(\beta, \sigma^{2})$ plane (a) in the $(\beta, s_{e})$ plane (b) in the $(\beta, \sigma^{2})$ plane.}
\label{fig1}
\end{figure}

\begin{figure}[H]
\centering
\subfloat[]{\includegraphics[width=.4\textwidth]{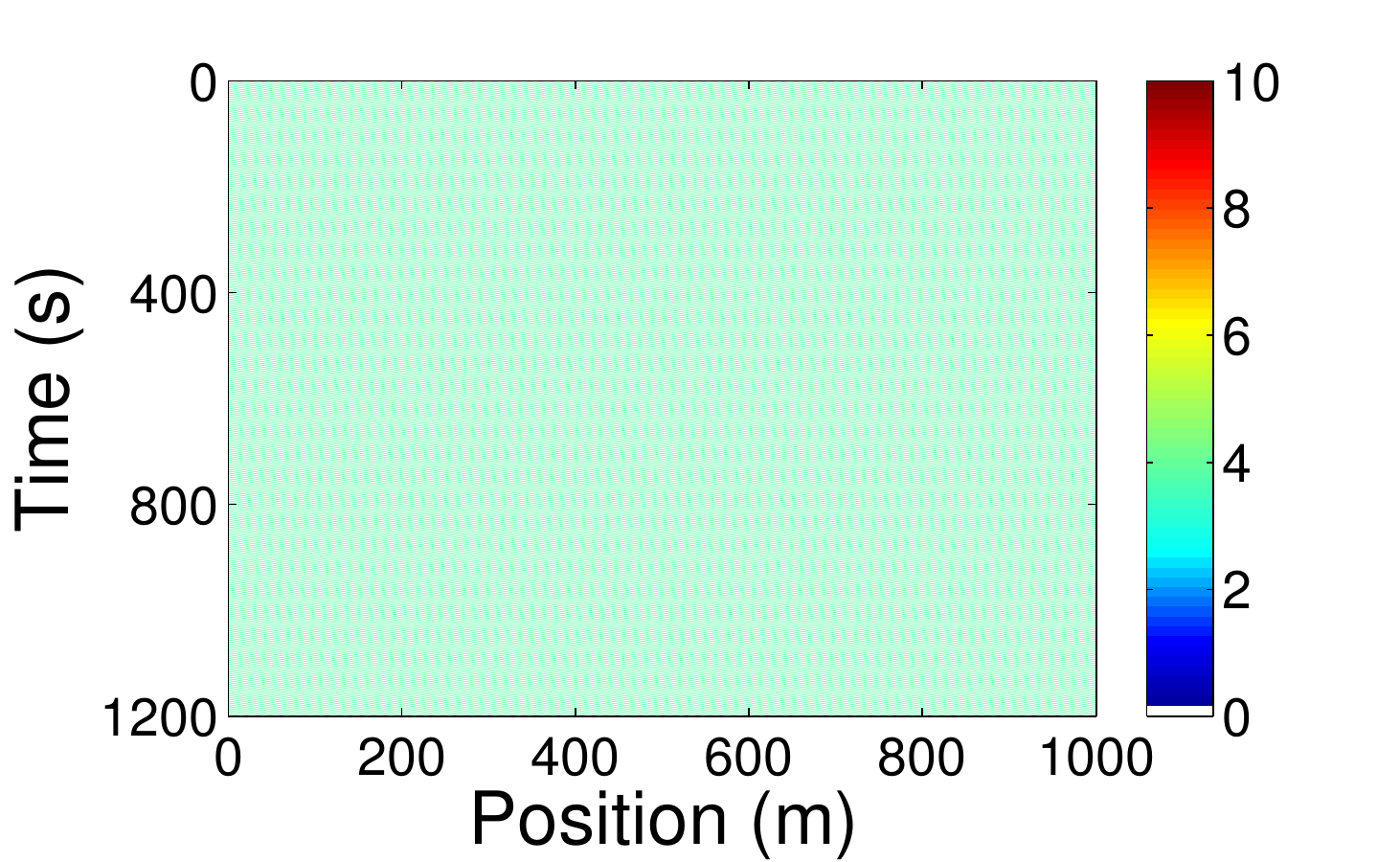}} 
\subfloat[]{\includegraphics[width=.4\textwidth]{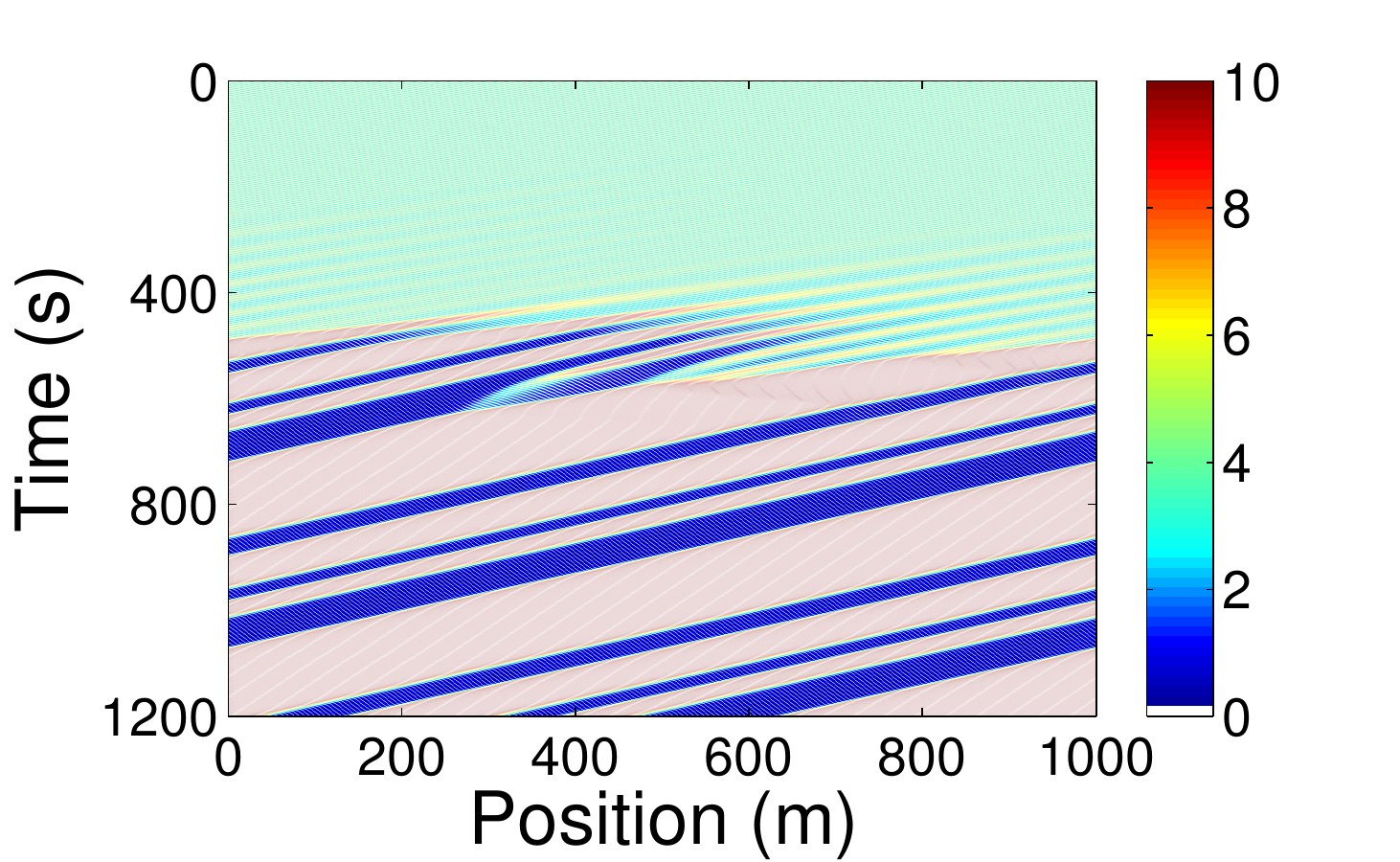}}\\
\subfloat[]{\includegraphics[width=.4\textwidth]{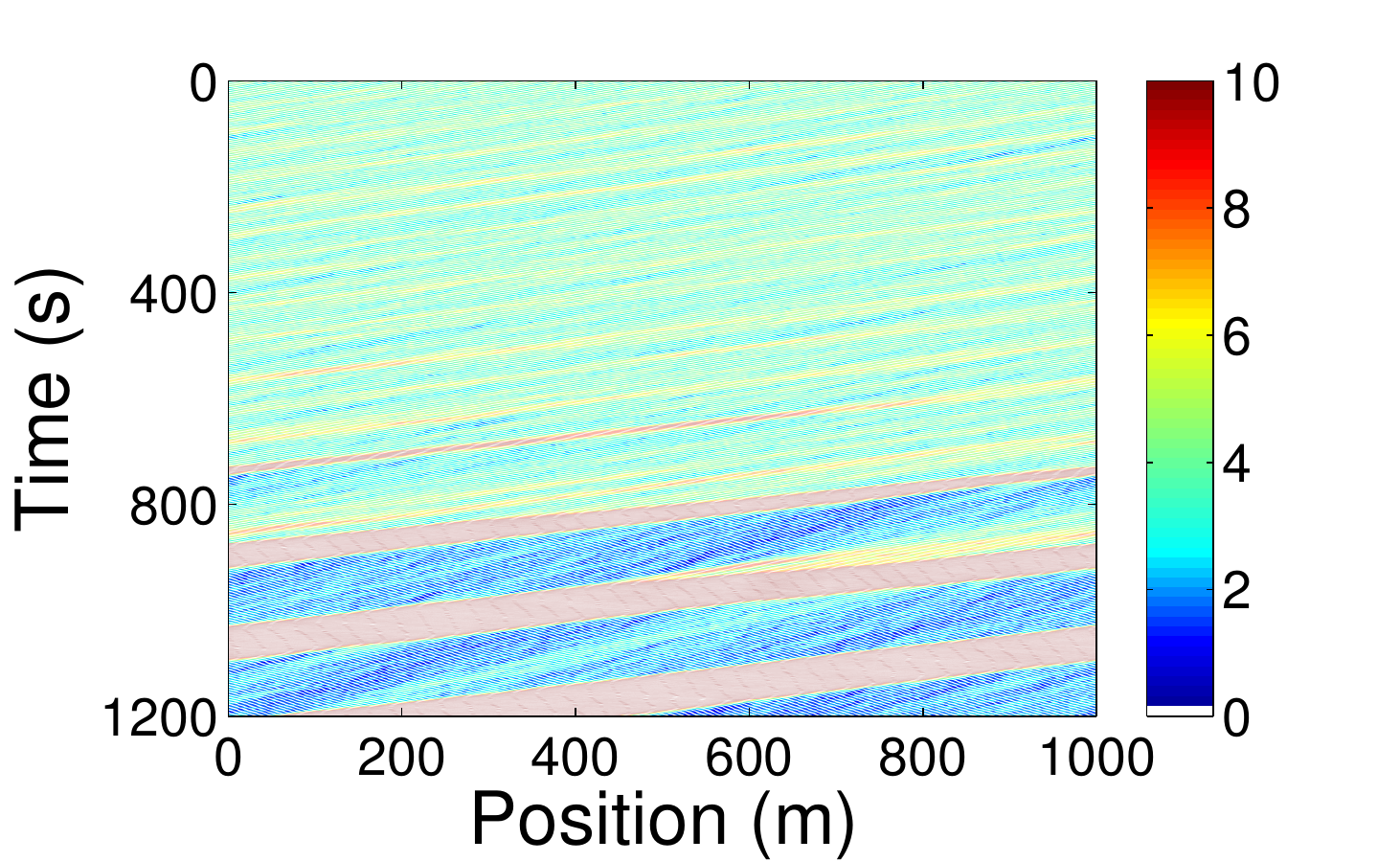}}
\subfloat[]{\includegraphics[width=.4\textwidth]{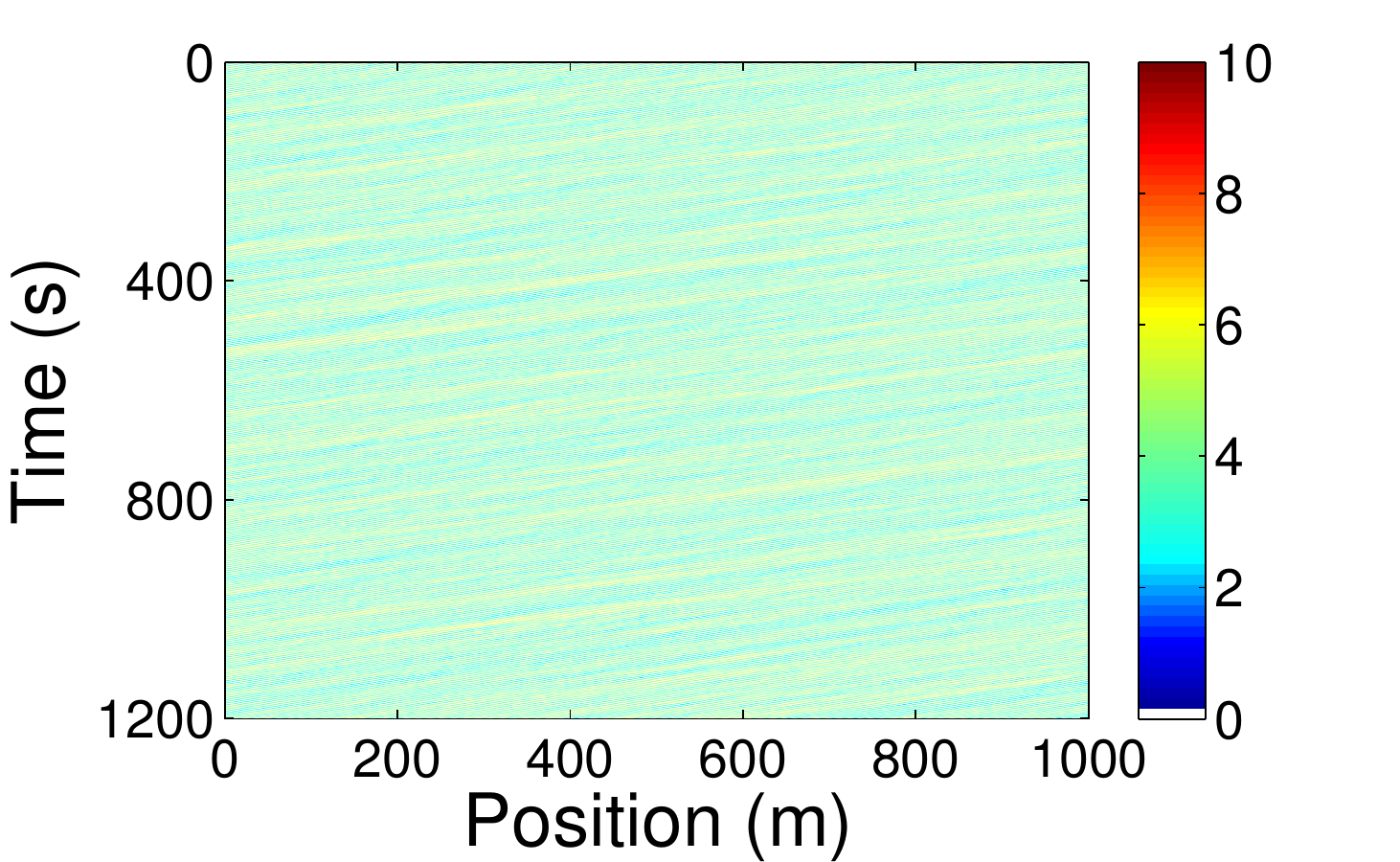}}\\
\caption{Time-space diagrams corresponding to stable and unstable regime of the SOVM (a) Stable for $\sigma^2 = 0 \ \mathrm{m/s^2}$, $\beta = 1.35 \ \mathrm{s^{-1}}$ (b) Unstable for $\sigma^2=0\ \mathrm{m/s^2}$, $\beta=1 \ \mathrm{s^{-1}}$ (c) Unstable for $\sigma^2=0.5 \ \mathrm{m/s^2}$, $\beta=1.35 \ \mathrm{s^{-1}}$ (d) Stable for $\sigma^2=0.5 \ \mathrm{m/s^2}$, $\beta=1.6 \ \mathrm{s^{-1}}$. The velocity unit in the color bar is $\ \mathrm{(m/s)}$.} 
\label{fig2}
\end{figure}

Next, we present the stability diagram corresponding to the SOVM for $k=2$ and $s_{c}=10$ m. Figure~\ref{fig1}(a) depicts the stability phase diagram in the $(\beta,s_{e})$ plane of both the deterministic case and the stochastic case. The stability phase diagram shows that stochasticity destabilizes the traffic flow, especially when the spacing is low, which is in agreement with empirical observations. Moreover, the stability phase diagram shows that the unstable region extends with the increase of the dissipation term $\sigma$ especially when the spacing is low. To visualize the effect of stochasticity in inducing traffic oscillations, we display in Figure~\ref{fig2}, the time-space diagrams corresponding to a spacing $s_{e}=13.33\ \mathrm{m}$ and different values of $\beta$. For the deterministic case, Figures~\ref{fig2}(a) and (b) show how traffic becomes unstable by decreasing the parameter $\beta$. For the stochastic case, Figure 2(d) shows that traffic is stable while fluctuations around the equilibrium persist for $\beta=1.6 \ \mathrm{s^{-1}}$. Figure~\ref{fig2}(c) shows how stochasticity can destabilize the traffic flow for $\beta=1.35 \ \mathrm{s^{-1}}$ which corresponds to stable traffic in the deterministic case. 

 
Figure~\ref{fig1} shows that theoretical analysis is in agreement with simulation. It is worth mentioning that the difference between the theoretical curve and simulation enlarges with decreasing the spacing, which is probably due to the dominating effect of non-linearities. In the limit, $s_{e}\rightarrow 0$, the theoretical prediction diverges; however, the non-linear effects dominate to bound the unstable regime \citep{wang1}. 

Moreover, Figure~\ref{fig1}(a) exhibits a comparison between the stability condition derived by \cite{ngod} (equation (40)) with the stability condition derived in this work (equation (39)). Clearly, the formula derived by \cite{ngod} deviates significantly from the one derived in this work and numerical simulation especially when the spacing is low. For a more complete comparison, we show in Figure~\ref{fig1}(b) the stability diagram in the $(\beta, \sigma^{2})$ plane by using equations (39) and (40). The numerical simulation shows that the stability condition corresponding to equation (39) outperforms the stability condition given by equation (40). After rewriting the formula (40) as the following:

\begin{equation} 
\beta \leq \frac{2{V'}^2}{V'-\mu^2}
\end{equation}

\noindent one can see that the stability condition derived by \cite{ngod} will be divided by zero for a given low spacing. As a result, the stability boundary diverges as shown in Figure~\ref{fig1}(a,b). Indeed, for a given low spacing, the derivative $V'$ is small (since the variation of equation (38) is slow for low spacing values). Thus, $V'$ would be equal to the noise factor $\mu^2$. Our analysis suggests that Lyapunov method is more suited to studying the stability of stochastic traffic models. 

The problem in the stability condition in equation (40) is related to the adopted derivation method. According to this method, the mean square stability of equation (7) is equivalent to searching the eigenvalues of $\Re{(A)}+0.5R^2$. This method has been proven only for one-dimensional stochastic differential equations. On the other hand, the eigenvalues-based stability analysis might be suitable only when the matrices $A$ and $R$ commute for exponential stability \citep{maob}. Indeed, in the last case, the solution of equation (7) has been explicitly derived, which implies the sufficient stability condition in \cite{ngod}. However, for general matrices $A$ and $R$, the eigenvalue analysis might be cumbersome or not suitable for studying the stability of stochastic differential equations, especially for a complex-valued matrix $A$.

\begin{figure}[H]
\centering
{\includegraphics[width=.8\textwidth]{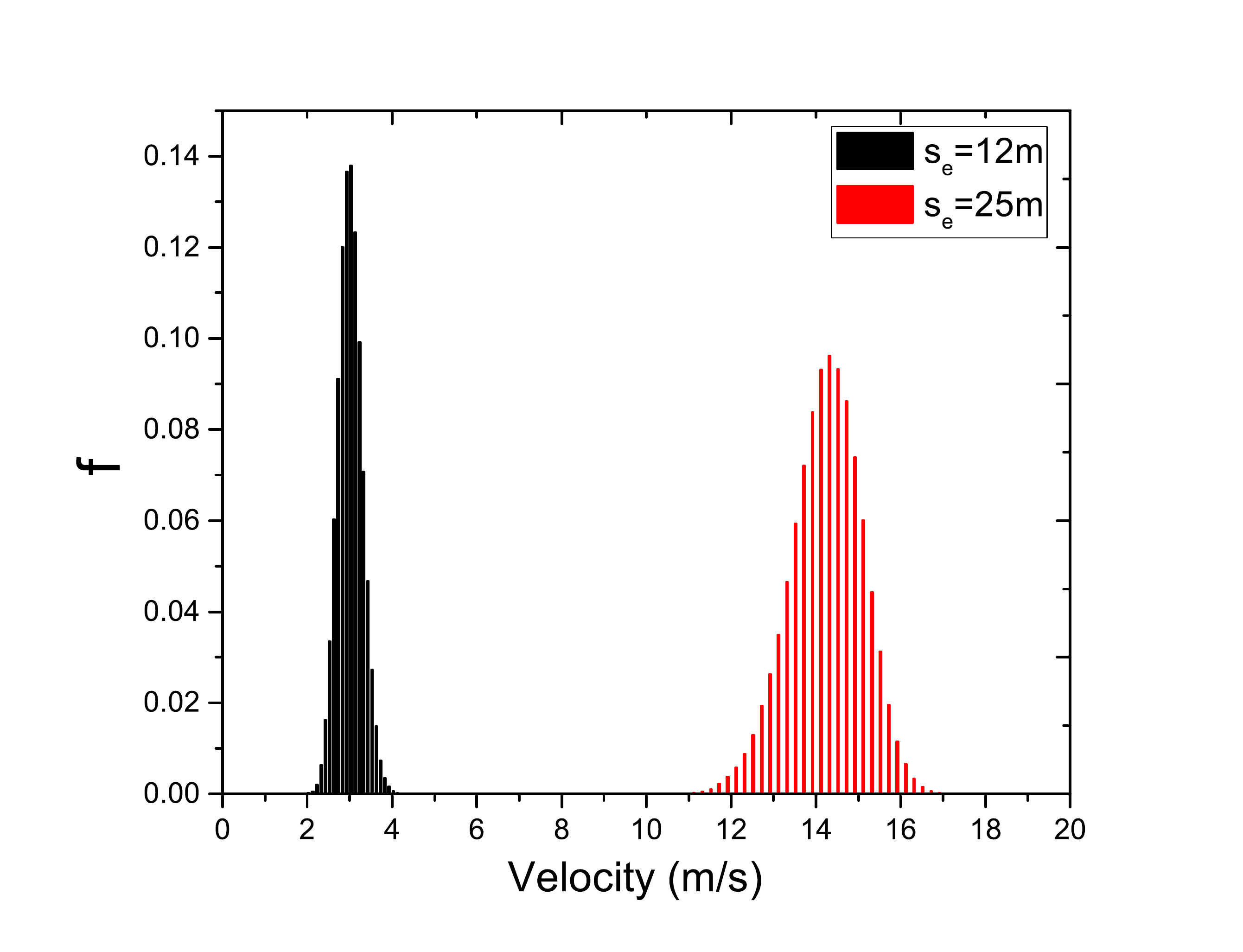}} 
\caption{Velocity distributions for a stable traffic and two spacing values $s_e$ in a stochastic environment where $\sigma^2=0.09 \ \mathrm{m/s^2}$ and $\beta=0.6 \ \mathrm{s^{-1}}$.}
\label{fig3}
\end{figure}

Next, we present in Figure~\ref{fig3}, the velocity distributions for two spacing values in stable traffic. Both distributions exhibit a normal distribution which becomes sharper when the spacing is low. This is qualitatively in agreement with empirical observations \citep{hel2}.

\subsection{The stochastic full velocity difference model}

In the OVM, accidents might occur for a wide range of realistic values of $\beta$. To circumvent this shortcoming, \cite{jiang4} considered the sensitivity of drivers to the velocity difference and proposed the FVDM. The stochastic version of the FVDM (SFVDM) reads:

\begin{equation} 
dv_{n}(t)=\beta(V_{op}(s_{n}(t))-v_{n}(t))dt+\lambda \Delta v dt+\sigma \sqrt{v_{n}(t)} {dW}_{n}
\end{equation}

\noindent where $\Delta v= v_{n} - v_{l}$,  $v_{l}$ is the velocity of the leader of the vehicle $n$.

From equations (21) and (42), the stability condition of the SFVDM is given by: 

\begin{equation} 
V'<\frac{1}{2}(\beta+2\lambda)(1-\frac{\mu^2}{2\beta})
\end{equation}

\noindent Note that in the deterministic case $(\mu=0)$, $V'<\frac{1}{2}(\beta+2\lambda)$ \citep{jiang4}. 

\begin{figure}[H]
\centering
{\includegraphics[width=.8\textwidth]{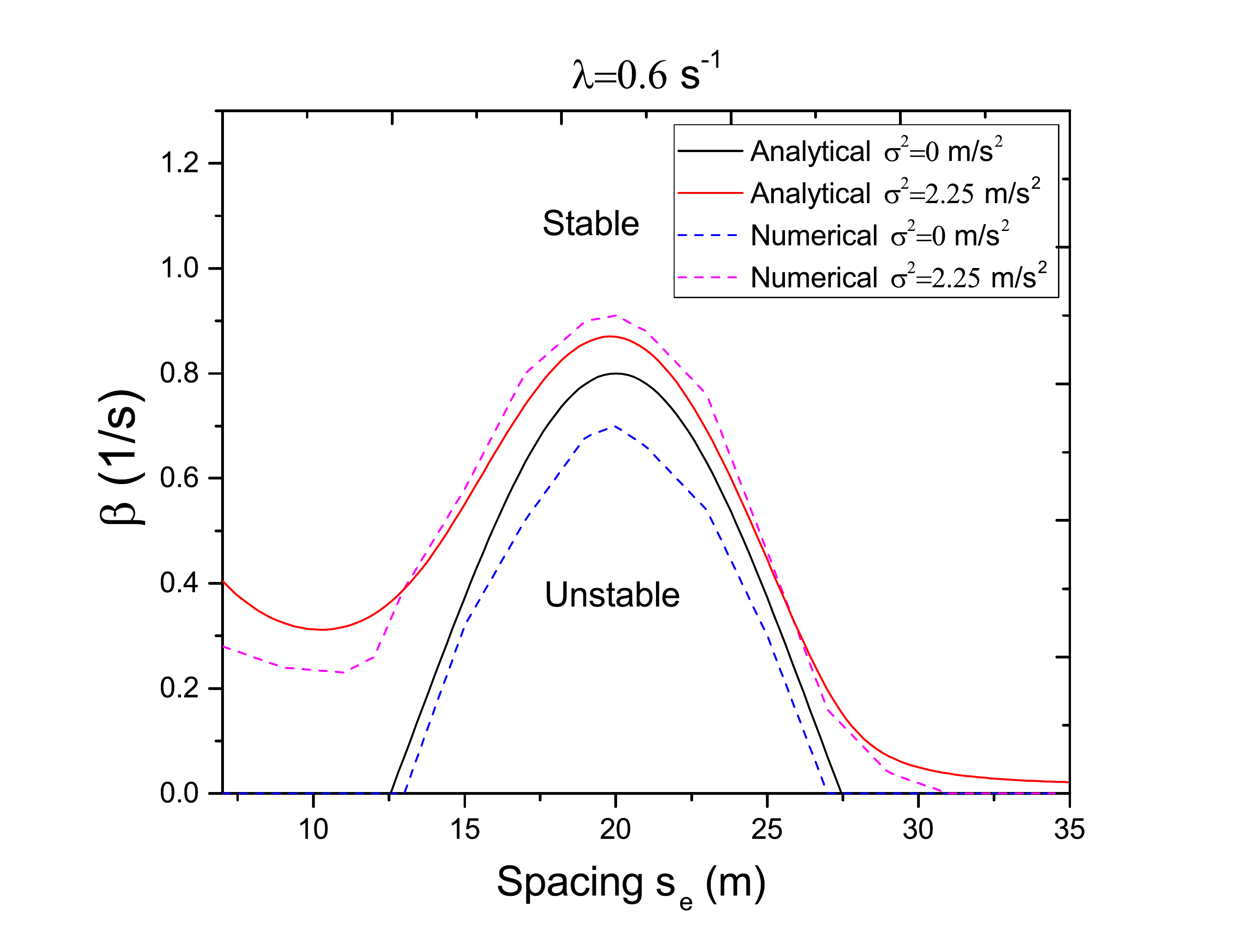}}
\caption{Stability phase diagram of the SFVDM in the $(\beta, s_{e})$ plane.}
\label{fig4}
\end{figure}

The relation (43) suggests that the impact of stochasticity becomes more pronounced when taking into account the sensitivity of drivers to the velocity difference through the parameter $\lambda$. To visualize the impact of stochasticity in the SFVDM, we plot in Figure~\ref{fig4}, the stability phase diagram in the $(\beta,s_{e})$ plane for a fixed value of $\lambda=0.6 \ \mathrm{s^{-1}}$.  

\begin{figure}[H]
\centering
\subfloat[]{\includegraphics[width=.4\textwidth]{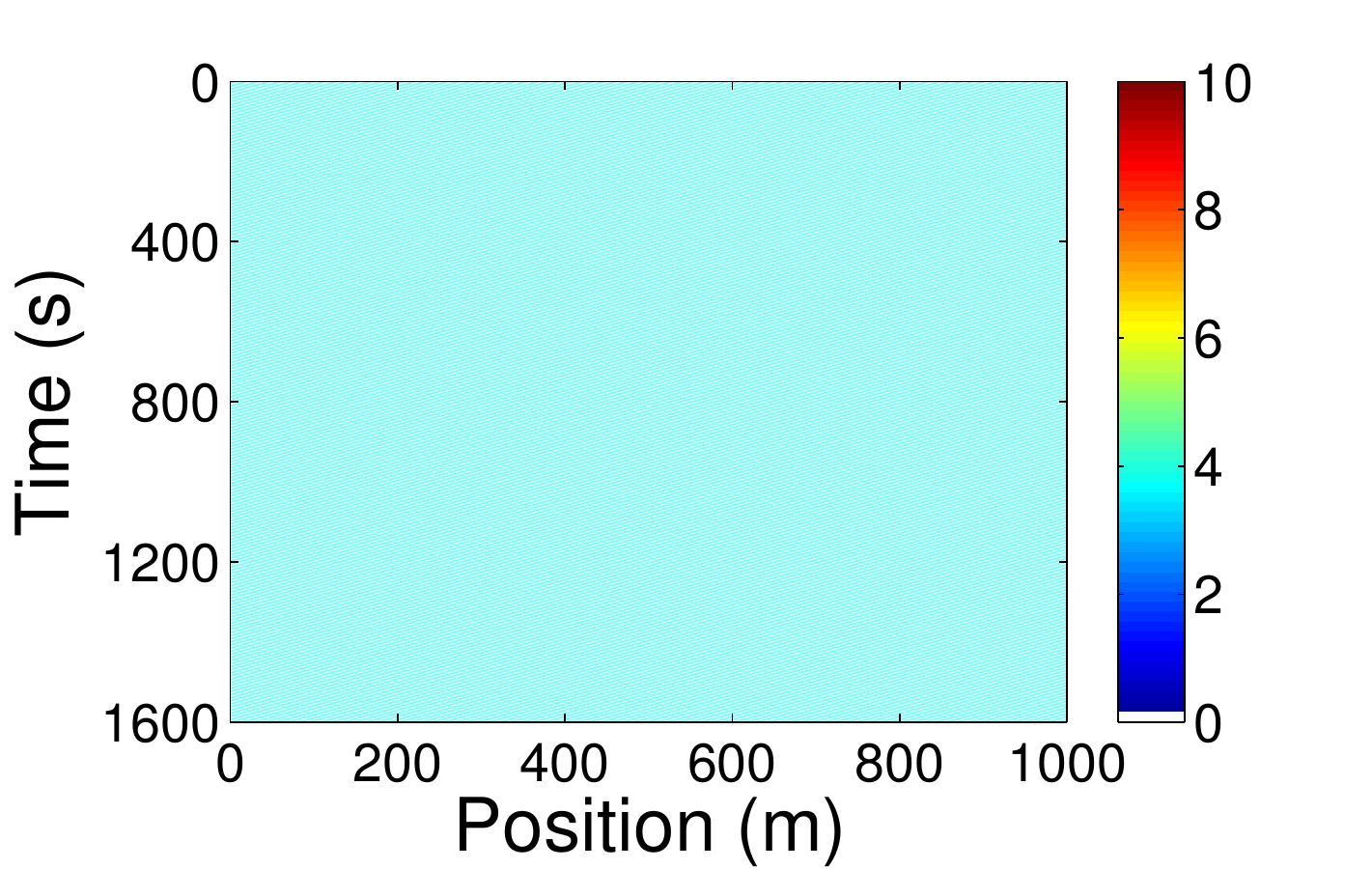}}
\subfloat[]{\includegraphics[width=.4\textwidth]{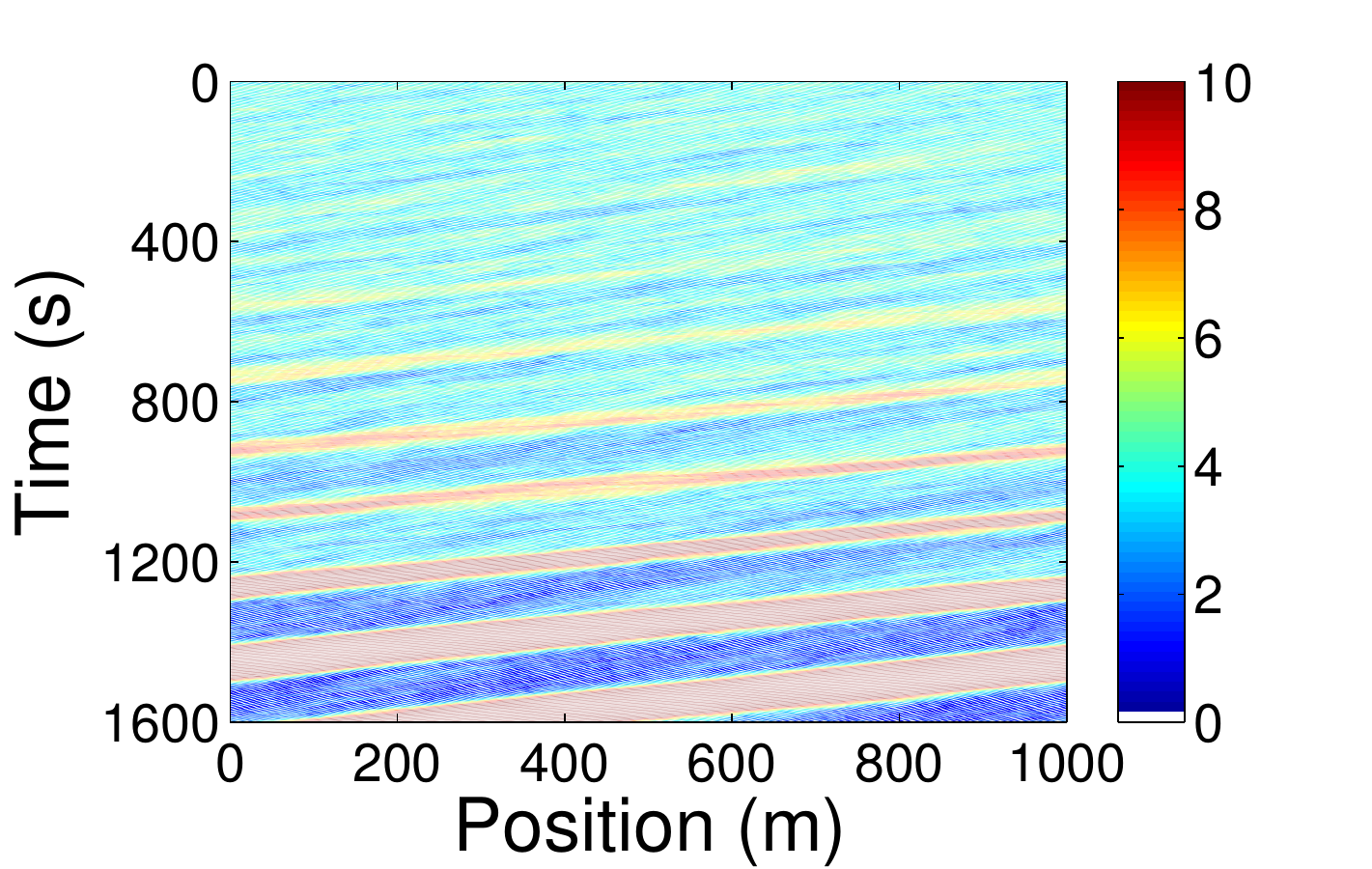}}\\
\caption{Time-space diagrams corresponding to stable and unstable regime of the SFVDM model and $s_{e}=13.33\ \mathrm{m}$ (a) Stable for $\sigma^2=0  \ \mathrm{m/s^2}$, $\beta=0.2  \ \mathrm{s^{-1}}$ (b) Unstable for $\sigma^2=0.36 \ \mathrm{m/s^2}$, $\beta=0.2  \ \mathrm{s^{-1}}$. The velocity unit in the color bar is $\ \mathrm{(m/s)}$.}
\label{fig5}
\end{figure}

\begin{figure}[H]
\centering
{\includegraphics[width=.8\textwidth]{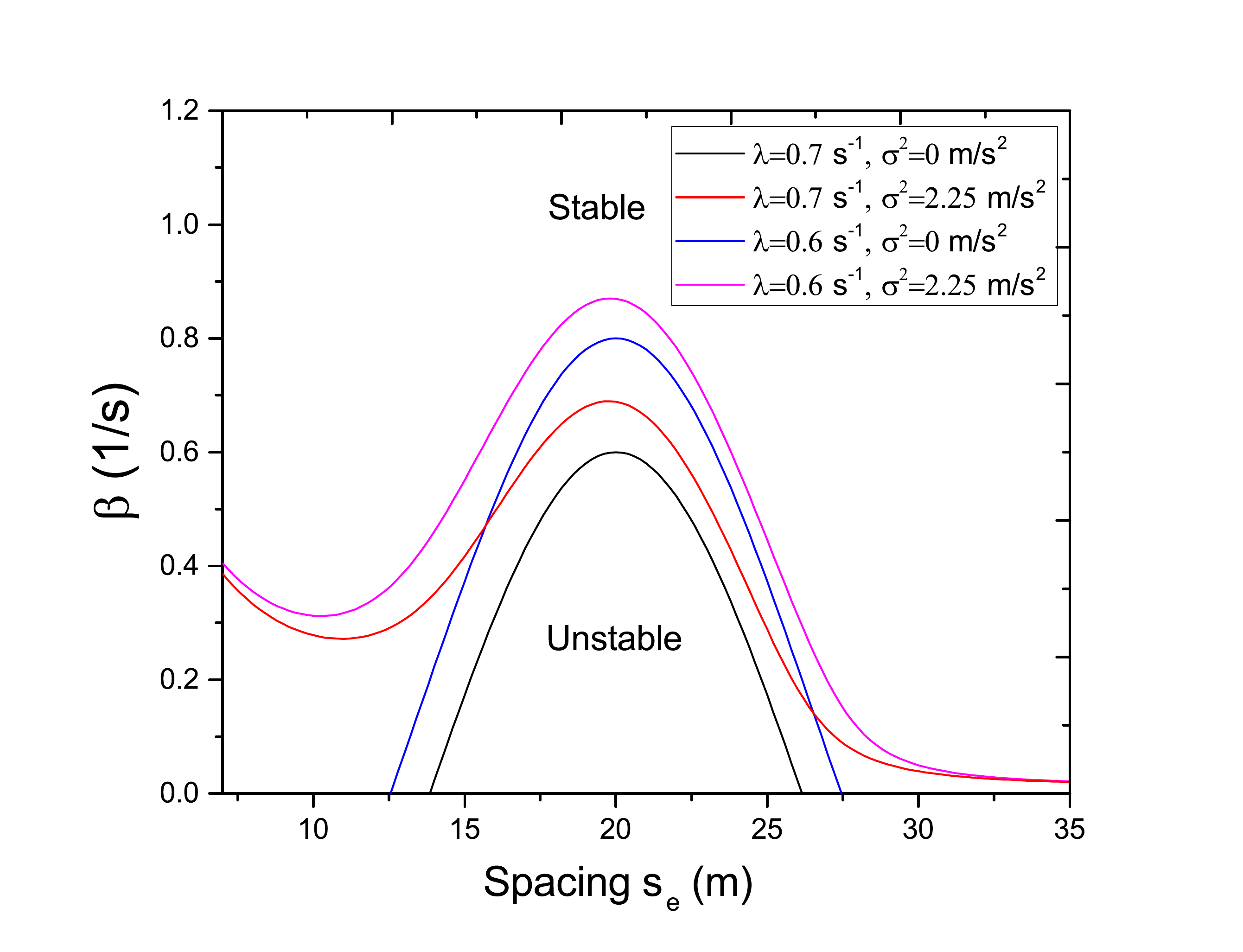}} 
\caption{Stability phase diagram of the SFVDM for two values of the velocity difference sensitivity $\lambda$.}
\label{fig6}
\end{figure}

Figure~\ref{fig4} shows that stochasticity destabilizes traffic flow significantly for $\sigma^2=2.25 \ \mathrm{m/s^2}$, especially for low spacing values. The destabilizing effect of stochasticity for high spacing values is not negligible. Moreover, the theoretical stability phase diagram is in good agreement with numerical simulation. The time-space diagrams in Figure~\ref{fig5} show how the presence of stochasticity can induce traffic instability for fixed values of $\lambda=0.6 \ \mathrm{s^{-1}}$ and $\beta=0.2 \ \mathrm{s^{-1}}$. Finally, to visualize the effect of the sensitivity parameter $\lambda$, we have plotted in Figure~\ref{fig6}, the stability diagram corresponding to two values of $\lambda$. One can see that a lower sensitivity enlarges the unstable region for both the deterministic and the stochastic cases.


\subsection{The stochastic intelligent driver model}

The intelligent driver model (IDM) was developed by \cite{hel1}. The stochastic IDM (SIDM) considered in this work reads: 

\begin{equation} 
dv_{n}(t)=a[1-{(\frac{v}{V_{max}})}^{\delta}- (\frac{s^{*}(v_{n},\Delta v)}{s_{n}})^{2}]+\sigma \sqrt{v_{n}(t)} {dW}_{n}
\end{equation}

\noindent where $s^{*}$ is the desired gap and given by:

\begin{equation} 
s^{*} (v_{n},\Delta  v)=s_{0}+v_{n} T+\frac{v \Delta v}{2\sqrt{ab}},
\end{equation}

\noindent $a$ and $b$ are respectively the desired acceleration and deceleration of vehicles, $T$ is the time headway, $s_{0}$ is the minimum spacing, $s_{n}=x_{l}-x_{n}-l$ is the inter-vehicular gap between two successive vehicles and $l$ is the vehicle's length. For simplicity, we consider that $\delta \rightarrow \infty$, thus, the equilibrium velocity reads: 

\begin{equation} 
v_{e}=\min(V_{max},\frac{s_{e}-s_{0}}{T}),
\end{equation}

The stability condition of the present stochastic version of the IDM is given by equation (35) where $\tilde{\alpha}_{1}$, $\tilde{\alpha}_{2}$ and $\tilde{\alpha}_{3}$ read:

\begin{equation} 
\tilde{\alpha}_{1}=\frac{2a{(s_{0}+Tv_{e})}^{2}}{s_{e}^3}
\end{equation}

\begin{equation} 
\tilde{\alpha}_{2}=\frac{-2aT{(s_{0}+Tv_{e})}}{s_{e}^2}
\end{equation}

\begin{equation} 
\tilde{\alpha}_{3}=\frac{-a v_{e}{(s_{0}+Tv_{e})}}{s_{e}^2 \sqrt{ab}}
\end{equation}

\noindent In the deterministic case, the stability condition becomes: 

\begin{equation} 
s_{e} <a T^{2} +  v_{e} T \sqrt{\frac{a}{b}}
\end{equation}

\begin{figure}[H]
\centering
{\includegraphics[width=.7\textwidth]{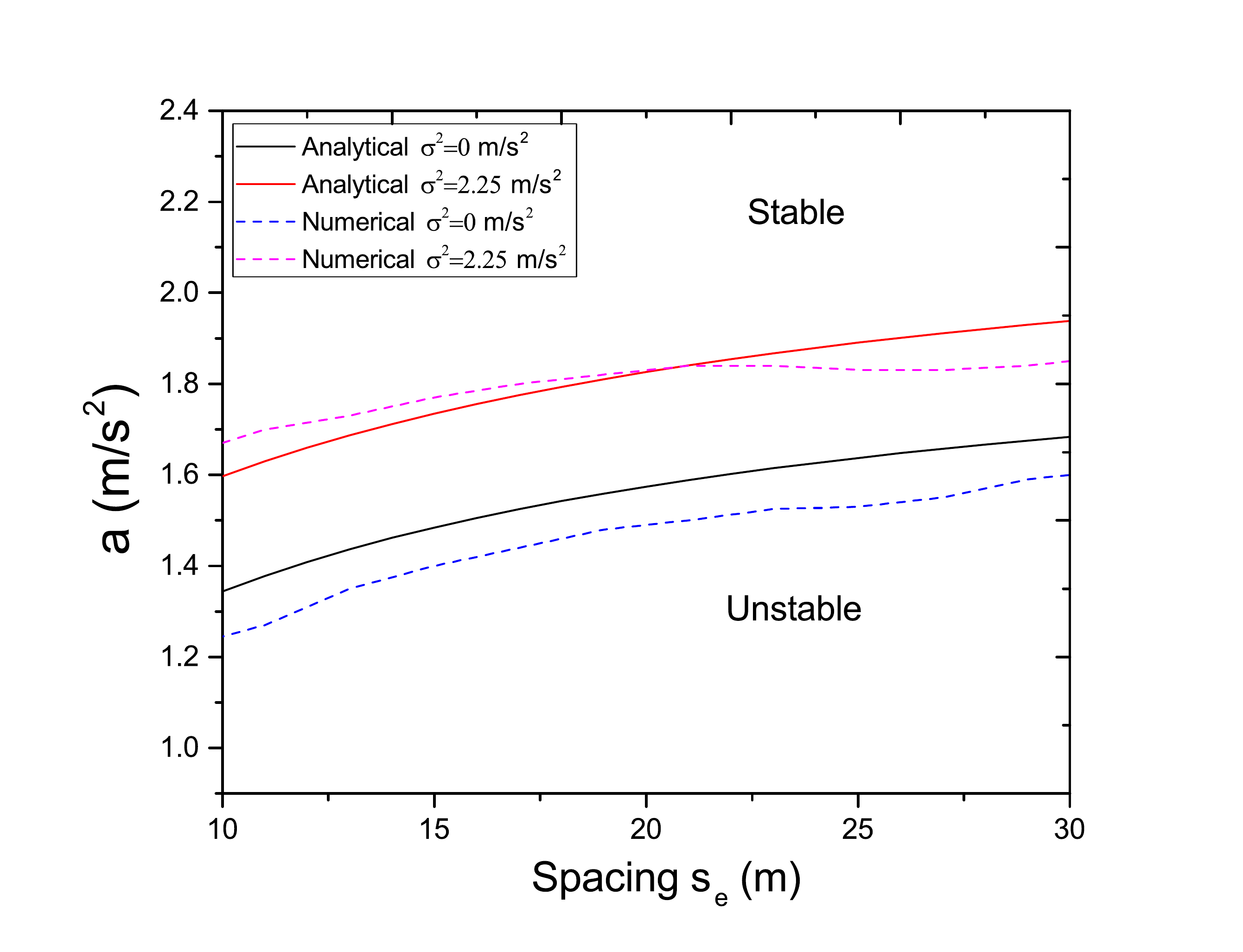}}\\ 
\caption{Stability phase diagram of the SIDM in the $(a,s_{e})$ plane.}
\label{fig7}
\end{figure}

\begin{figure}[H]
\centering
\subfloat[]{\includegraphics[width=.7\textwidth]{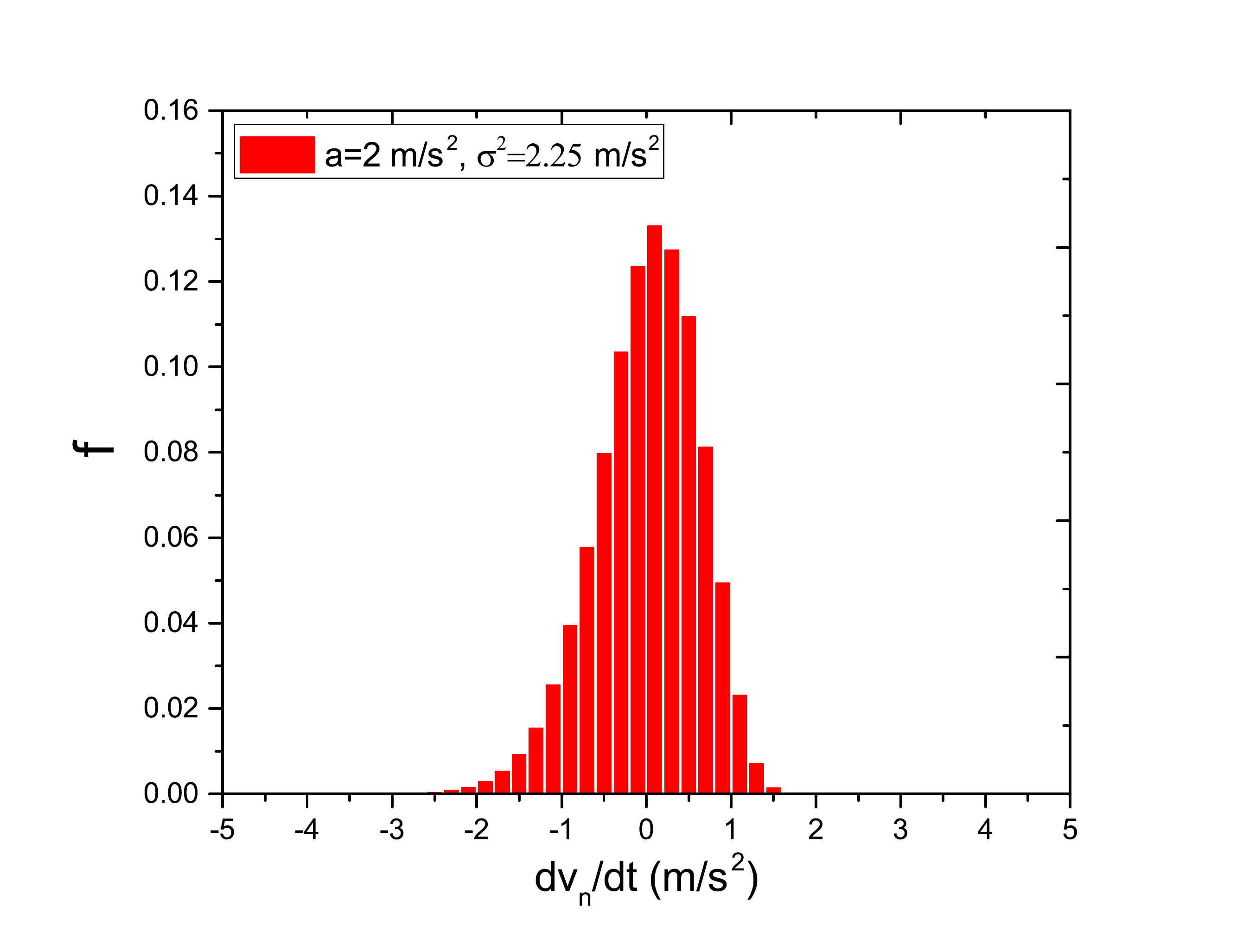}}\\  
\subfloat[]{\includegraphics[width=.7\textwidth]{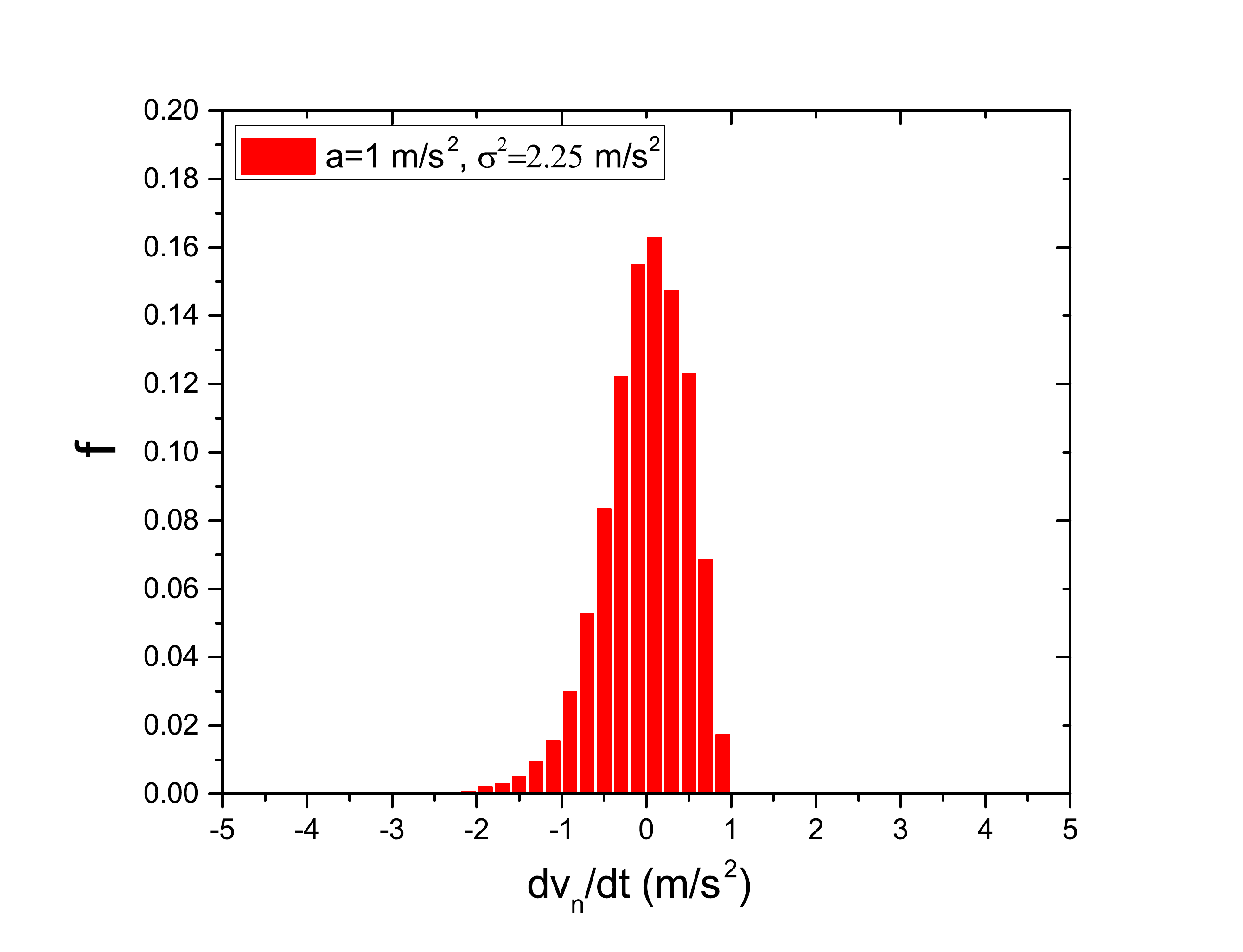}} 
\caption{Acceleration distributions of the SIDM for two acceleration parameter values (a) $a=2 \ \mathrm{m/s^{2}}$ (b) $a=1 \ \mathrm{m/s^{2}}$.}
\label{fig8}
\end{figure}

\begin{figure}[H]
\centering
\subfloat[]{\includegraphics[width=.4\textwidth]{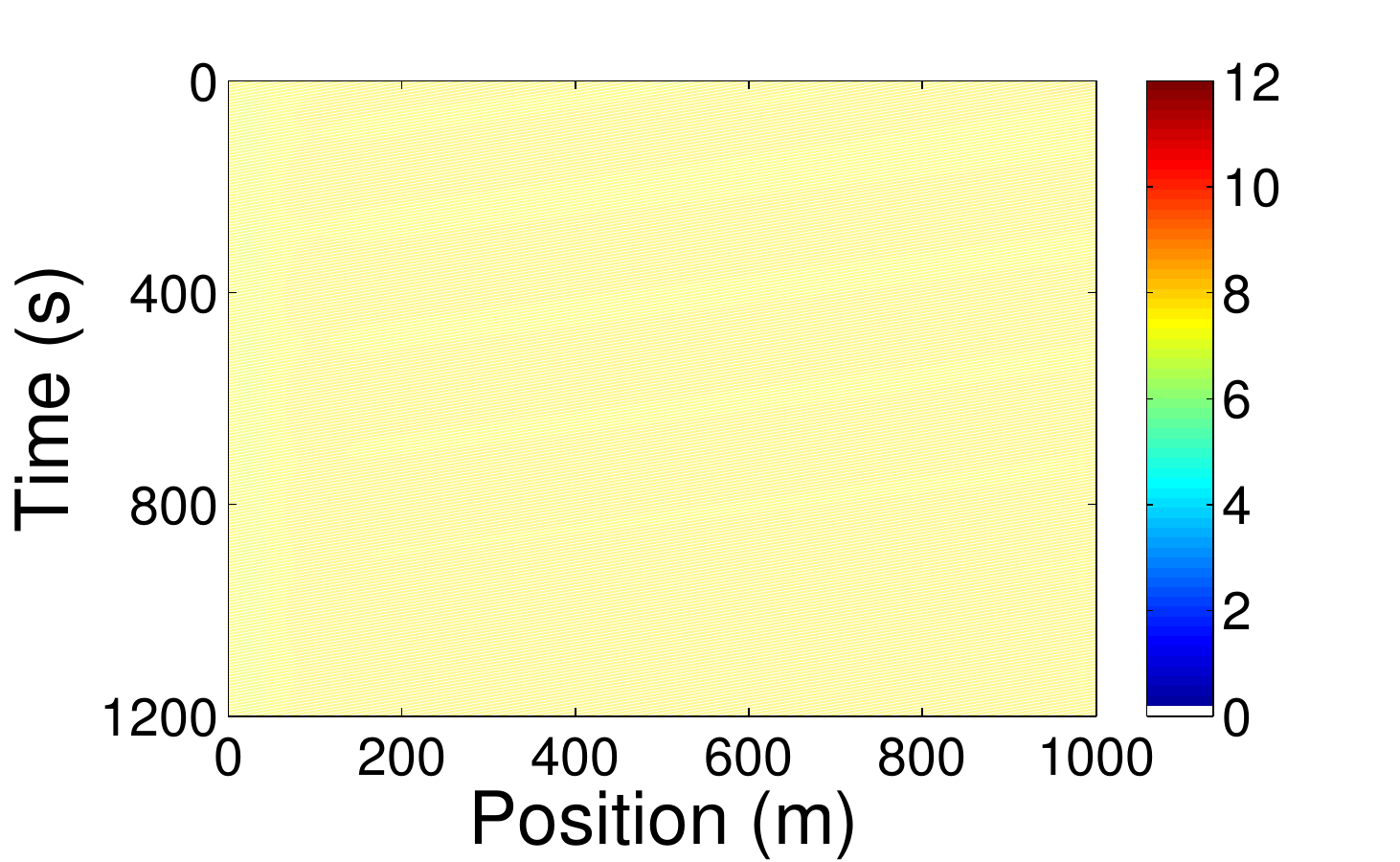}} 
\subfloat[]{\includegraphics[width=.4\textwidth]{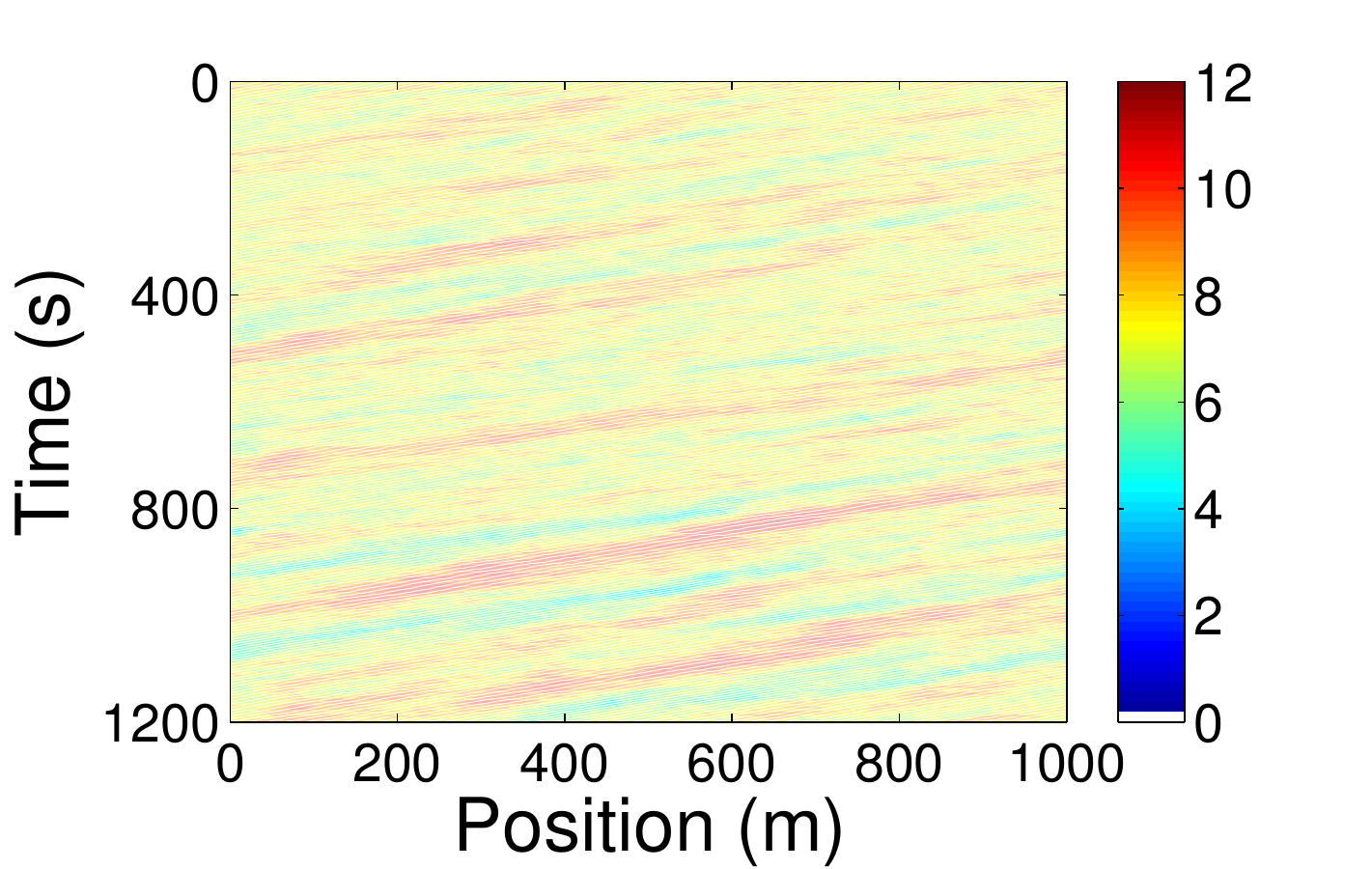}}\\
\caption{Time-space diagrams corresponding to stable and unstable regime of the SIDM model (a) Stable for $\sigma^2=0 \ \mathrm{m/s^2}$, $a=1.5 \ \mathrm{m/s^{2}}$ (b) Unstable for $\sigma^2=0.1 \ \mathrm{m/s^2}$, $a=1.5 \ \mathrm{m/s^{2}}$. The velocity unit in the color bar is $\ \mathrm{(m/s)}$.}
\label{fig9}
\end{figure}

Next, we plot in Figure~\ref{fig7}, the corresponding stability phase diagram in the $(a,s_{e})$ plane. Clearly, the presence of stochasticity destabilizes the traffic flow. Probably, the drivers need to be more aggressive in a stochastic environment to stabilize traffic flow. To show this, we display in Figure~\ref{fig8}(a,b) the acceleration distribution for a spacing $s_{e}=20\ \mathrm{m}$ and two acceleration parameter values $a$ before the instability occurs. A comparison between $a=2\ \mathrm{m/s^{2}}$ and $a=1 \ \mathrm{m/s^{2}}$ suggests that high accelerations are more frequent for $a=2\ \mathrm{m/s^{2}}$ than $a=1\ \mathrm{m/s^{2}}$. Next, the time-space diagram in Figure~\ref{fig9}(b) shows how stochasticity induces traffic oscillations while traffic is stable in the deterministic case (Figure~\ref{fig9}(a)). 

\begin{figure}[H]
\centering
\subfloat[]{\includegraphics[width=.7\textwidth]{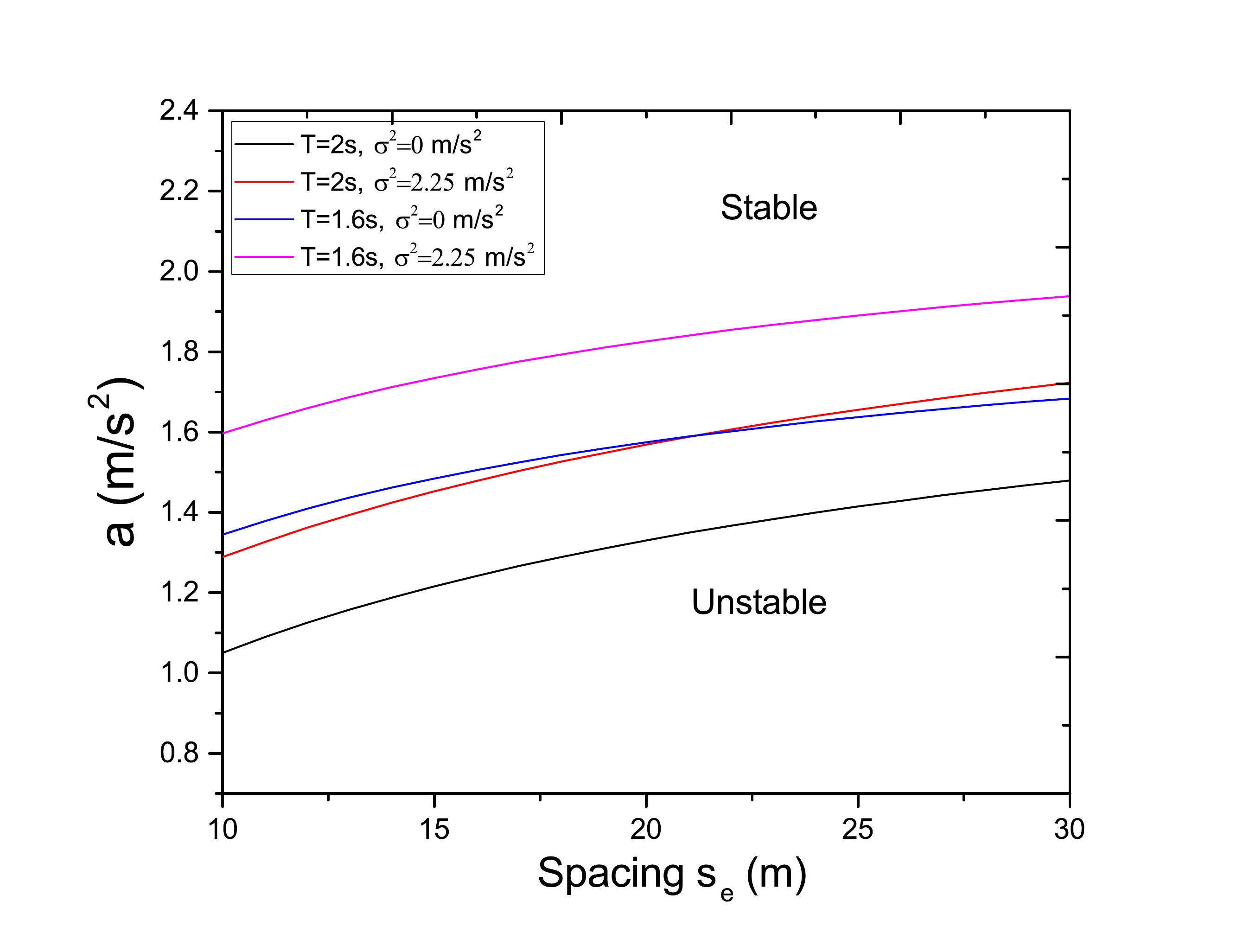}}\\ 
\subfloat[]{\includegraphics[width=.7\textwidth]{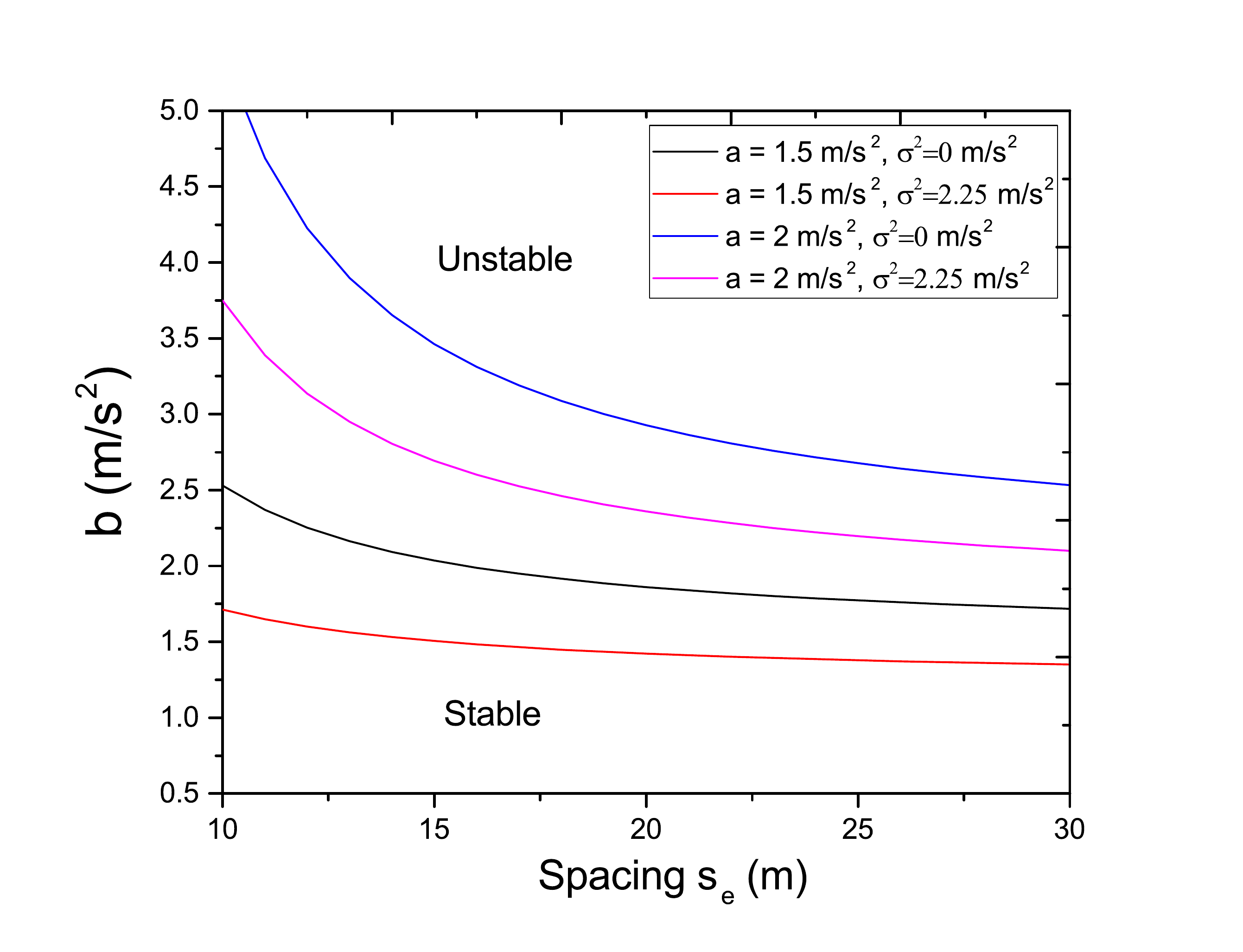}}
\caption{Stability phase diagram of the SIDM for different values of $T$ and deceleration parameter $b$ (a) Effect of the parameter $T$ (b) Effect of the parameter $b$.}
\label{fig10}
\end{figure}

Figure~\ref{fig10}(a) shows the stability phase diagram by varying the safe time headway $T$ for two acceleration parameter values $a$. It is shown that traffic flow is prone to be unstable when the safe time headway is relatively small, and the high acceleration parameter $a$ always plays a positive role. Indeed, small safe time headway may easily trigger the generation of perturbations. The presence of stochasticity increases the threshold of $T$ above which traffic becomes stable. Next, to visualize the effect of the deceleration parameter $b$ on traffic stability, we plot in Figure~\ref{fig10}(b) the stability diagram in the ($b,s_{e}$) plane for two constant values of $a$. Figure~\ref{fig10}(b) suggests that traffic is prone to be unstable when the deceleration parameter $b$ is relatively high. Indeed, a high deceleration parameter $b$ induces high deceleration patterns which might easily trigger perturbations. Clearly, the stochasticity has a destabilizing effect and requires $b$ to be smaller than the deterministic case for stable traffic.

It is worth mentioning that \cite{jiang3} developed a more plausible stochastic IDM called 2D-IDM. The proposed stochastic traffic model takes into account the fact that the velocity-spacing spans over a 2D plane. The 2D-IDM has shown not only qualitative but also quantitative agreement with experiments. In the 2D-IDM, the safe time headway is a stochastic quantity which implies (from equations (44) and (45)) that stochasticity is dependent on both the velocity and spacing in a non-linear way which makes the analytical analysis intractable.

\section{Calibration and validation against empirical data}

To assess the performance of the different stochastic traffic models and study to what extent they can reproduce the observed traffic instability, we calibrate and validate the stochastic traffic models and their corresponding deterministic case against empirical data. For this purpose, we first consider the circular road experiment performed by \cite{wu} and the NGSIM trajectory data. Finally, to assess the potential of the stochastic models to reproduce the concave growth pattern of traffic oscillations, we will consider the 51 car-platoon experiment conducted by \cite{jiang2}. 

\subsection {The circular road experiment}

Recently, \cite{wu} conducted circular road experiments with different traffic densities. The length of the road is $L=260$ m, and the average vehicle's length is approximately $l=5\ \mathrm{m}$. Table~\ref{t1} presents the traffic densities and the used time interval. More details can be found in \cite{wu}.

\begin{table*}[h] \footnotesize
	\caption{ Density and calibration time interval of each experiment. }
\centering
	 \begin{tabular}{cccccc}
\textbf{Experiment}	& \textbf{Density (veh/m)} & \textbf{Calibration time interval (s)}  \\

     \hline
         A			& 0.077   & [50,400] \\
				
		\hline
    B			 & 0.077 & [50,400] \\
		\hline
    C		  & 0.085 &  [50,350] \\
		\hline
		  D		  & 0.081 &  [50,450] \\
		\hline
		E		  & 0.073 & [50,300]  \\
		\hline

\end{tabular}

	\label{t1}
\end{table*}

To carry out the calibration task, we consider the typical stochastic traffic models studied in the previous section and their corresponding deterministic case, namely, the SOVM, the SFVDM, the SIDM, the OVM, the FVDM, and the IDM. The parameters to be calibrated for each traffic model are reported in Table~\ref{t2}.

The performance index used in this work is a summation of two terms. The first term deals with minimizing the error between simulations and observations for the speed time series, while the second term deals with the speed standard deviation. This method is similar to the one adopted by \cite{zheng2}. We present in Appendix B the critical role of the speed standard deviation term in increasing the calibration quality of stochastic traffic models. Hence, in this work, we use the genetic algorithm \citep{mat} to minimize the following error between simulations and empirical data:

\begin{equation}
I^{2} =\frac{1}{NT_{m}} \sum_{i=1}^{i=T_{m}}\sum_{k=1}^{k=N} {(v_{ik} - \hat{v}_{ik})}^{2} + \frac{1}{N}\sum_{k=1}^{k=N}{(s_{k} - \hat{s}_{k})}^{2}  
\end{equation} 

\noindent $T_{m}$ is the maximum time step, $N$ is the number of vehicles, $v_{ik}$ is the simulated velocity of a vehicle $k$ at time step i for one initial configuration, $\hat{v}_{ik}$ is the measured velocity of a vehicle $k$ at time step $i$, $s_{k}$ and $\hat{s}_{k}$ are respectively the simulated and the measured velocity standard deviation of a vehicle $k$ during the simulation time. The performance index in equation (51) aims to minimize both the velocity and the velocity standard deviation errors. We average the performance index over 20 initial configurations. We use the densities corresponding to experiments A, C, and E for calibration and experiments B and D for validation.    

Table~\ref{t2} reports the calibration result of each traffic model. Table~\ref{t3} shows the corresponding performance index and squared speed standard deviation error. The calibration result of each stochastic model in Table 2 indicates that the level of stochasticity $\sigma$ is non-negligible. Moreover, from Table~\ref{t3}, one can see that the overall performance index corresponding to the stochastic models is, in the majority of cases, smaller than its deterministic counterpart. Nevertheless, the performance index corresponding to the deterministic models is slightly smaller than the stochastic models in three cases. For example, in experiment C, the performance index of the IDM is 0.63, which is smaller than 0.69 for the SIDM model. However, a comparison of the time-space diagrams in Figure~\ref{fig13}(e) and Figure~\ref{fig13}(f) shows that the development of traffic oscillations in the SIDM better reproduces the experimental observations. Similar results can be observed in the two other cases (the IDM/SIDM in experiment D, and the OVM/SOVM in experiment B). 

\begin{figure}[p]
\centering
\begin{tabular}{cc} 
\centering
&\subfloat[]{\includegraphics[width=.25\textwidth]{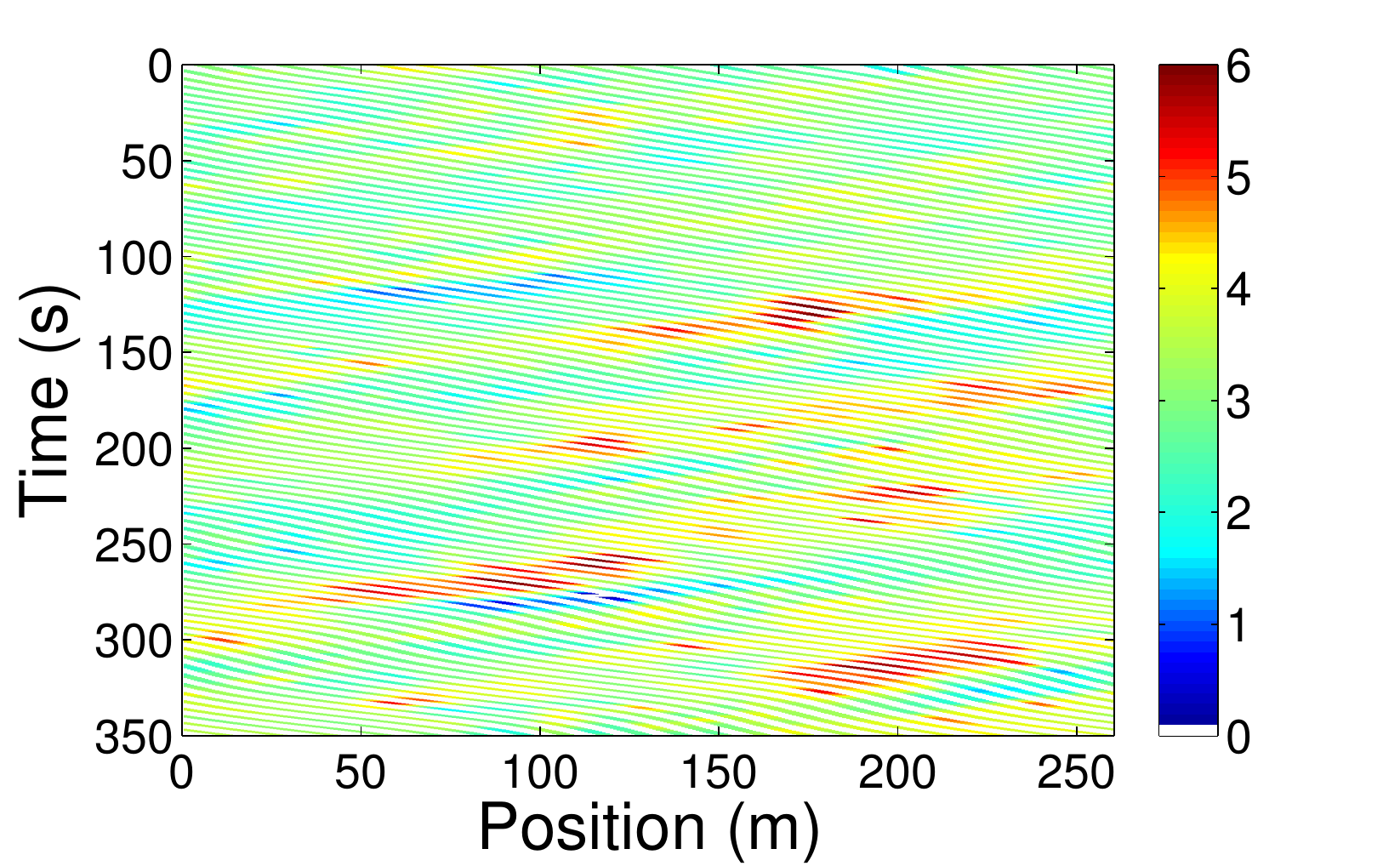}} 
\subfloat[]{\includegraphics[width=.25\textwidth]{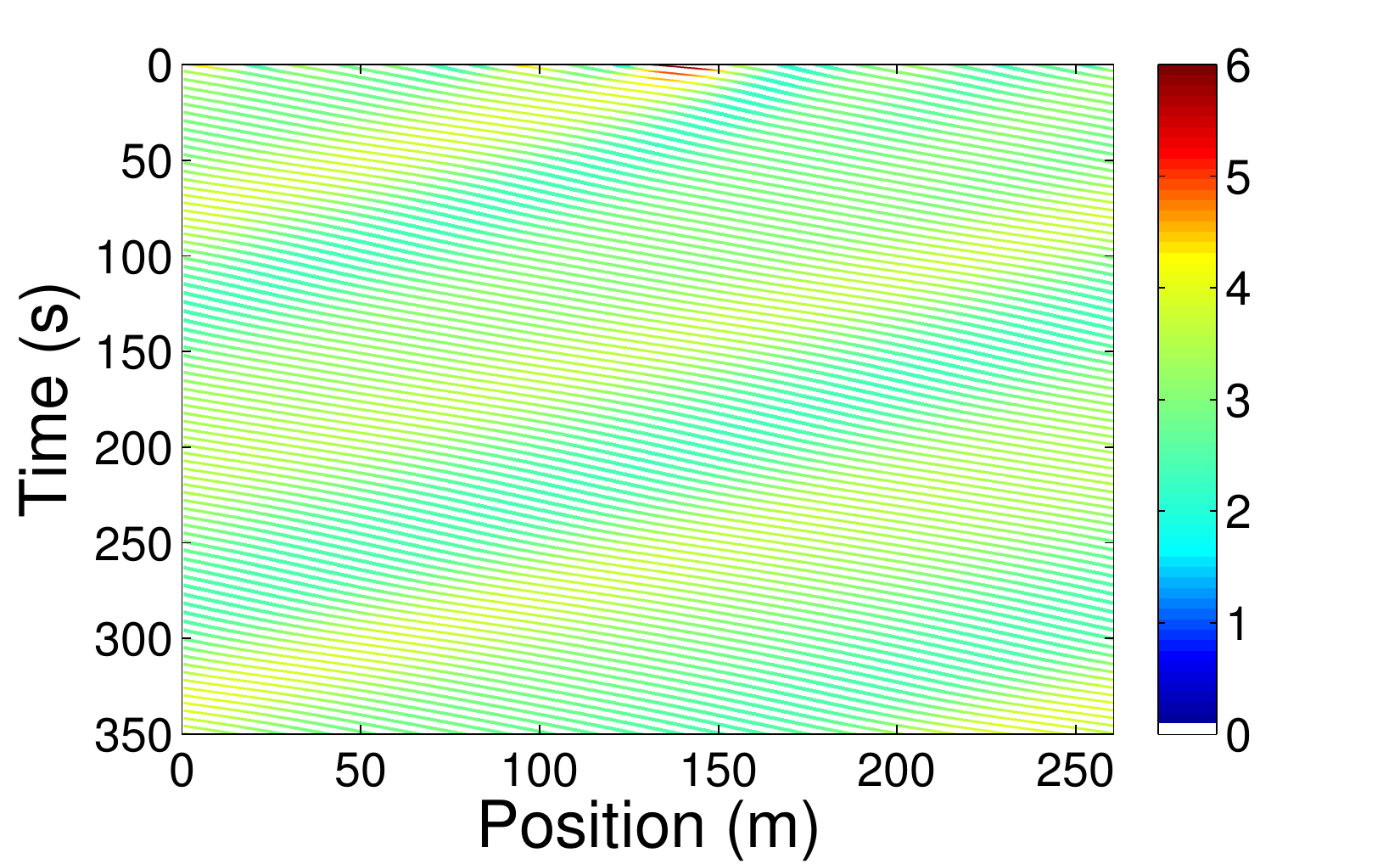}}
\subfloat[]{\includegraphics[width=.25\textwidth]{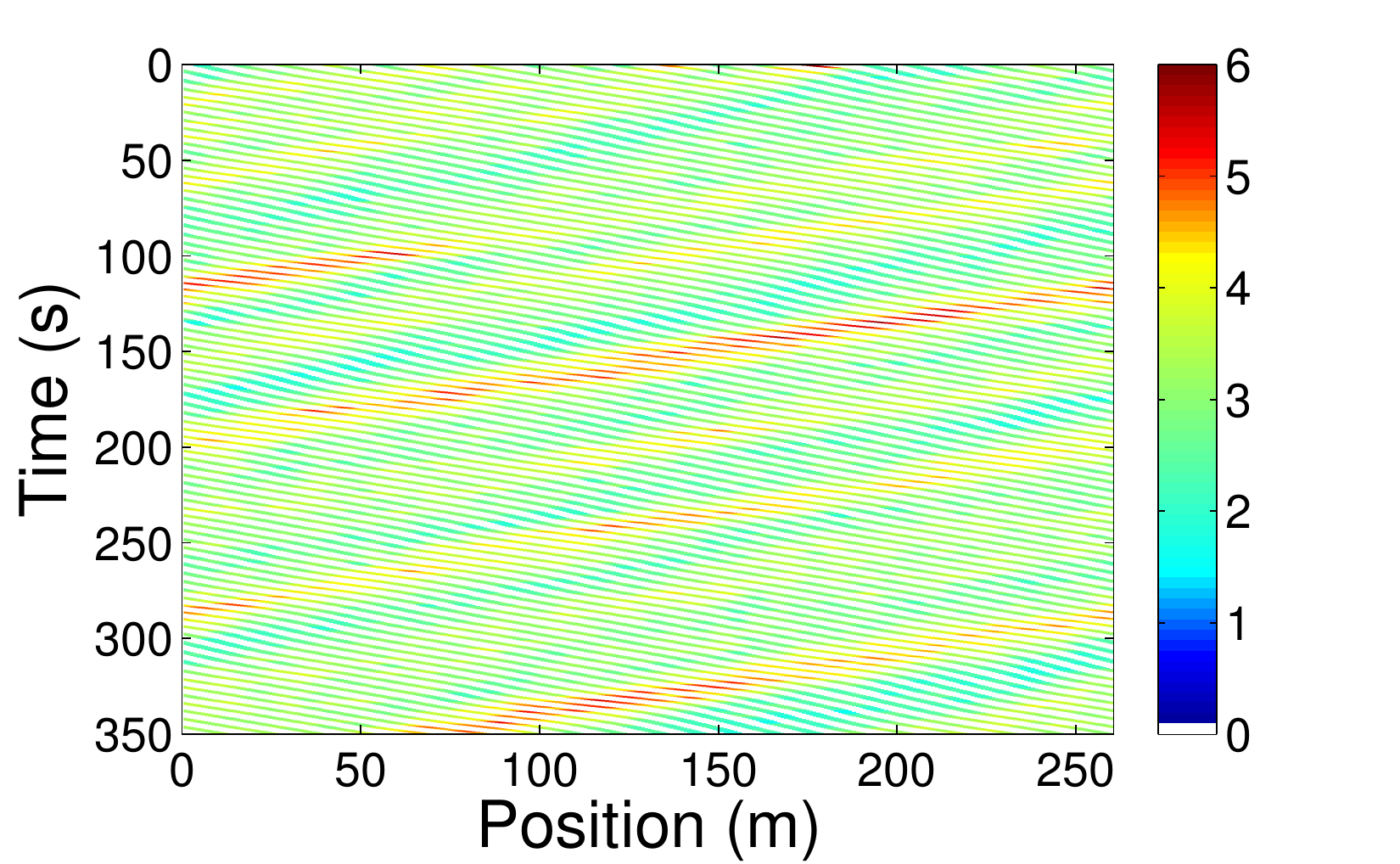}}\\
&\subfloat[]{\includegraphics[width=.25\textwidth]{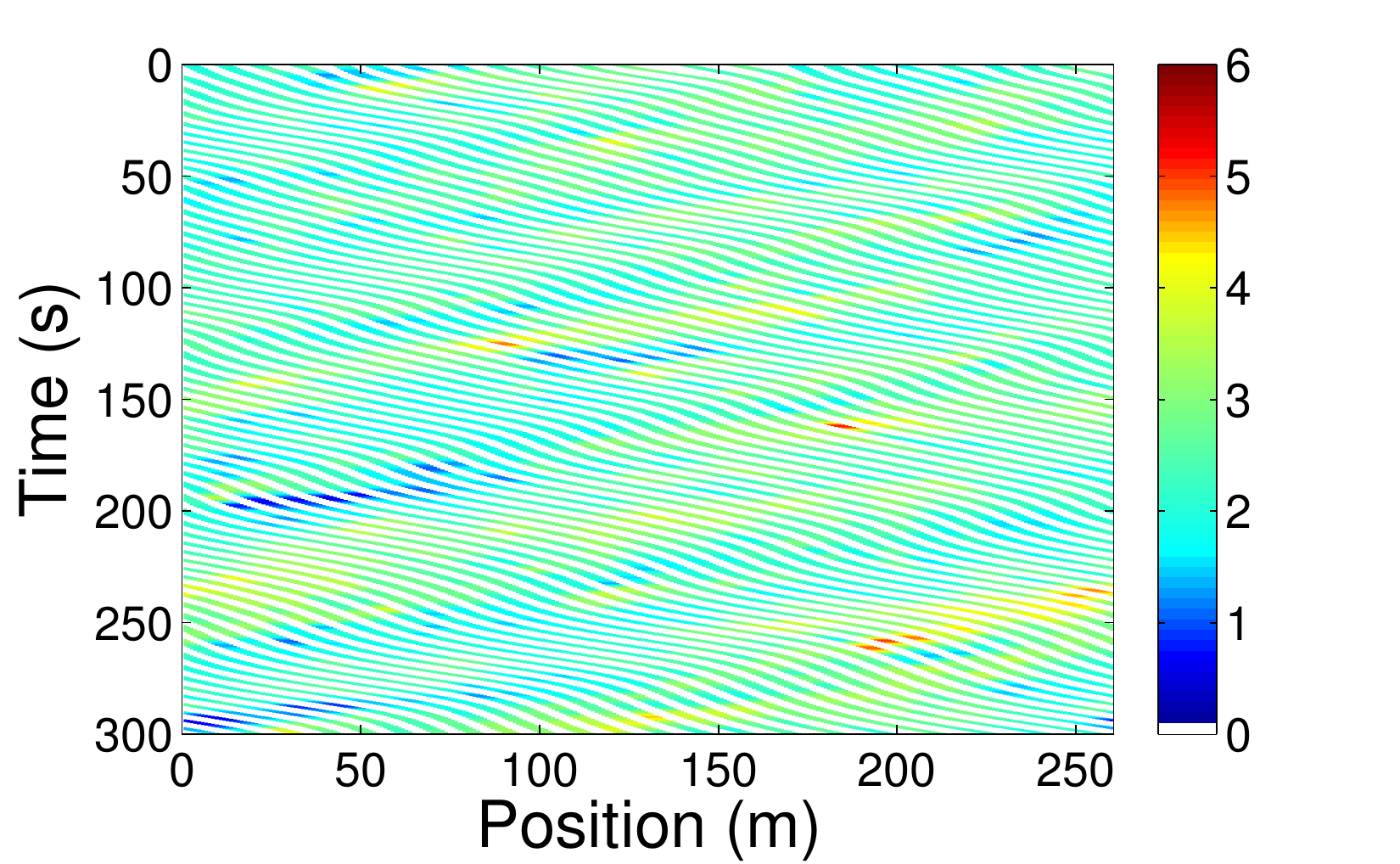}} 
\subfloat[]{\includegraphics[width=.25\textwidth]{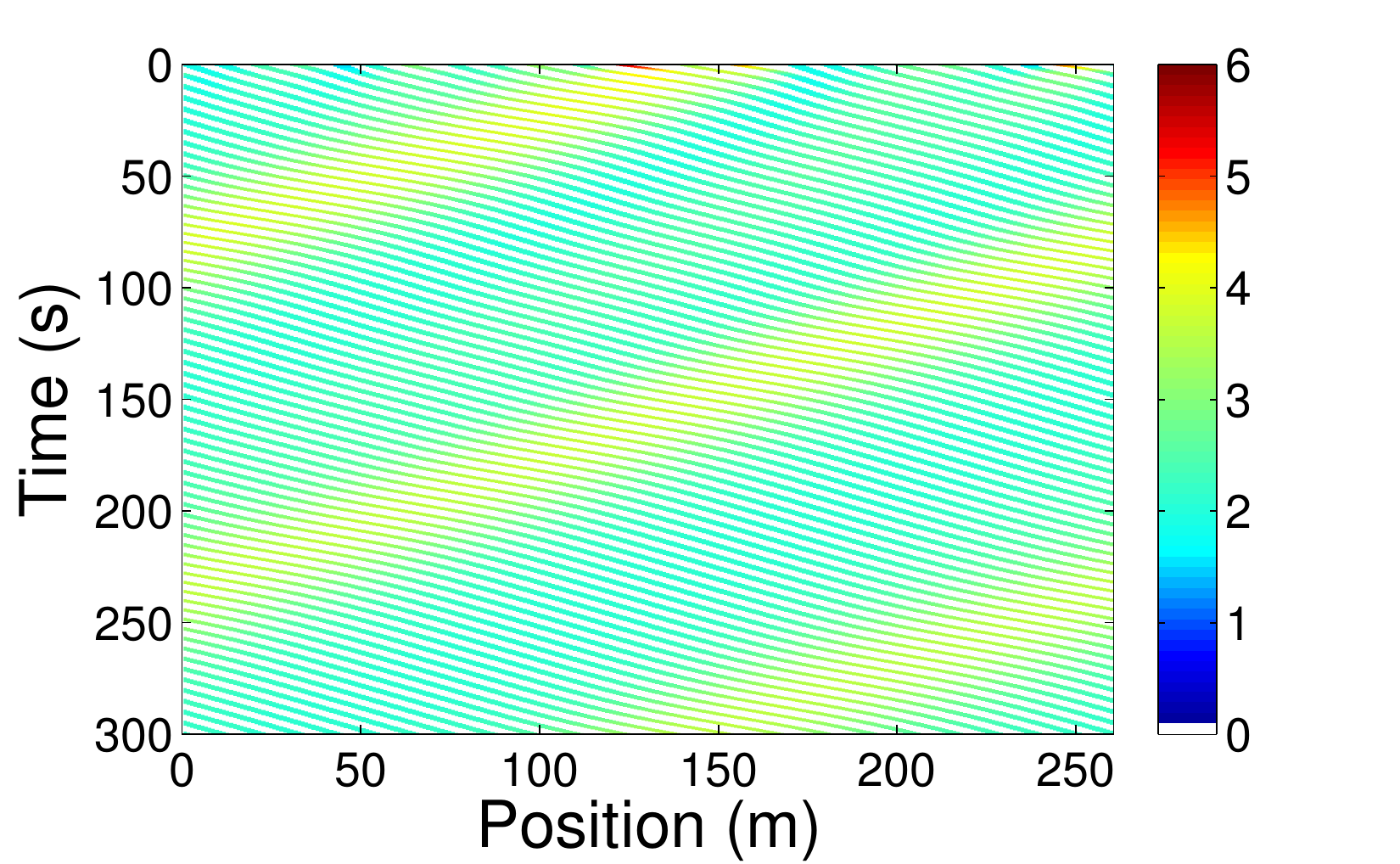}}
\subfloat[]{\includegraphics[width=.25\textwidth]{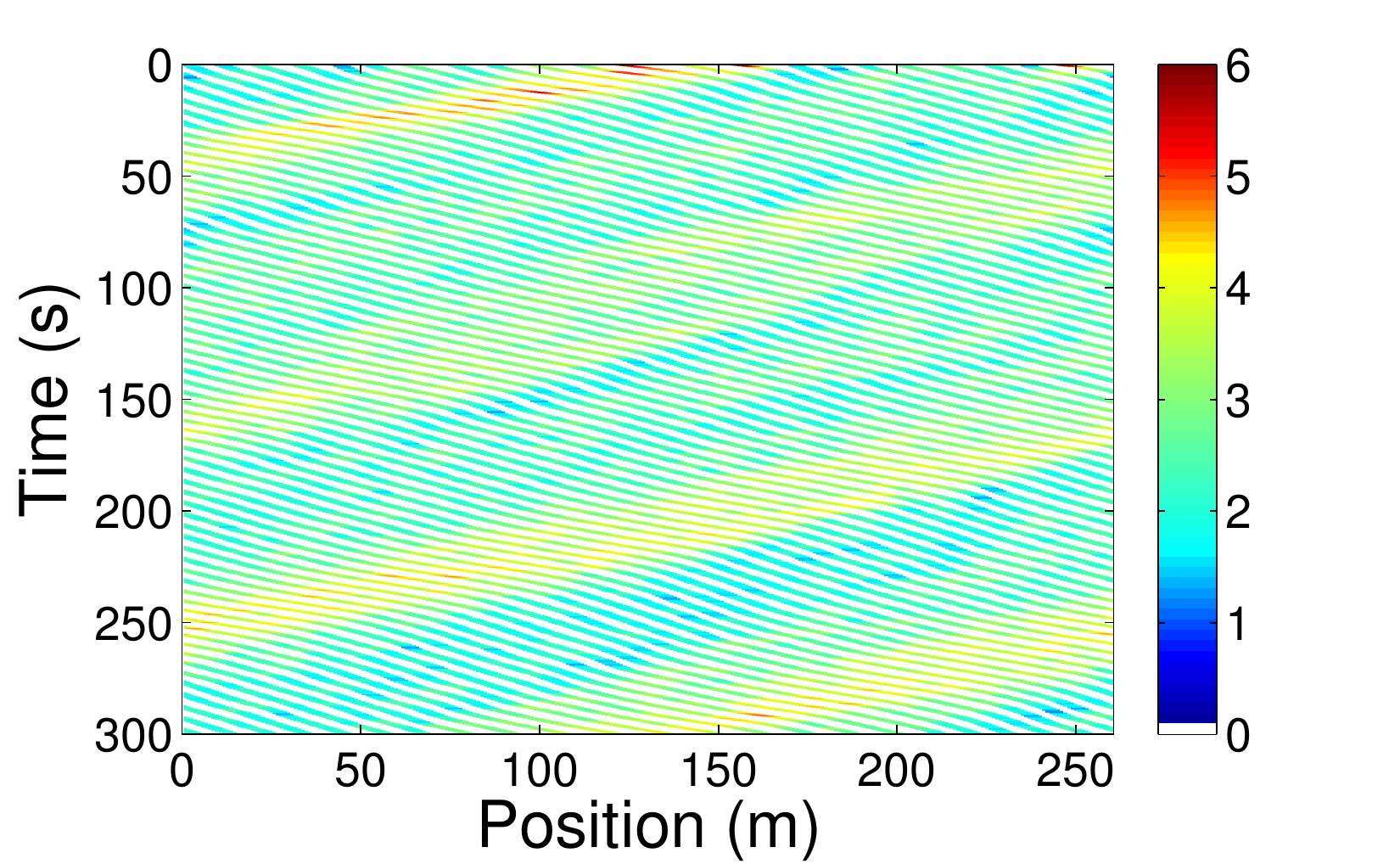}}\\
&\subfloat[]{\includegraphics[width=.25\textwidth]{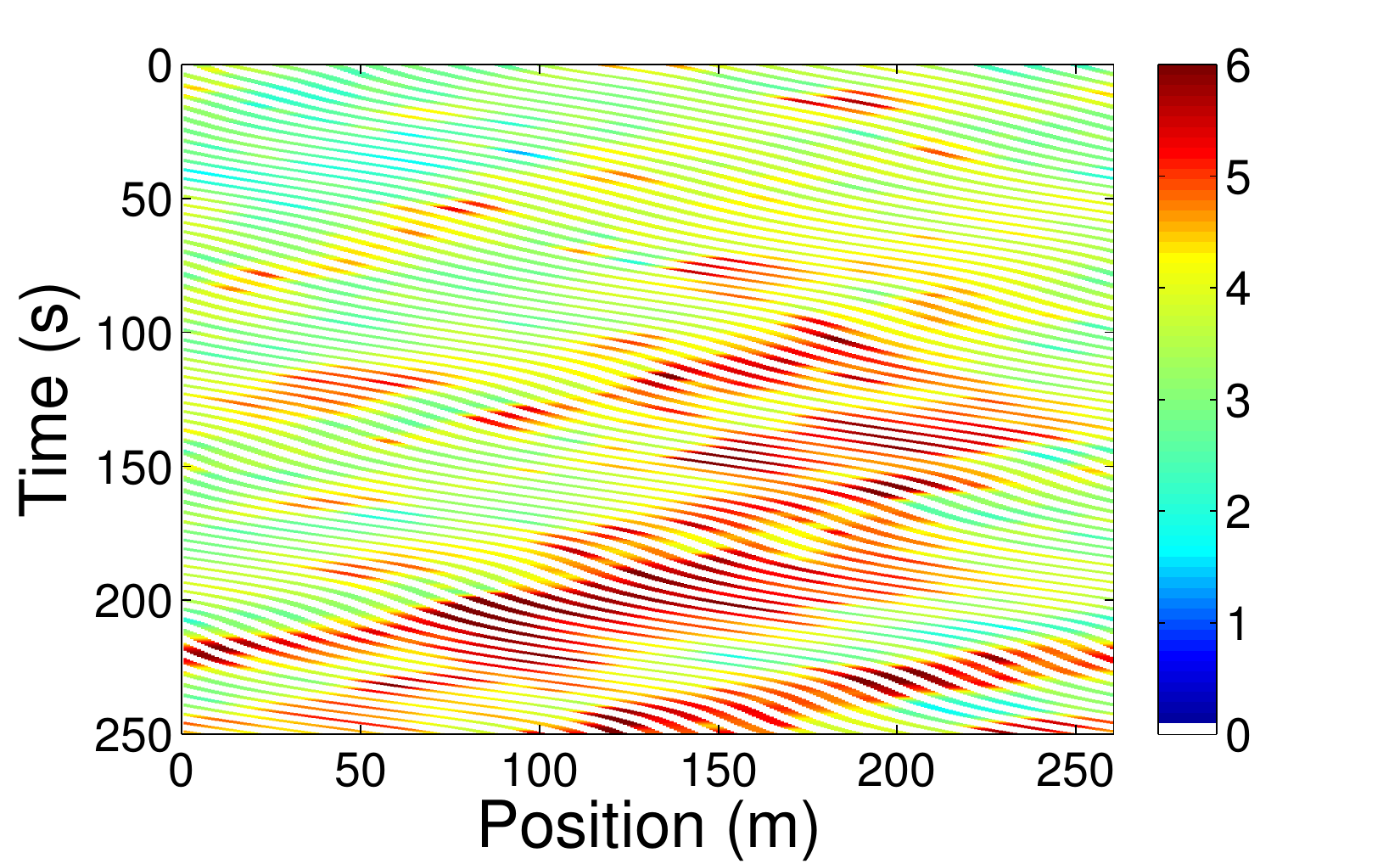}} 
\subfloat[]{\includegraphics[width=.25\textwidth]{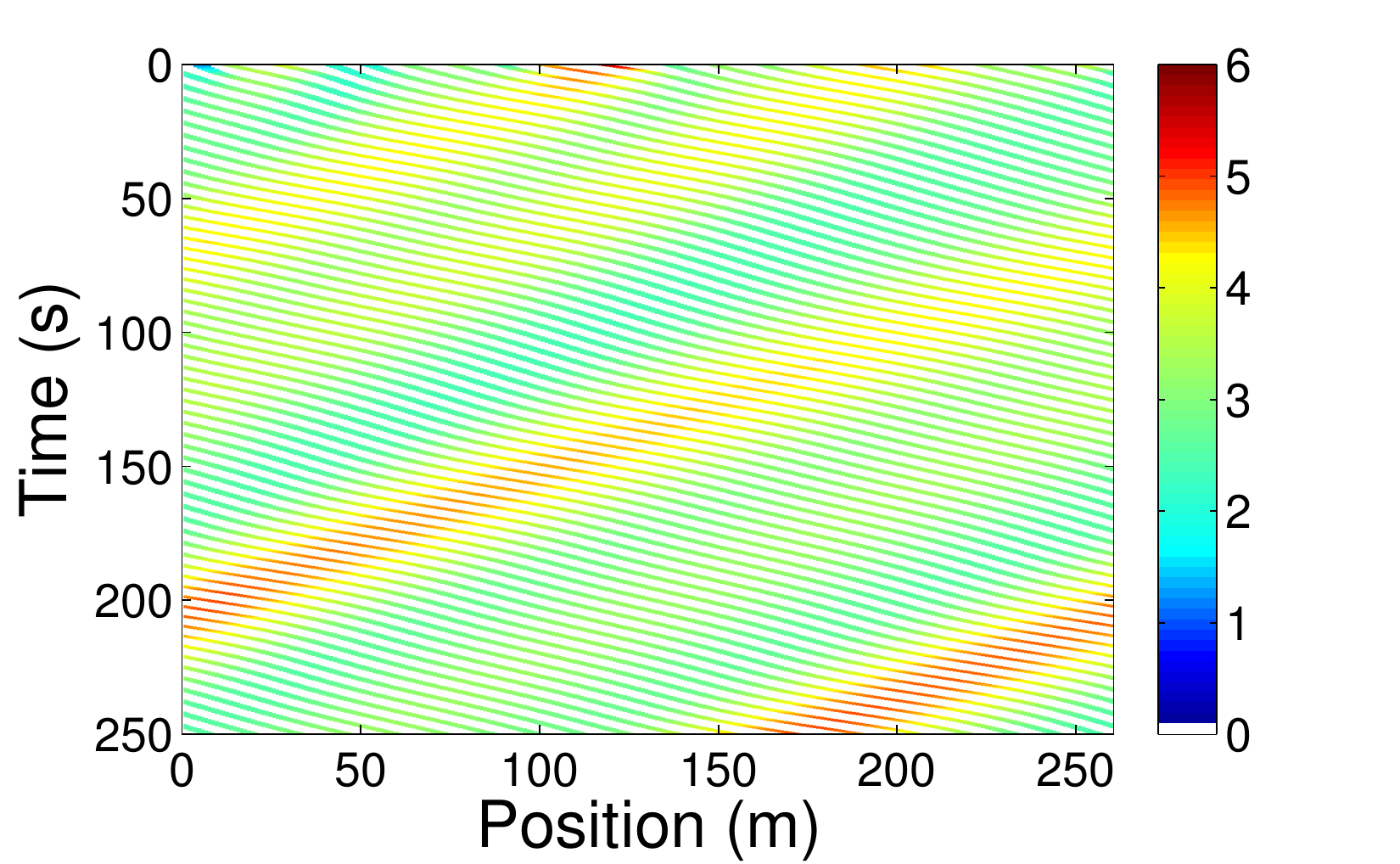}}
\subfloat[]{\includegraphics[width=.25\textwidth]{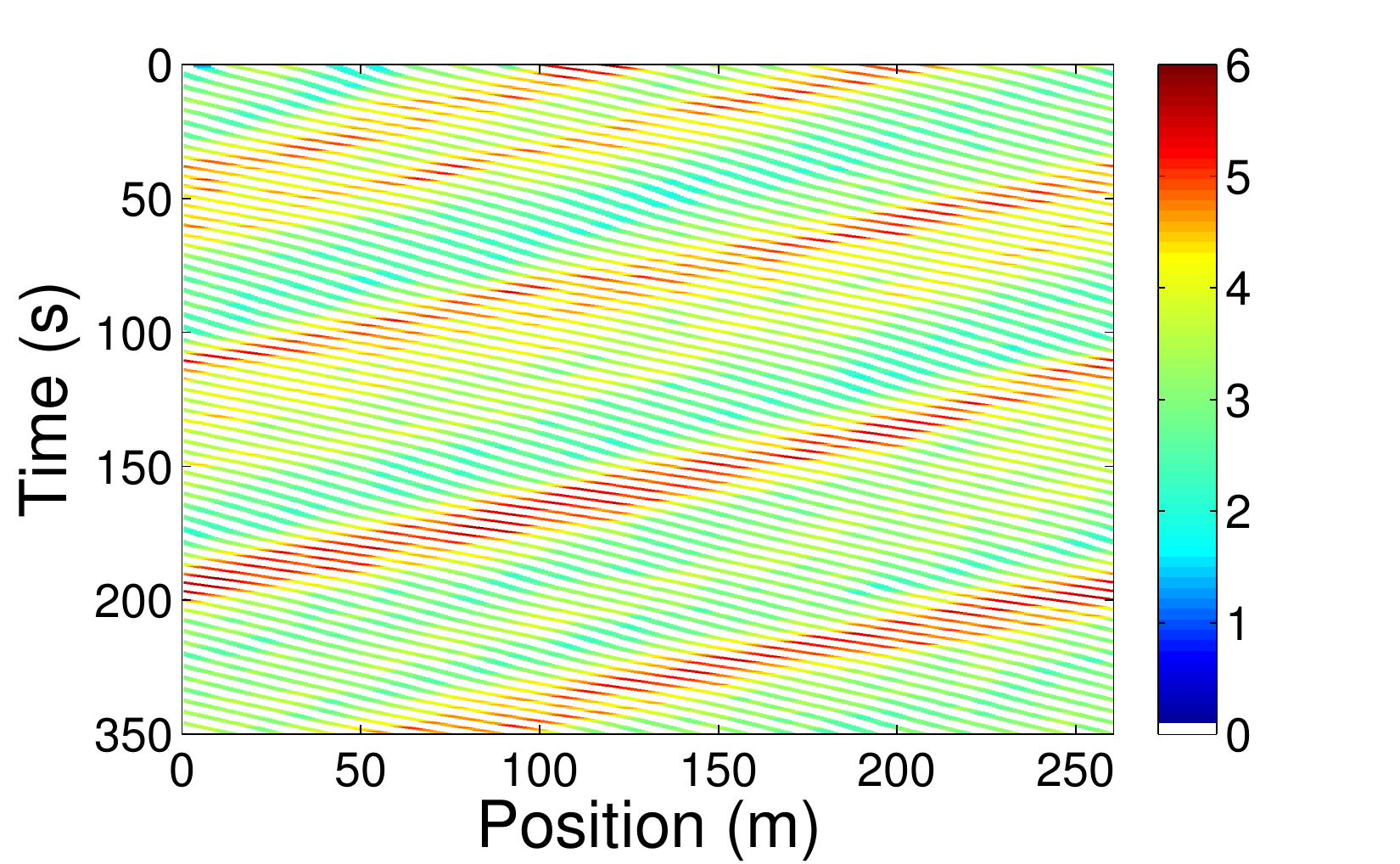}}\\
&\subfloat[]{\includegraphics[width=.25\textwidth]{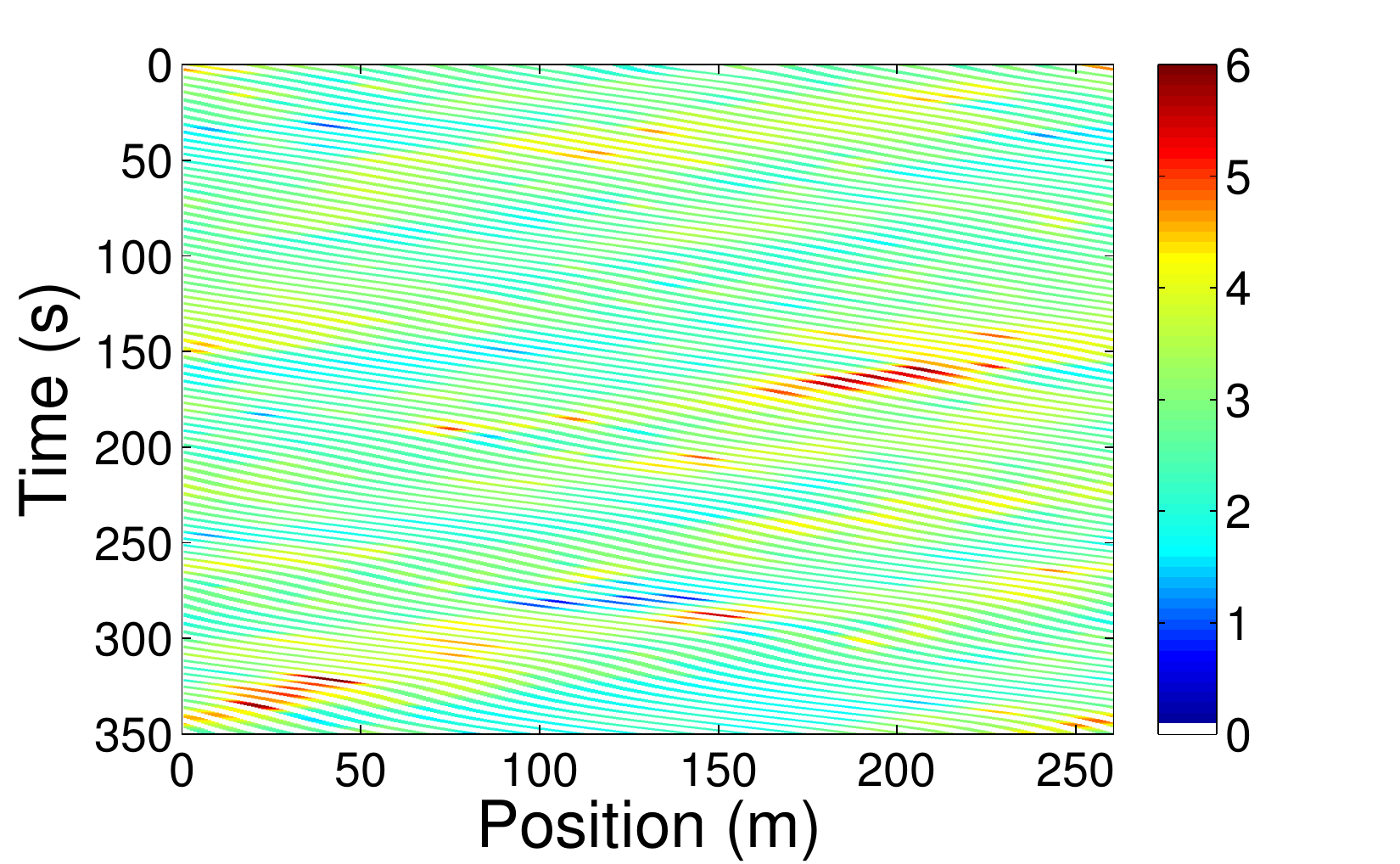}} 
\subfloat[]{\includegraphics[width=.25\textwidth]{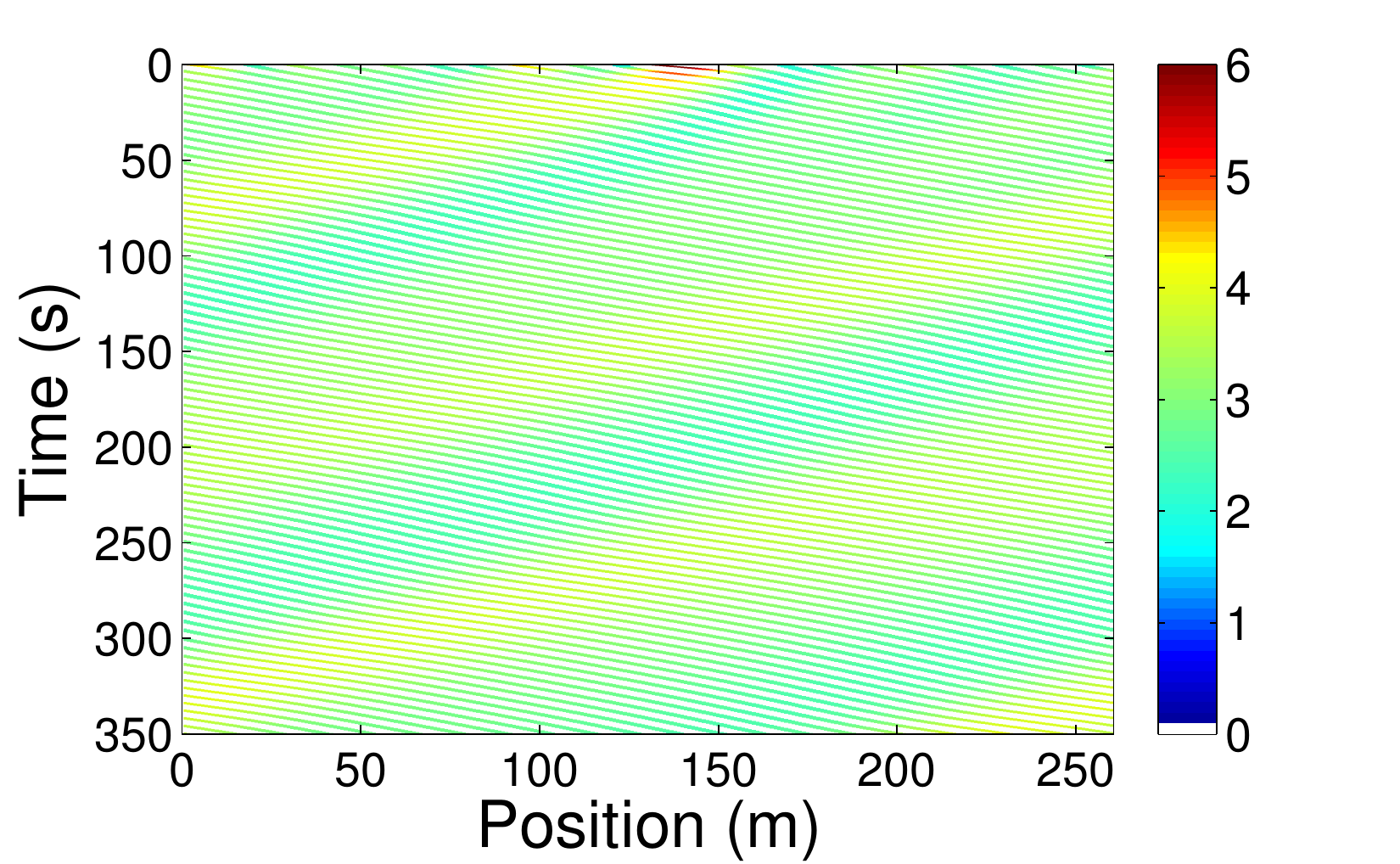}}
\subfloat[]{\includegraphics[width=.25\textwidth]{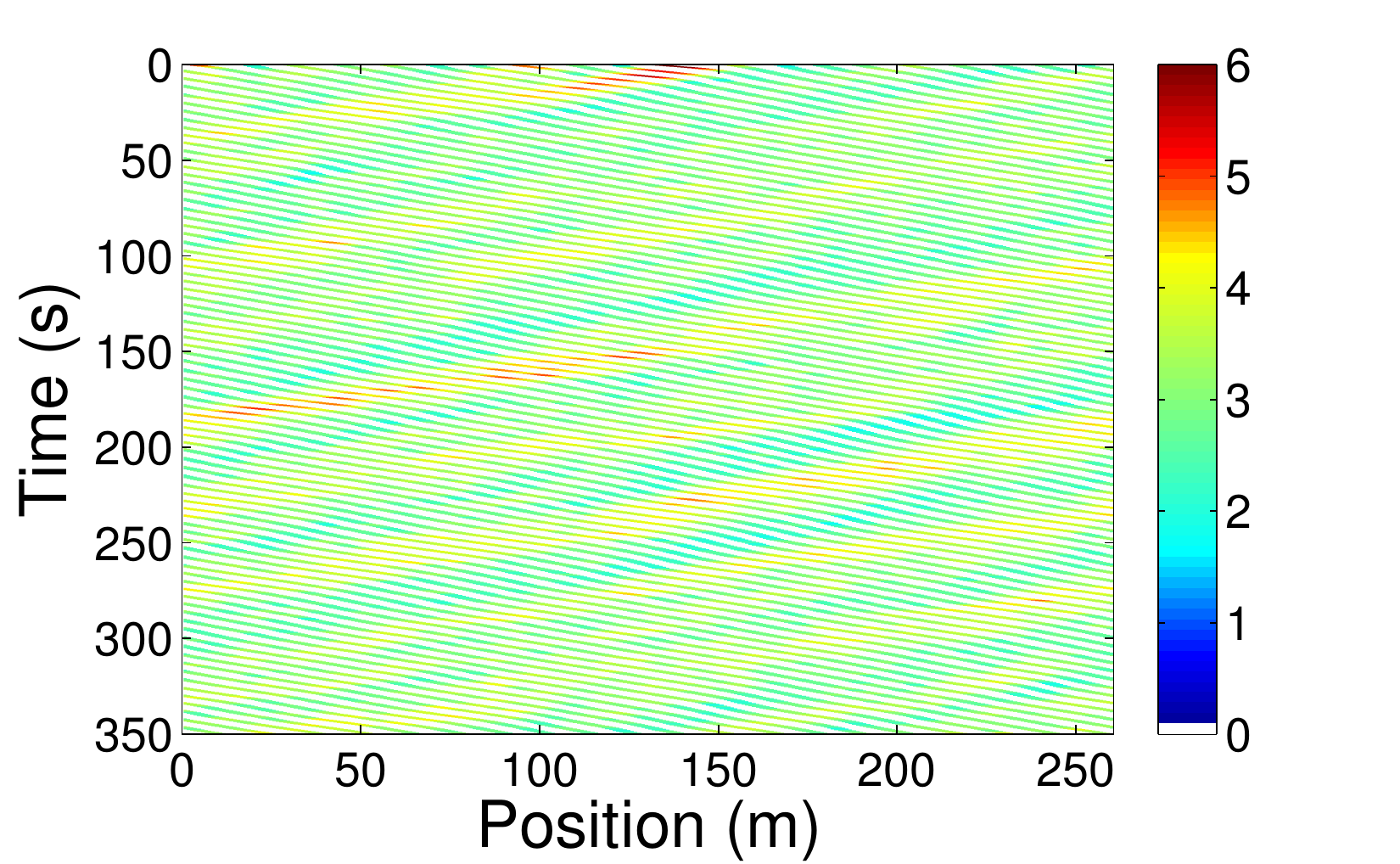}}\\
&\subfloat[]{\includegraphics[width=.25\textwidth]{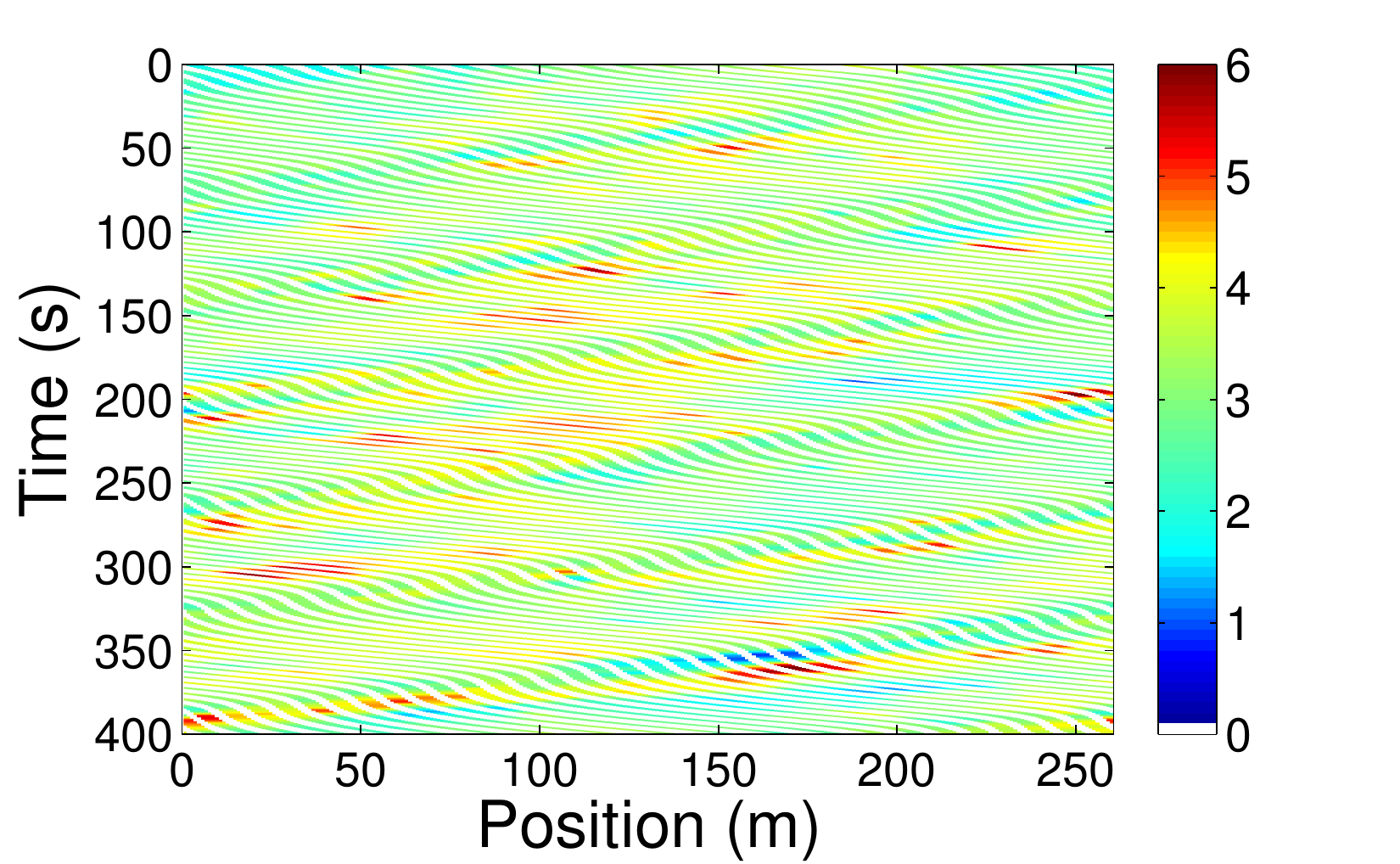}} 
\subfloat[]{\includegraphics[width=.25\textwidth]{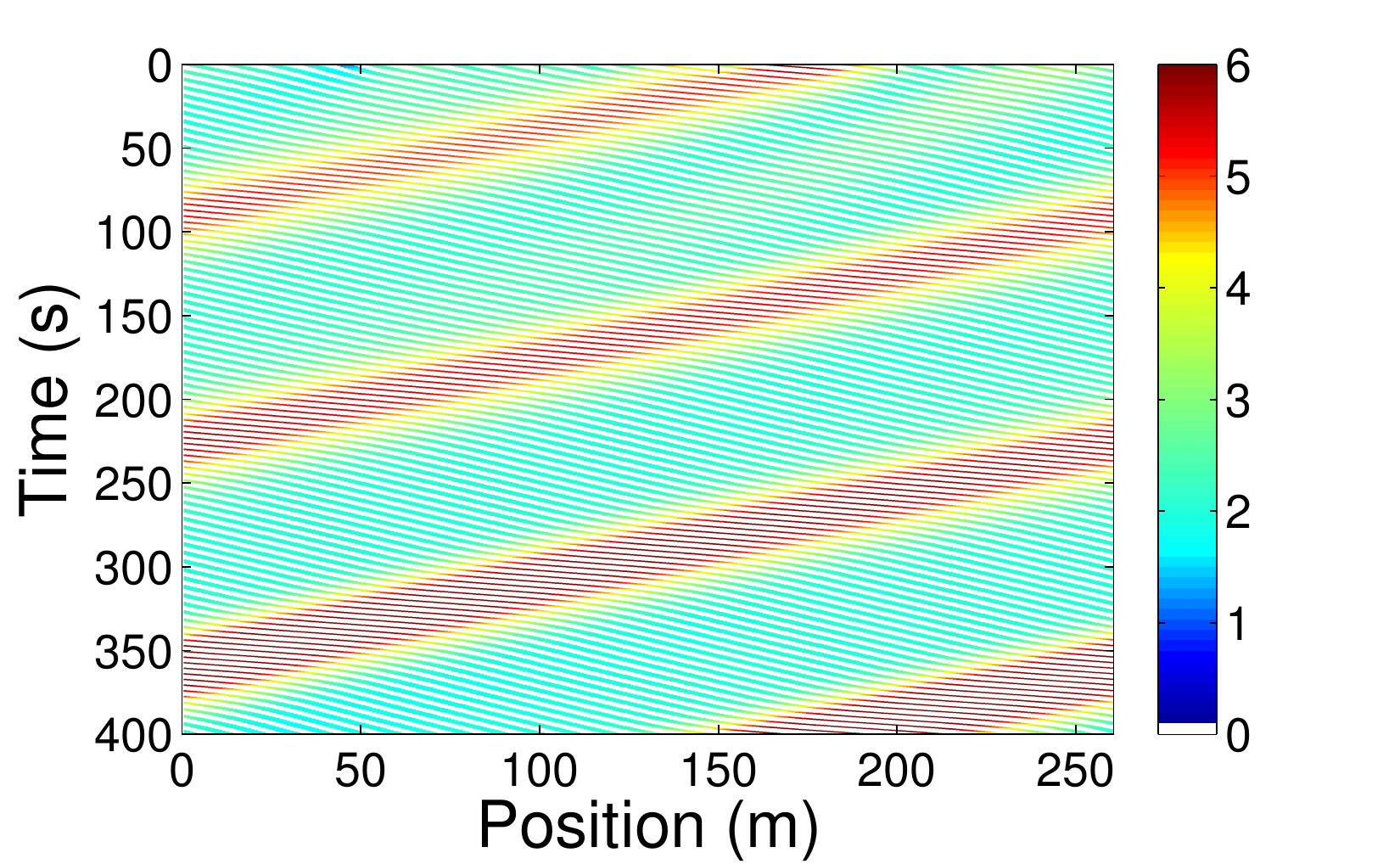}}
\subfloat[]{\includegraphics[width=.25\textwidth]{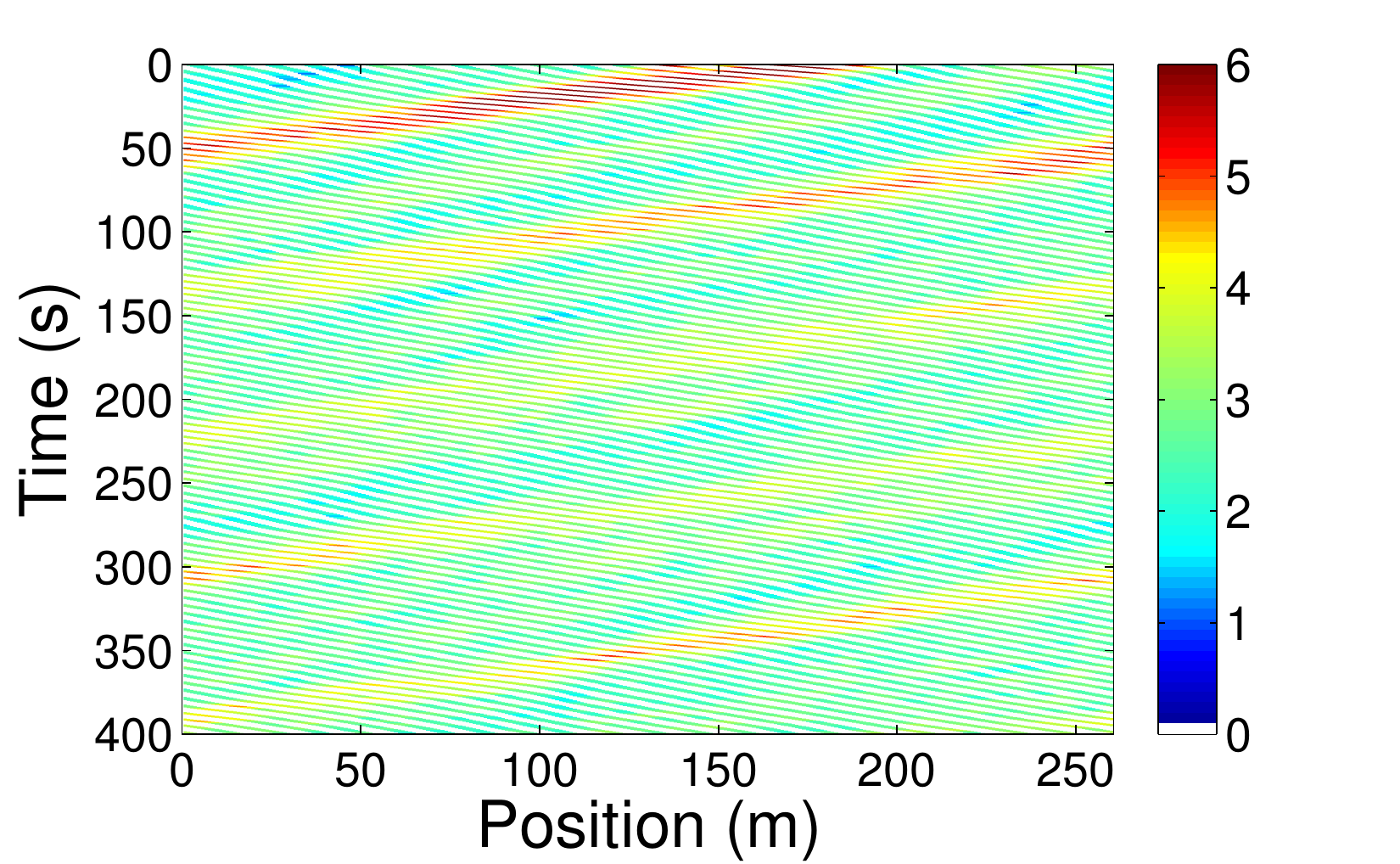}}\\
\end{tabular}
\caption{Time-space diagrams of the results of experiments A, C, E, B and D (left panel) and the corresponding simulations by using the OVM and the SOVM, respectively. The experiments A, C and E are used for calibration while experiments B and D are used for validation (a-c) Experiment A (d-f) Experiment C (g-i) Experiment E (j-l) Experiment B (m-o) Experiment D. The velocity unit in the color bar is $\mathrm{(m/s)}$.}
\label{fig11}
\end{figure}

\begin{figure}[p]
\centering
\begin{tabular}{cc} 
\centering
&\subfloat[]{\includegraphics[width=.25\textwidth]{expA}} 
\subfloat[]{\includegraphics[width=.25\textwidth]{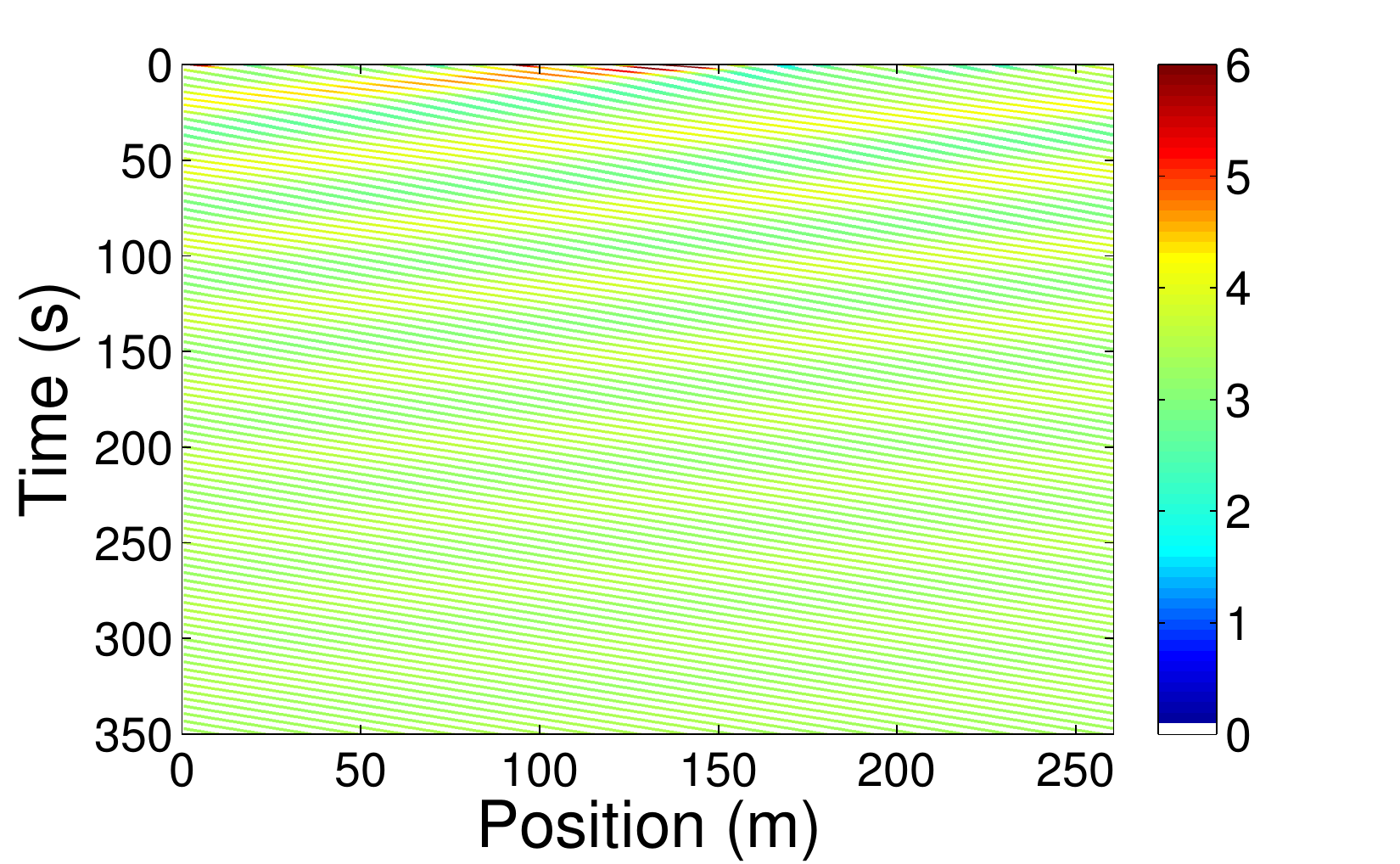}}
\subfloat[]{\includegraphics[width=.25\textwidth]{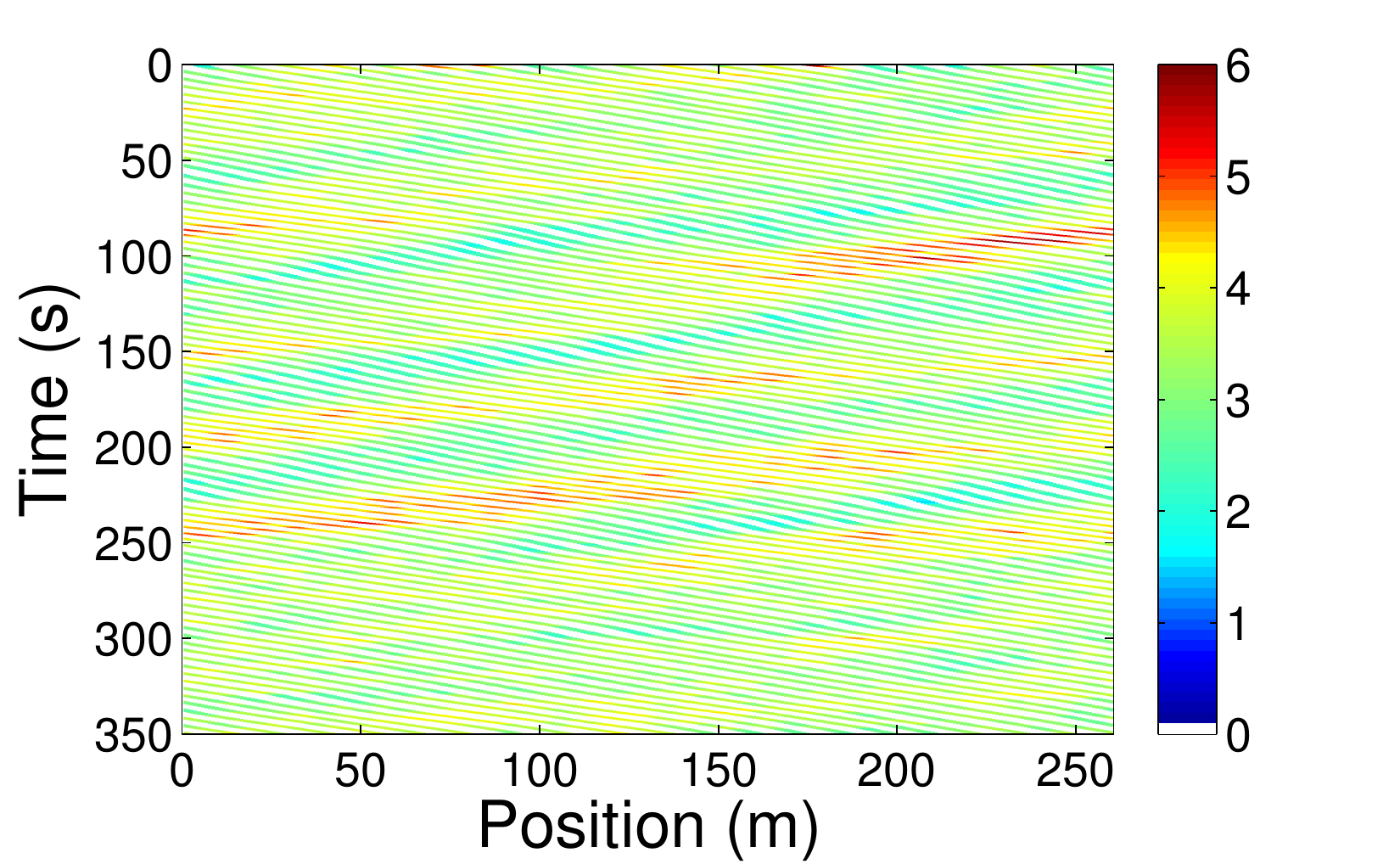}}\\
&\subfloat[]{\includegraphics[width=.25\textwidth]{expC}} 
\subfloat[]{\includegraphics[width=.25\textwidth]{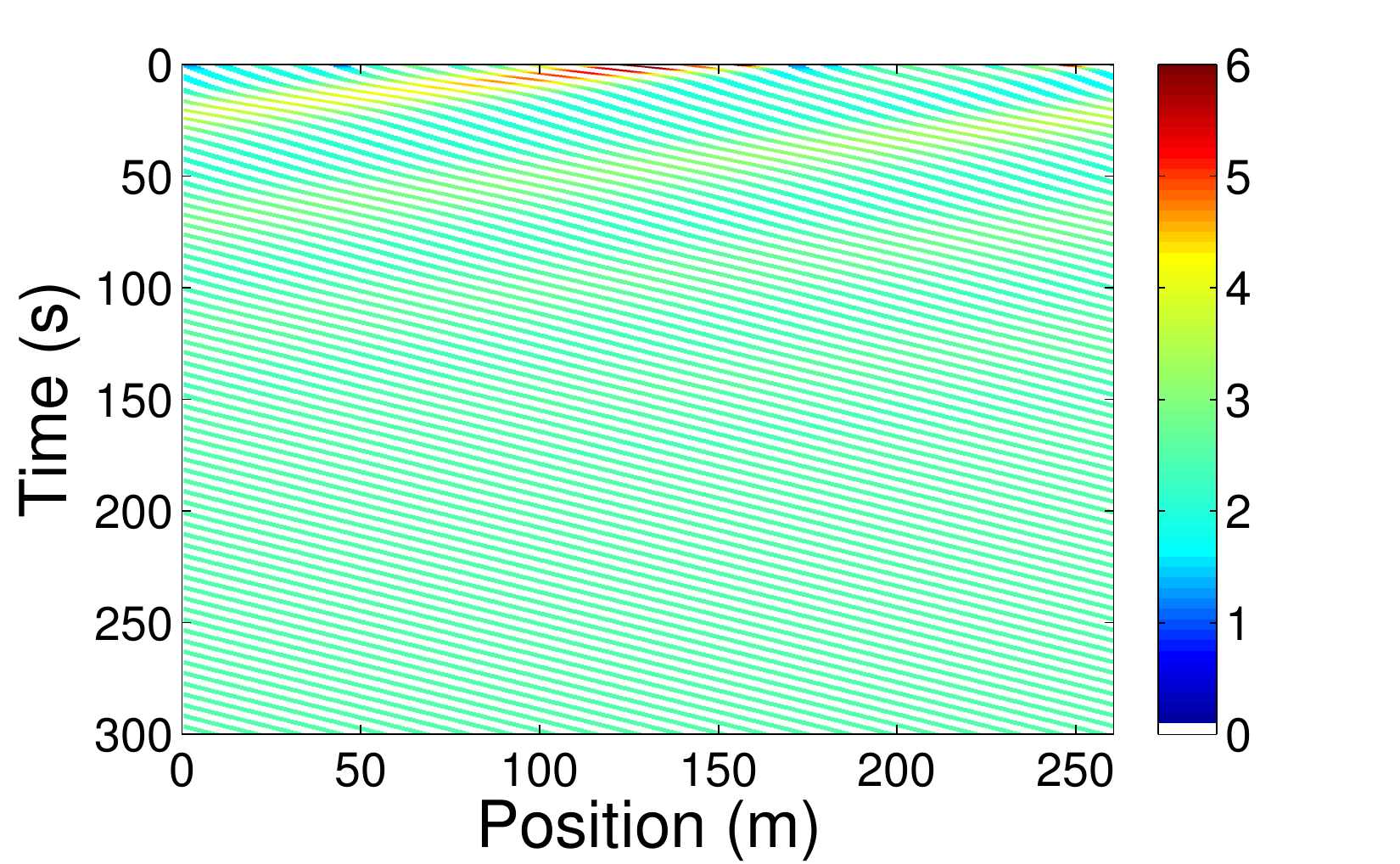}}
\subfloat[]{\includegraphics[width=.25\textwidth]{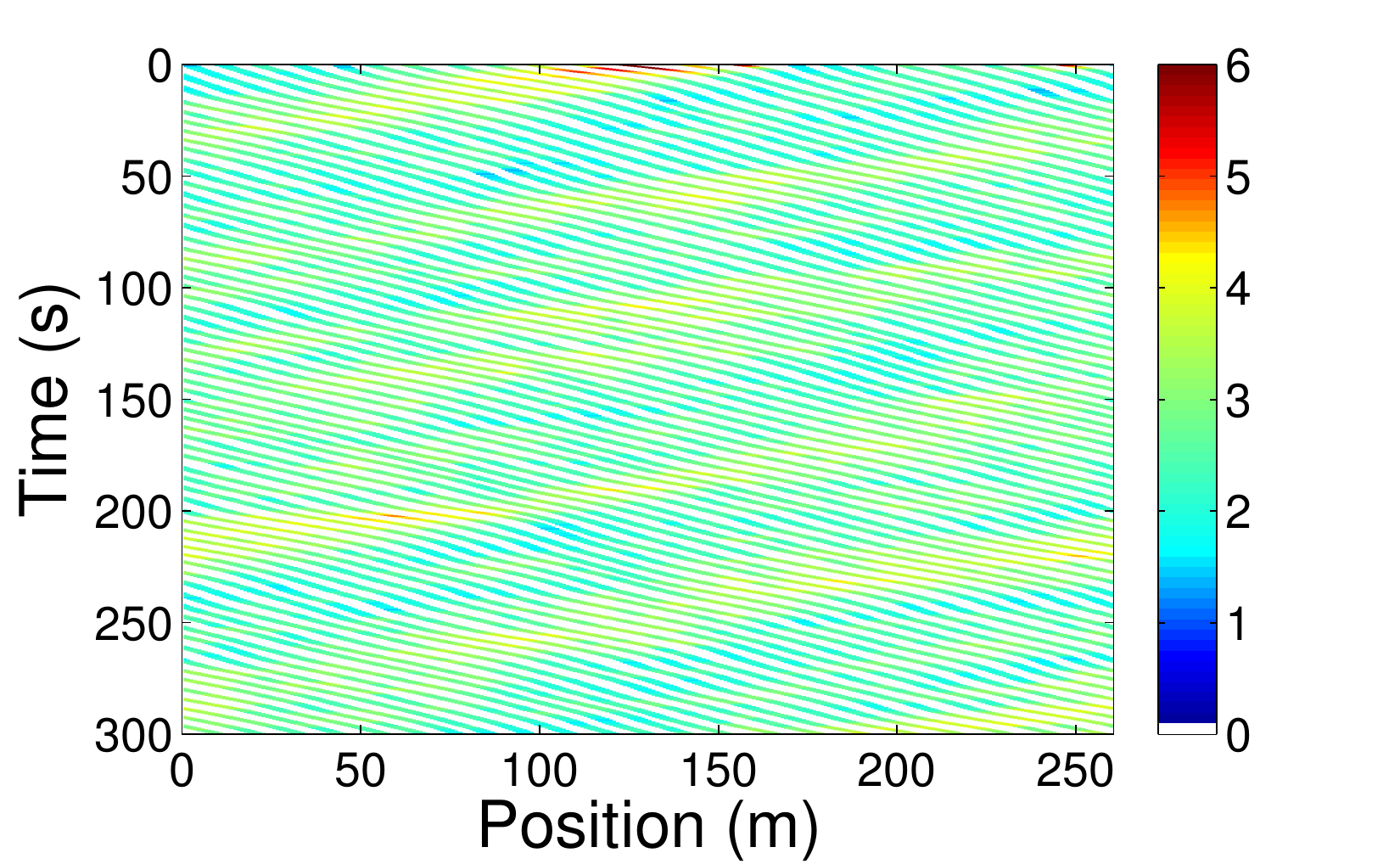}}\\
&\subfloat[]{\includegraphics[width=.25\textwidth]{expE}} 
\subfloat[]{\includegraphics[width=.25\textwidth]{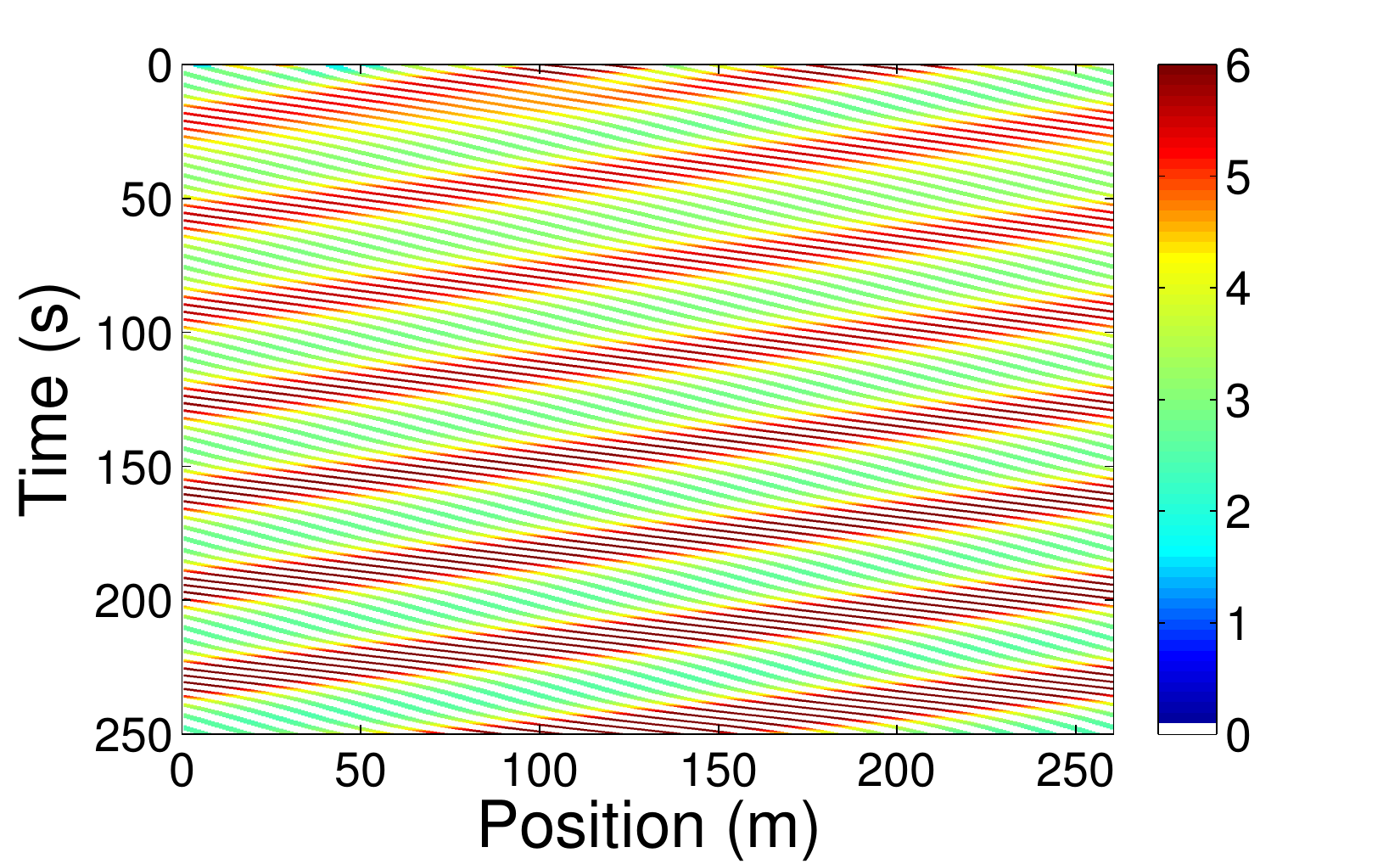}}
\subfloat[]{\includegraphics[width=.25\textwidth]{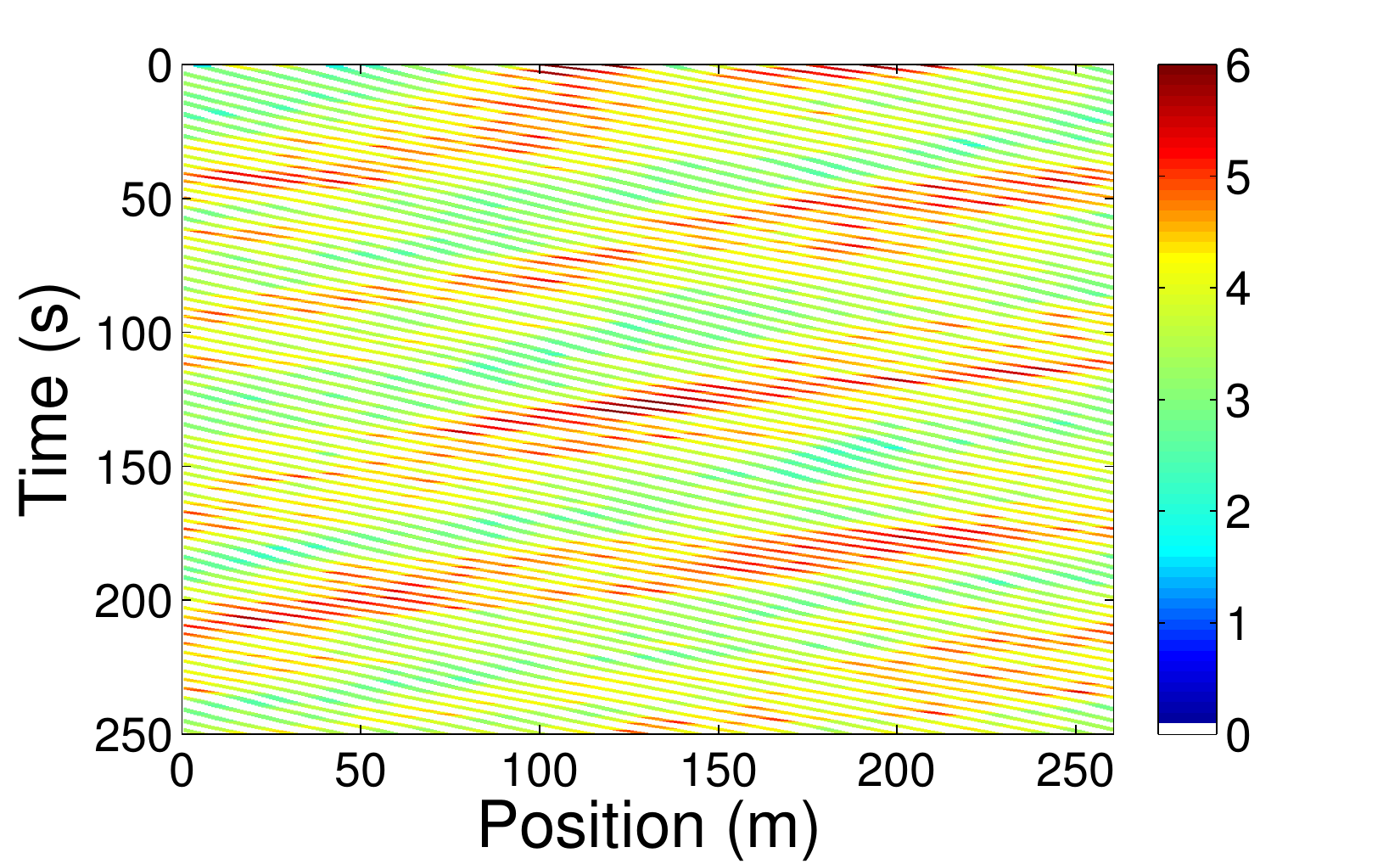}}\\
&\subfloat[]{\includegraphics[width=.25\textwidth]{expB}} 
\subfloat[]{\includegraphics[width=.25\textwidth]{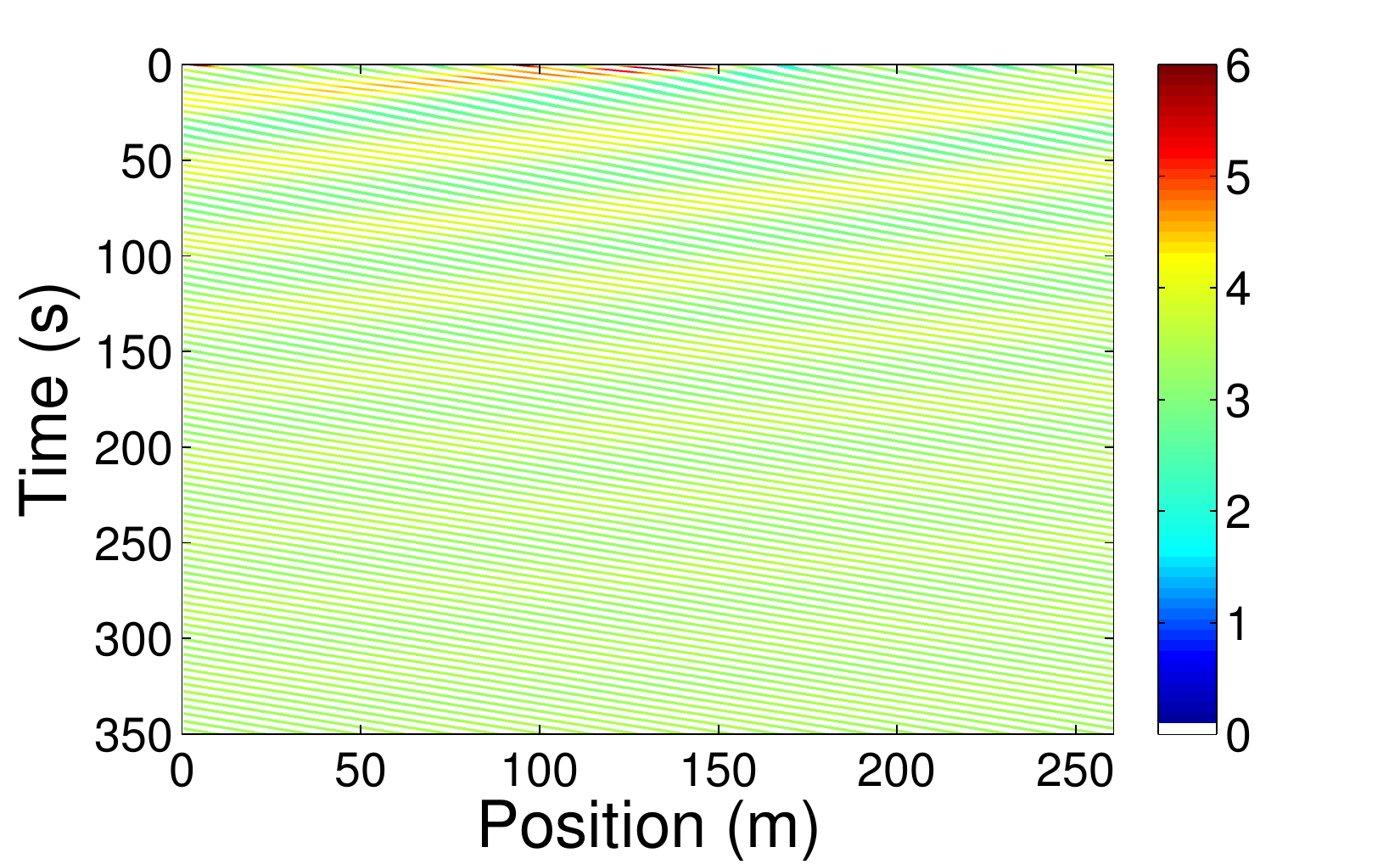}}
\subfloat[]{\includegraphics[width=.25\textwidth]{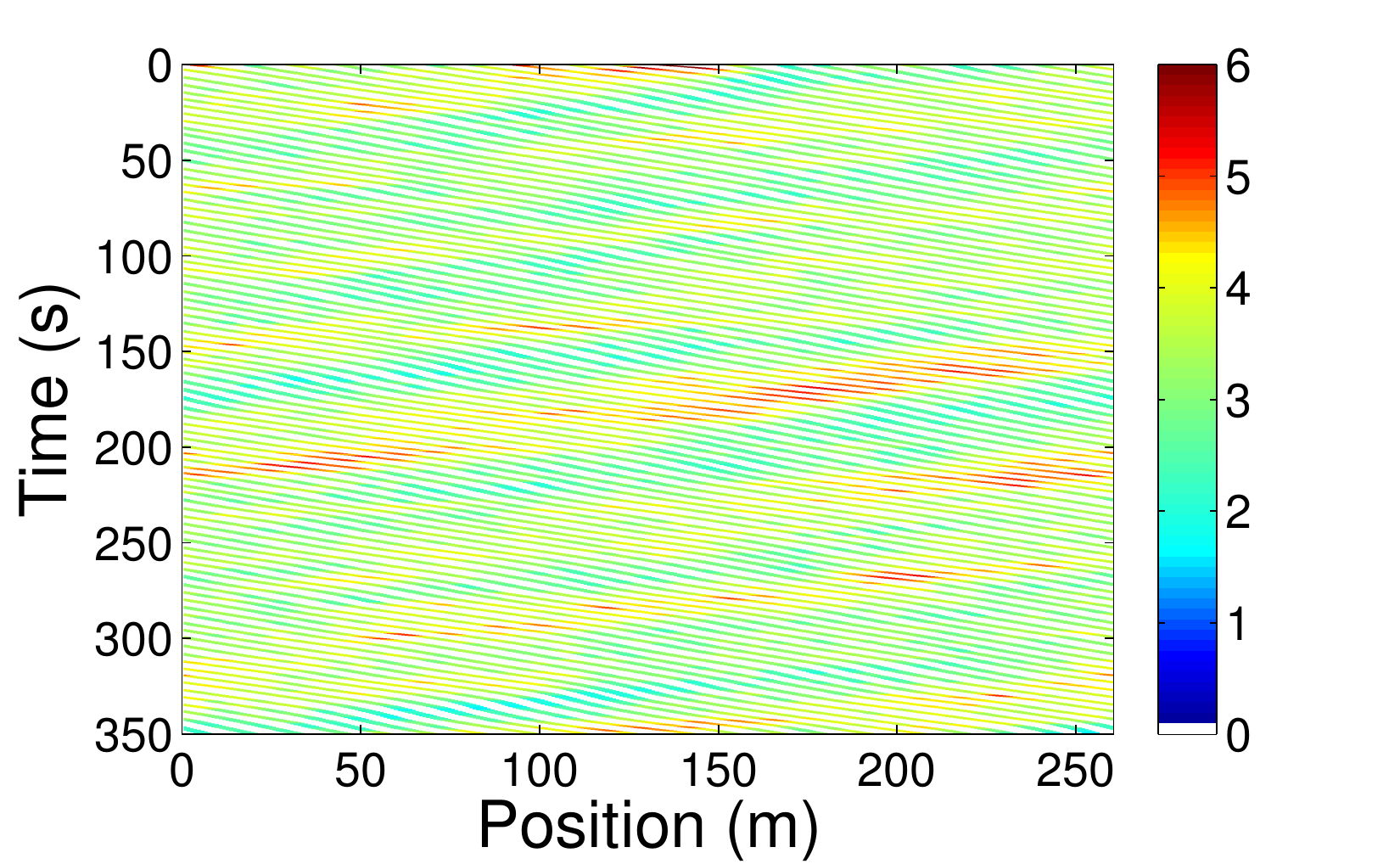}}\\
&\subfloat[]{\includegraphics[width=.25\textwidth]{expD}} 
\subfloat[]{\includegraphics[width=.25\textwidth]{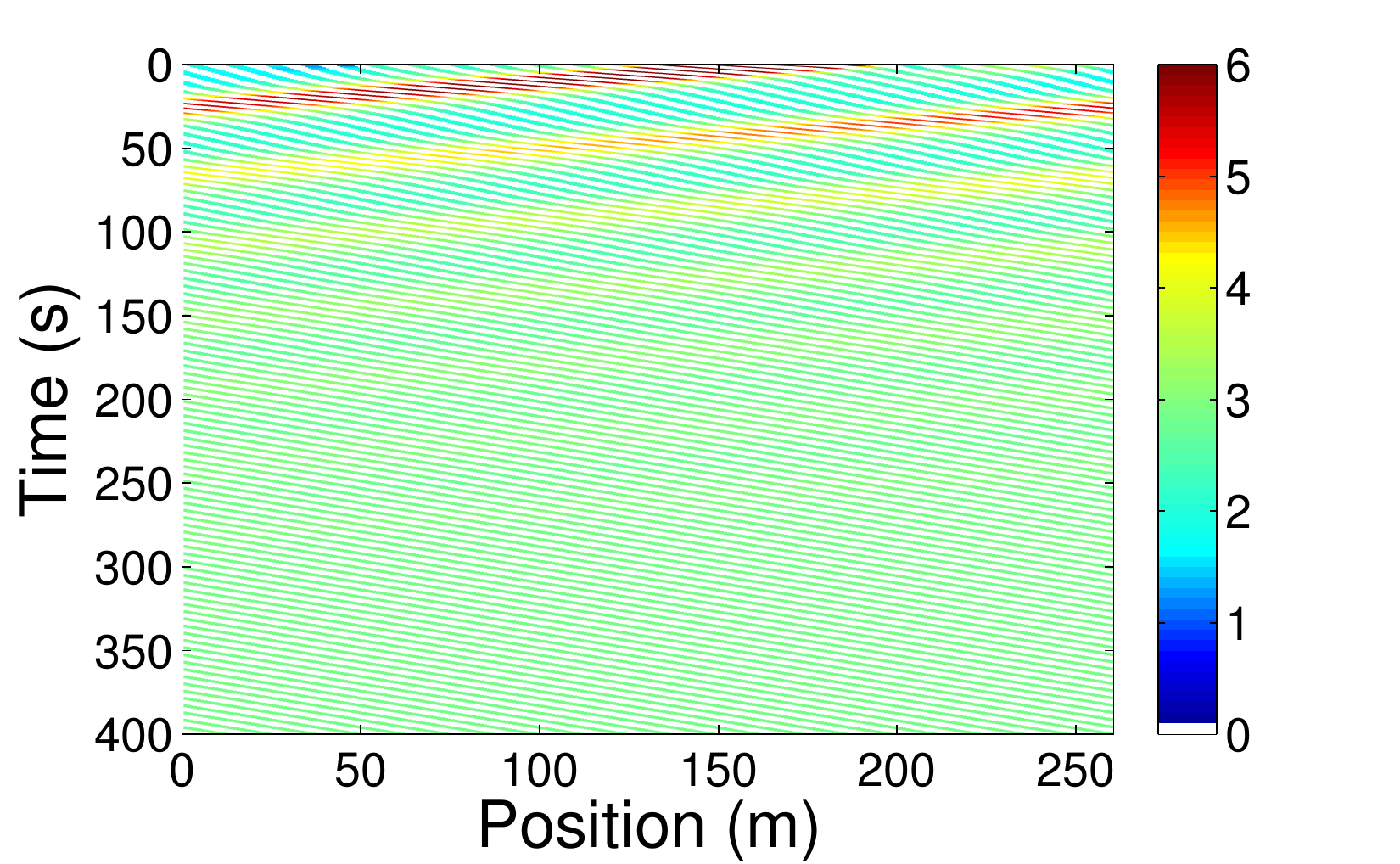}}
\subfloat[]{\includegraphics[width=.25\textwidth]{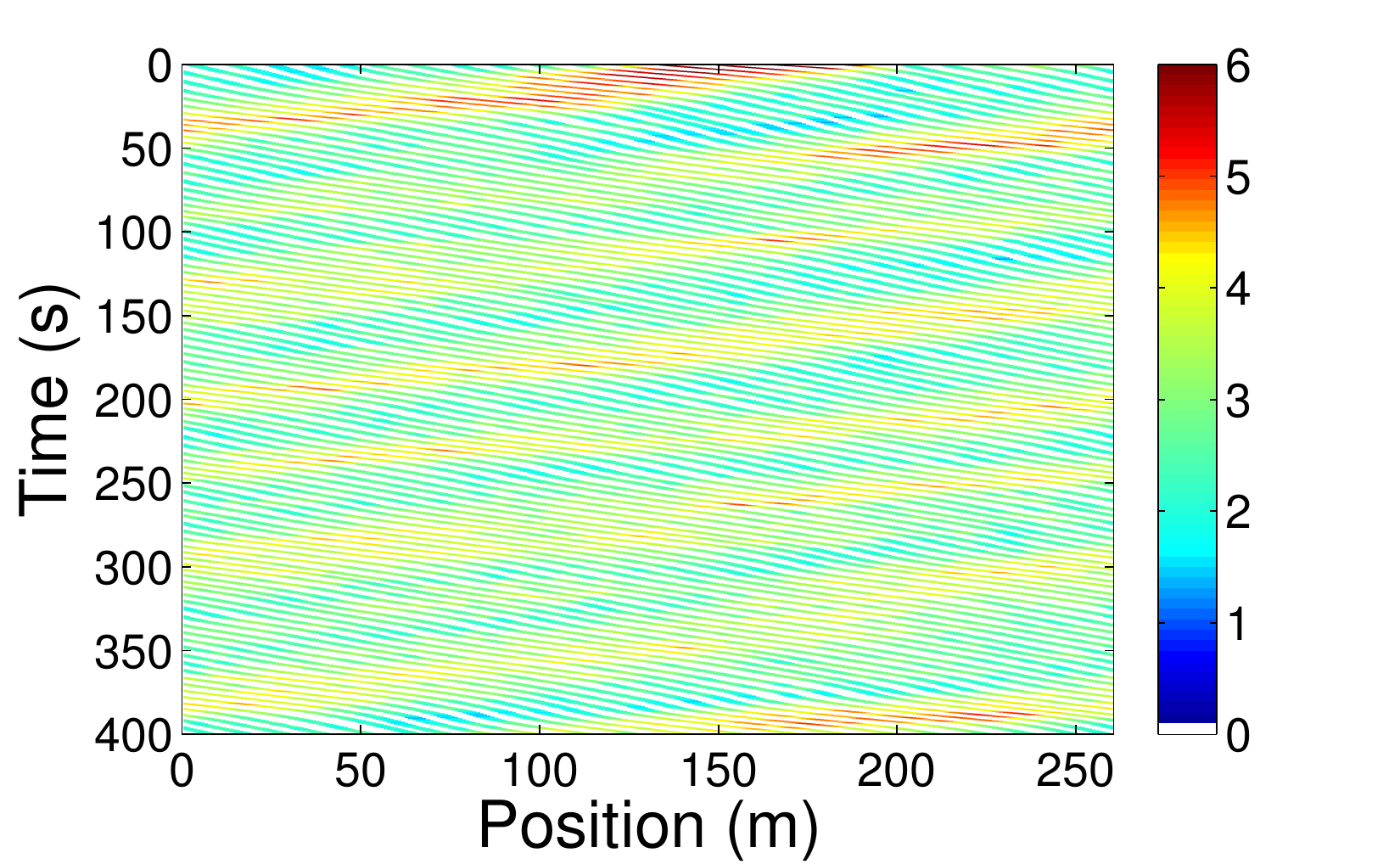}}\\
\end{tabular}
\caption{Time-space diagrams of the results of experiments A, C, E, B, and D (left panel) and the corresponding simulations by using the FVDM and the SFVDM, respectively. The experiments A, C, and E are used for calibration while experiments B and D are used for validation (a-c) Experiment A (d-f) Experiment C (g-i) Experiment E (j-l) Experiment B (m-o) Experiment D. The velocity unit in the color bar is $\mathrm{(m/s)}$.}
\label{fig12}
\end{figure}

\begin{figure}[p]  
\centering
\begin{tabular}{cc} 
\centering
&\subfloat[]{\includegraphics[width=.25\textwidth]{expA}} 
\subfloat[]{\includegraphics[width=.25\textwidth]{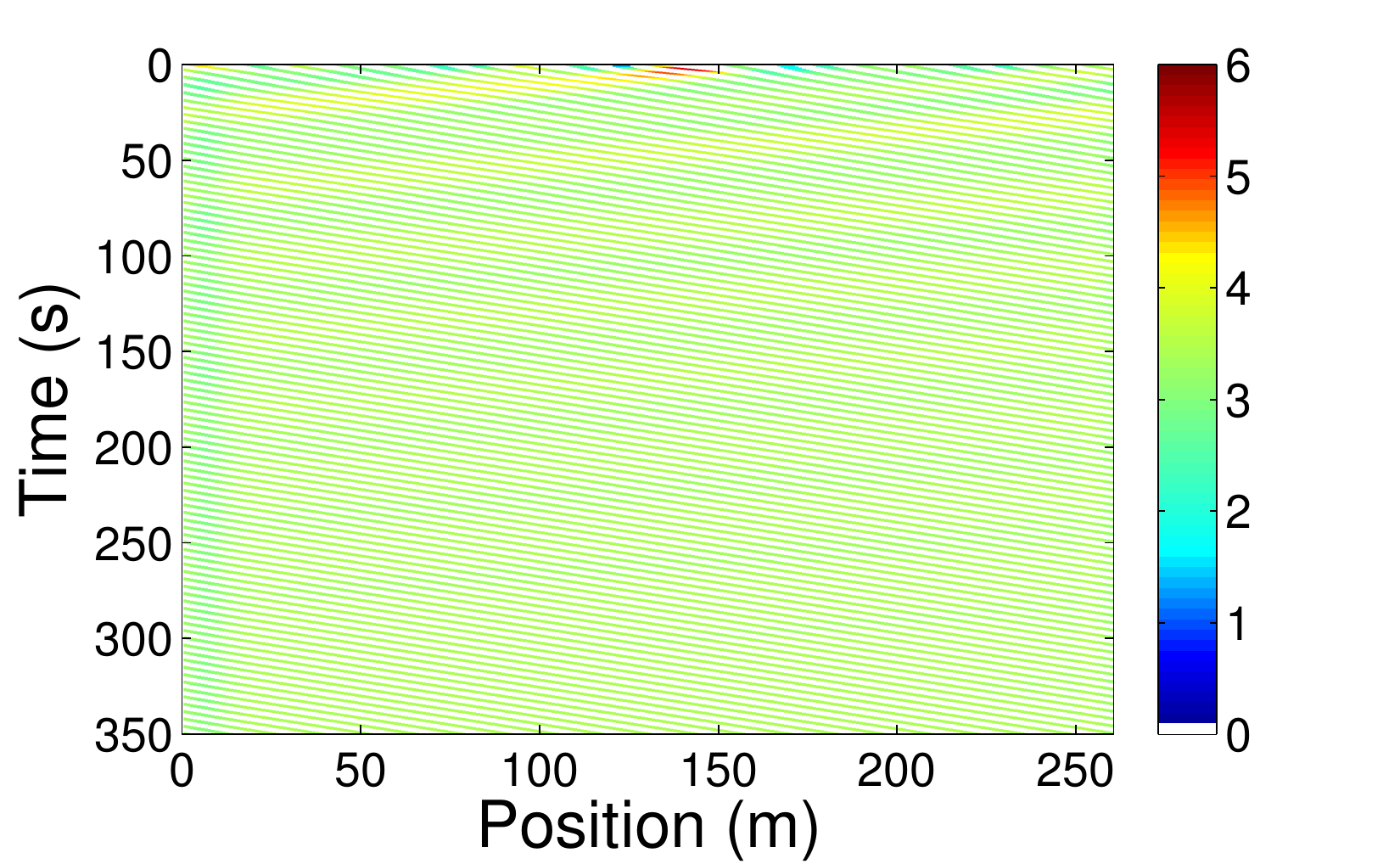}}
\subfloat[]{\includegraphics[width=.25\textwidth]{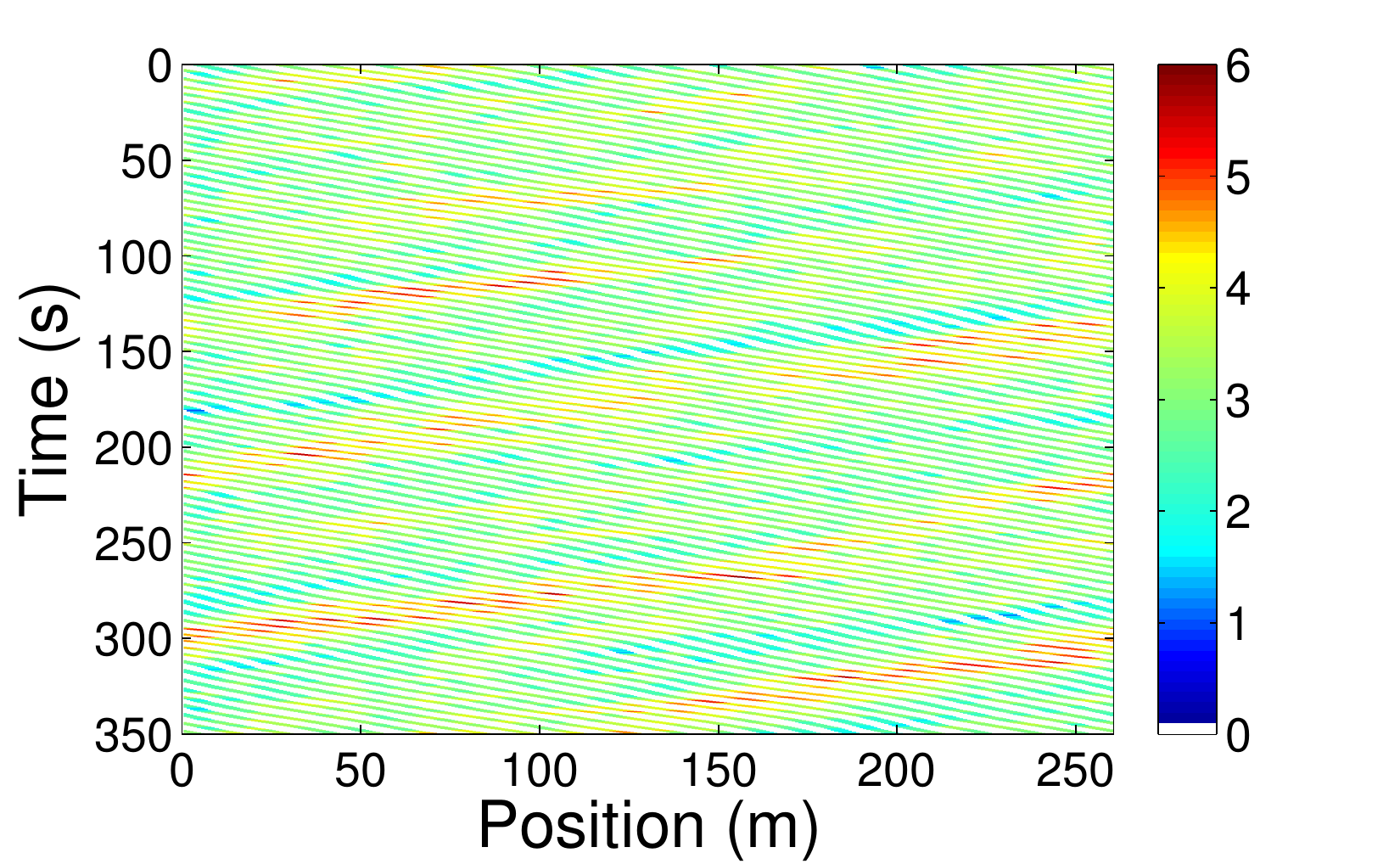}}\\
&\subfloat[]{\includegraphics[width=.25\textwidth]{expC}} 
\subfloat[]{\includegraphics[width=.25\textwidth]{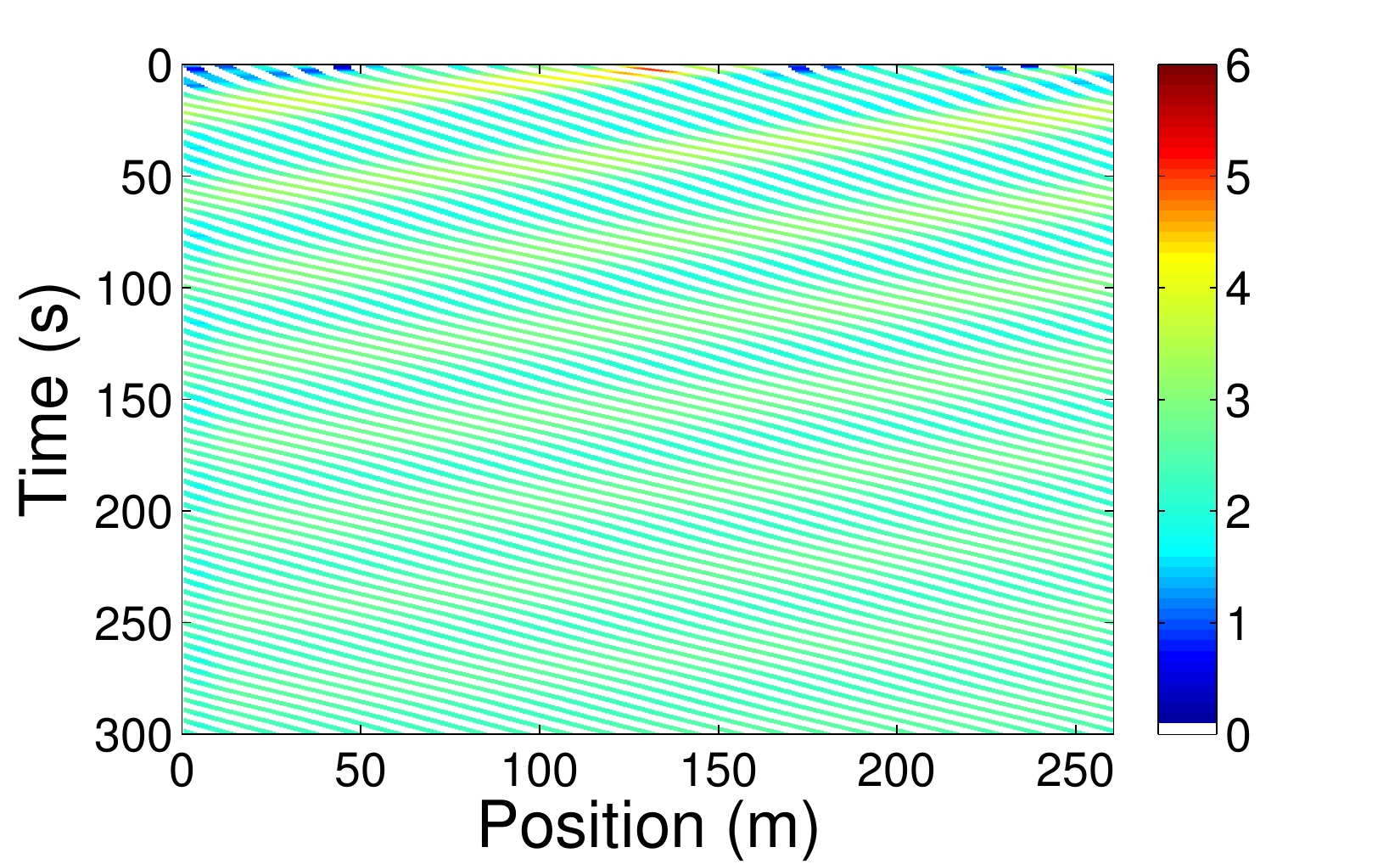}}
\subfloat[]{\includegraphics[width=.25\textwidth]{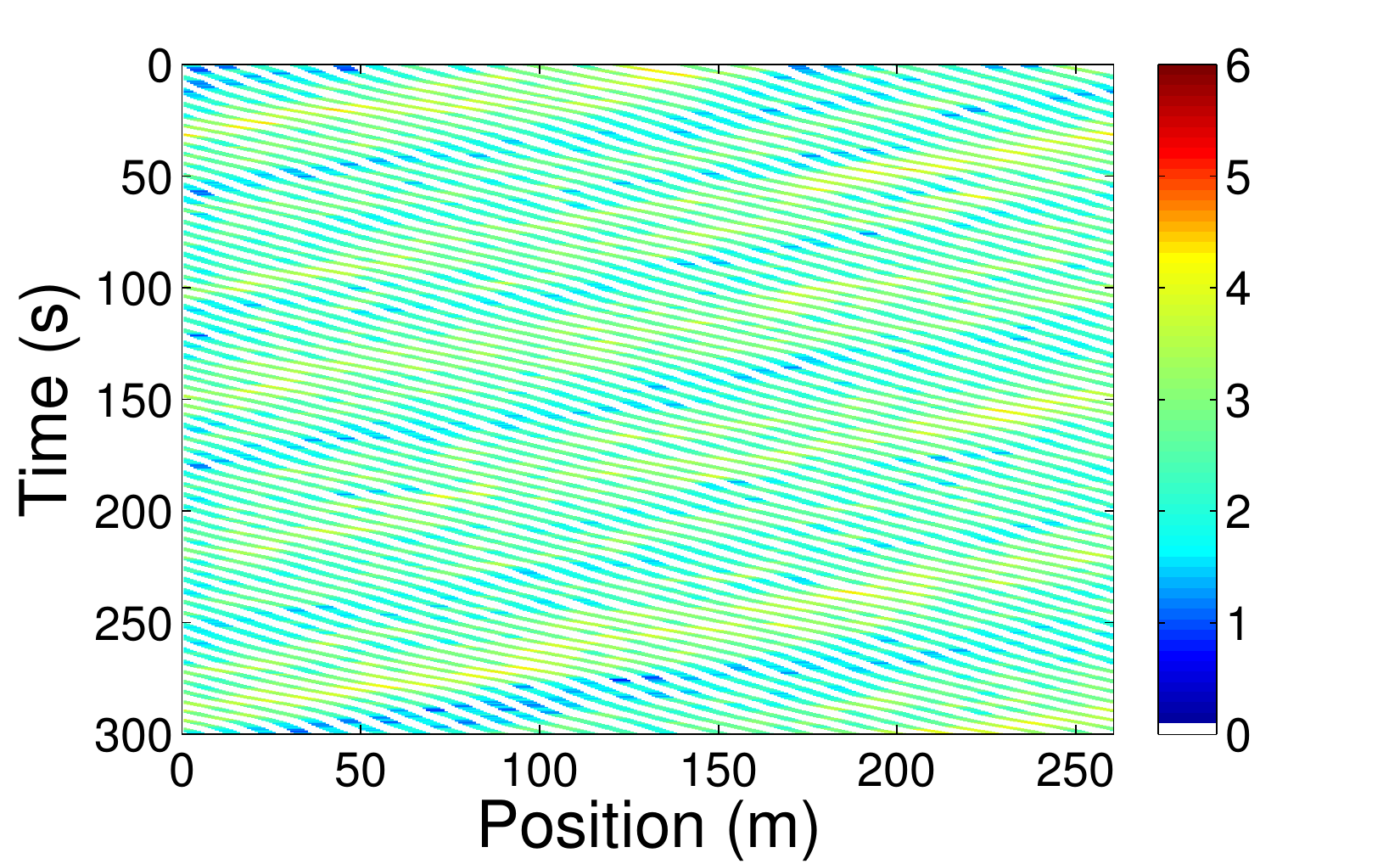}}\\
&\subfloat[]{\includegraphics[width=.25\textwidth]{expE}} 
\subfloat[]{\includegraphics[width=.25\textwidth]{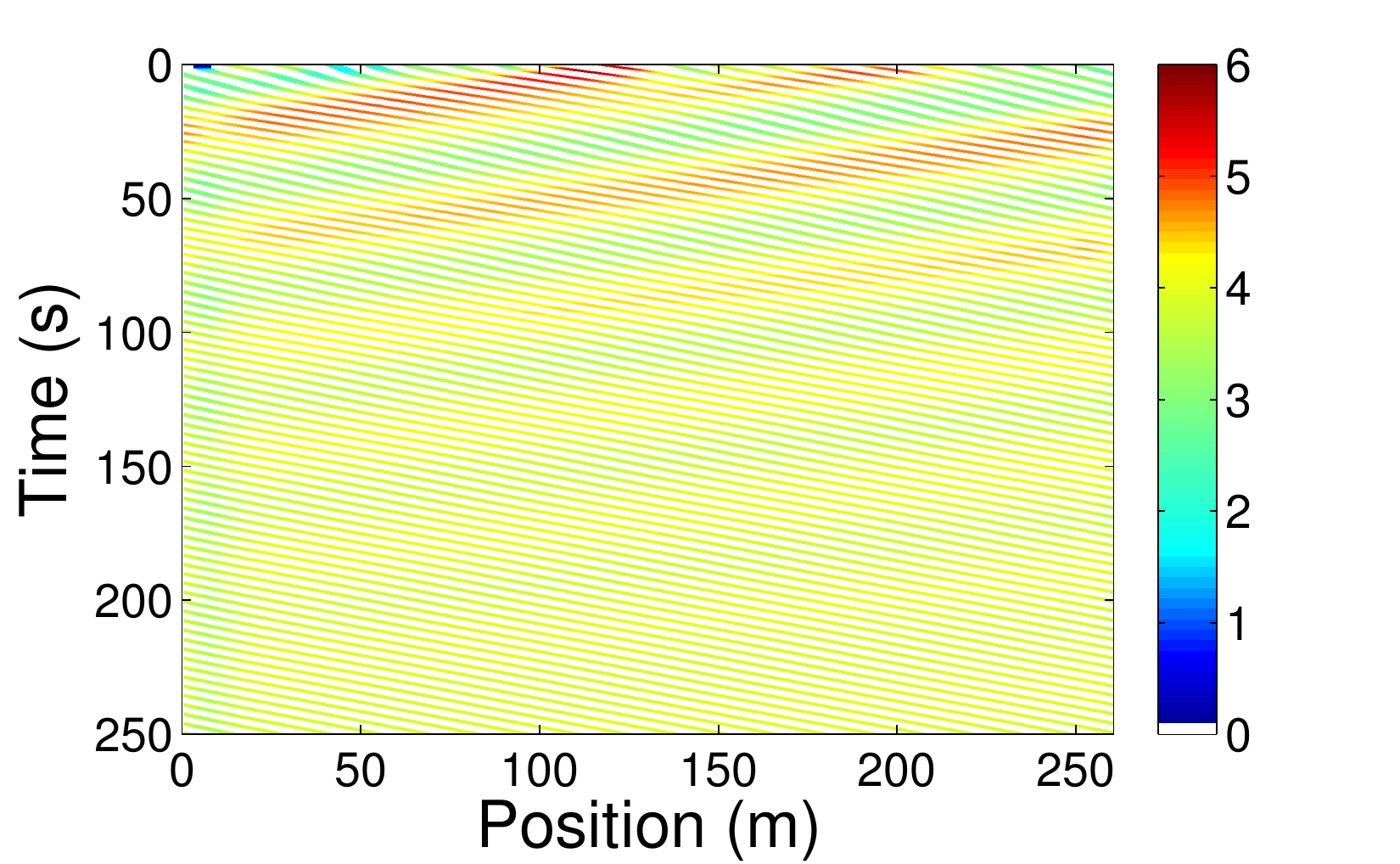}}
\subfloat[]{\includegraphics[width=.25\textwidth]{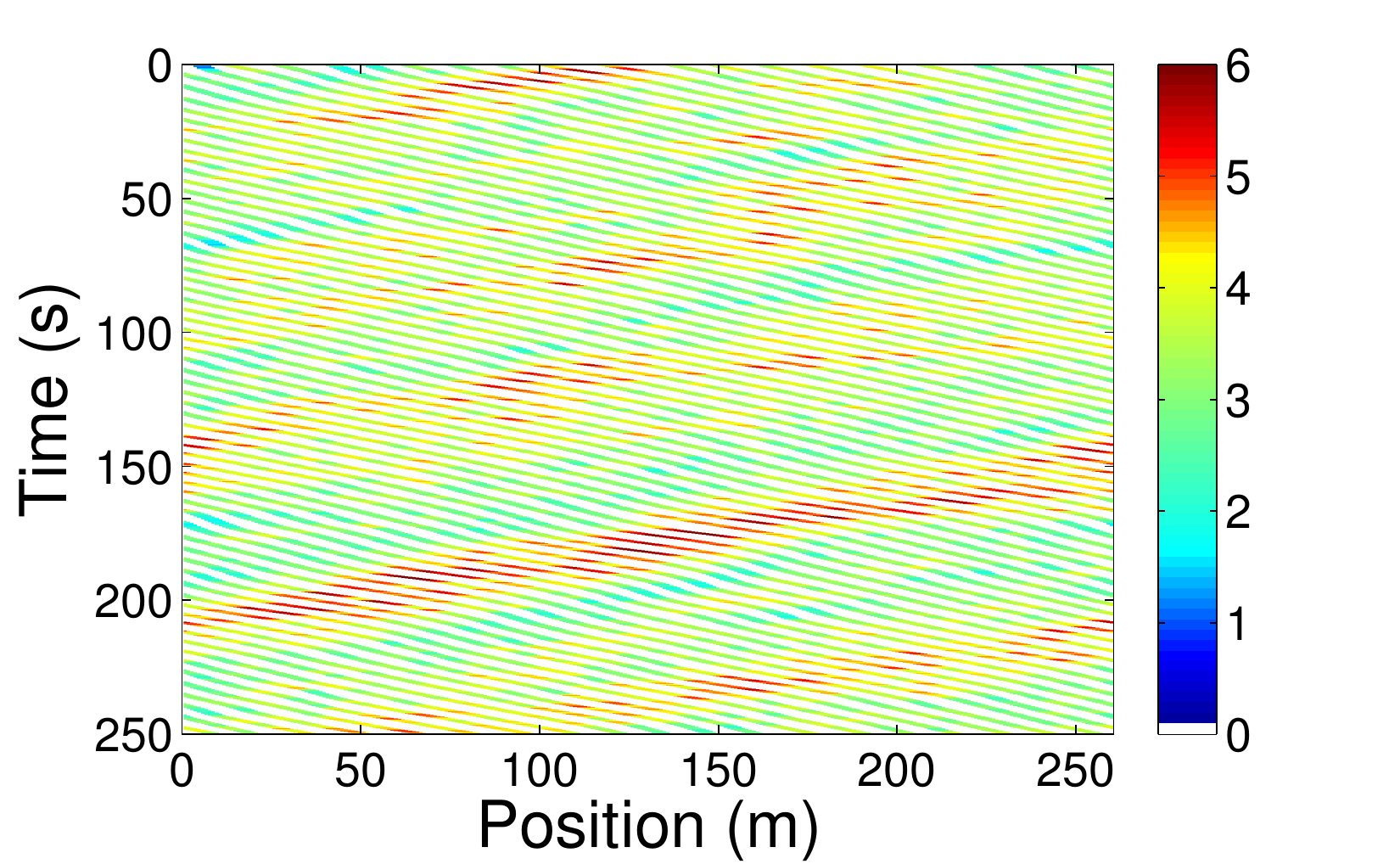}}\\
&\subfloat[]{\includegraphics[width=.25\textwidth]{expB}} 
\subfloat[]{\includegraphics[width=.25\textwidth]{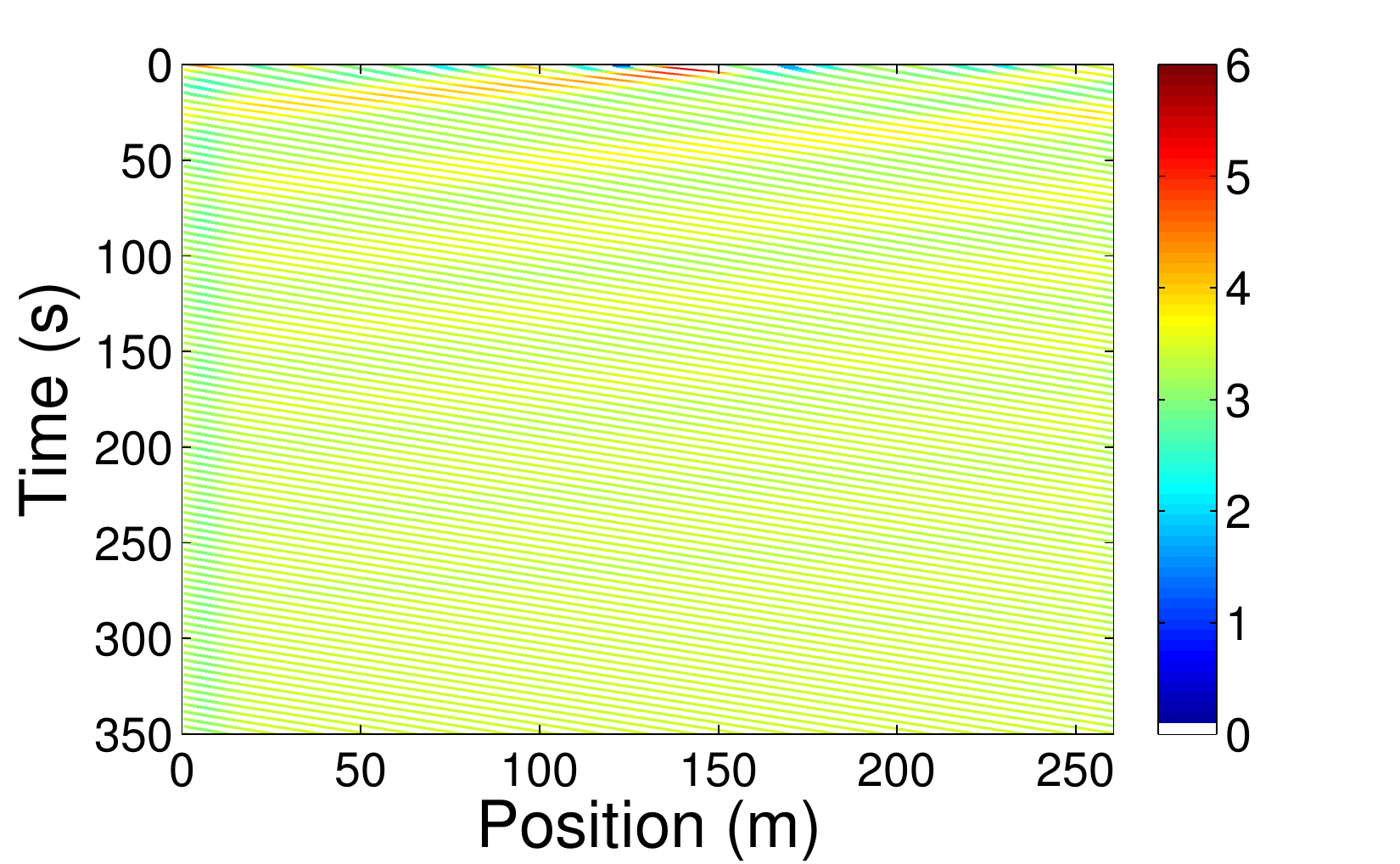}}
\subfloat[]{\includegraphics[width=.25\textwidth]{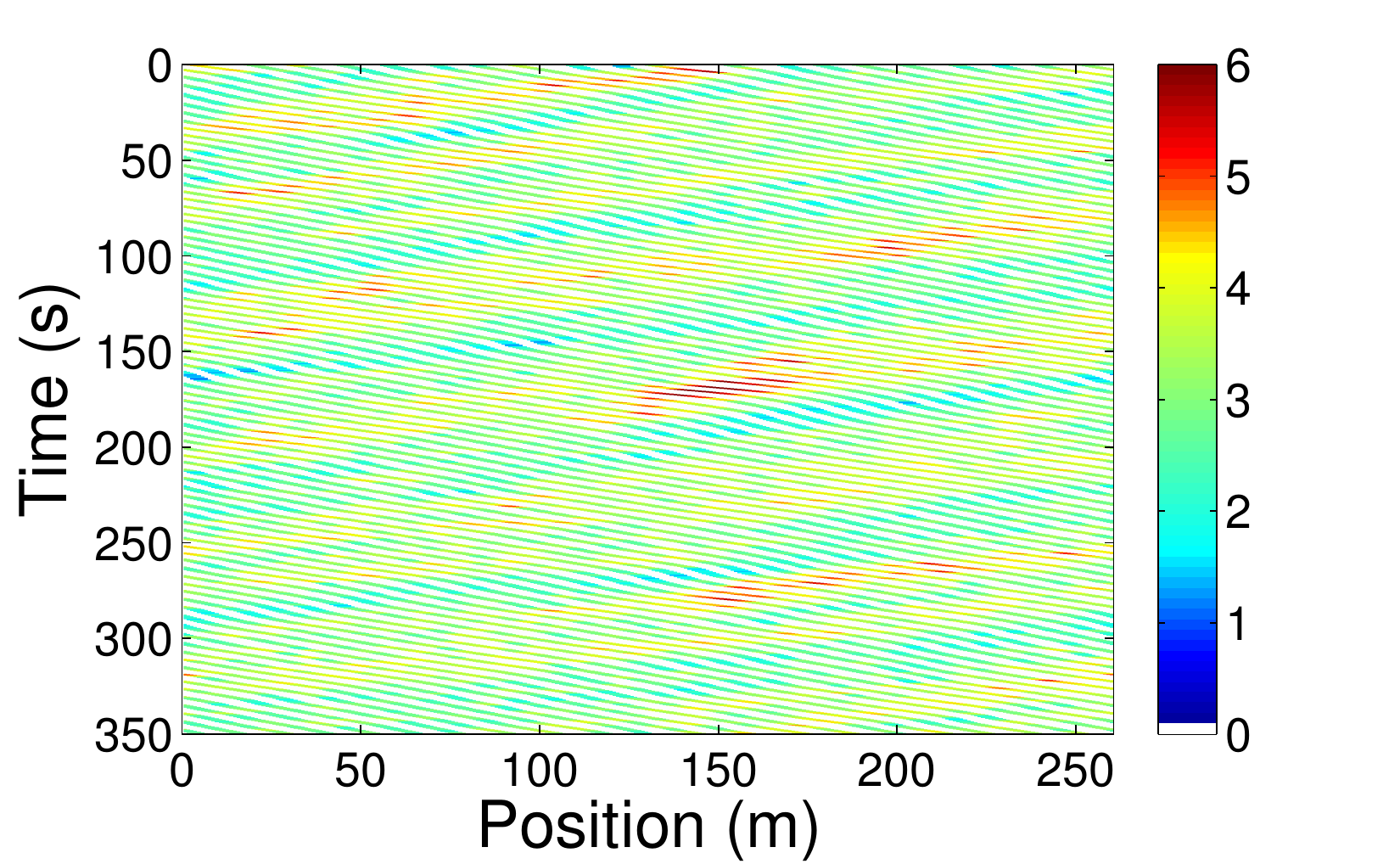}}\\
&\subfloat[]{\includegraphics[width=.25\textwidth]{expD}} 
\subfloat[]{\includegraphics[width=.25\textwidth]{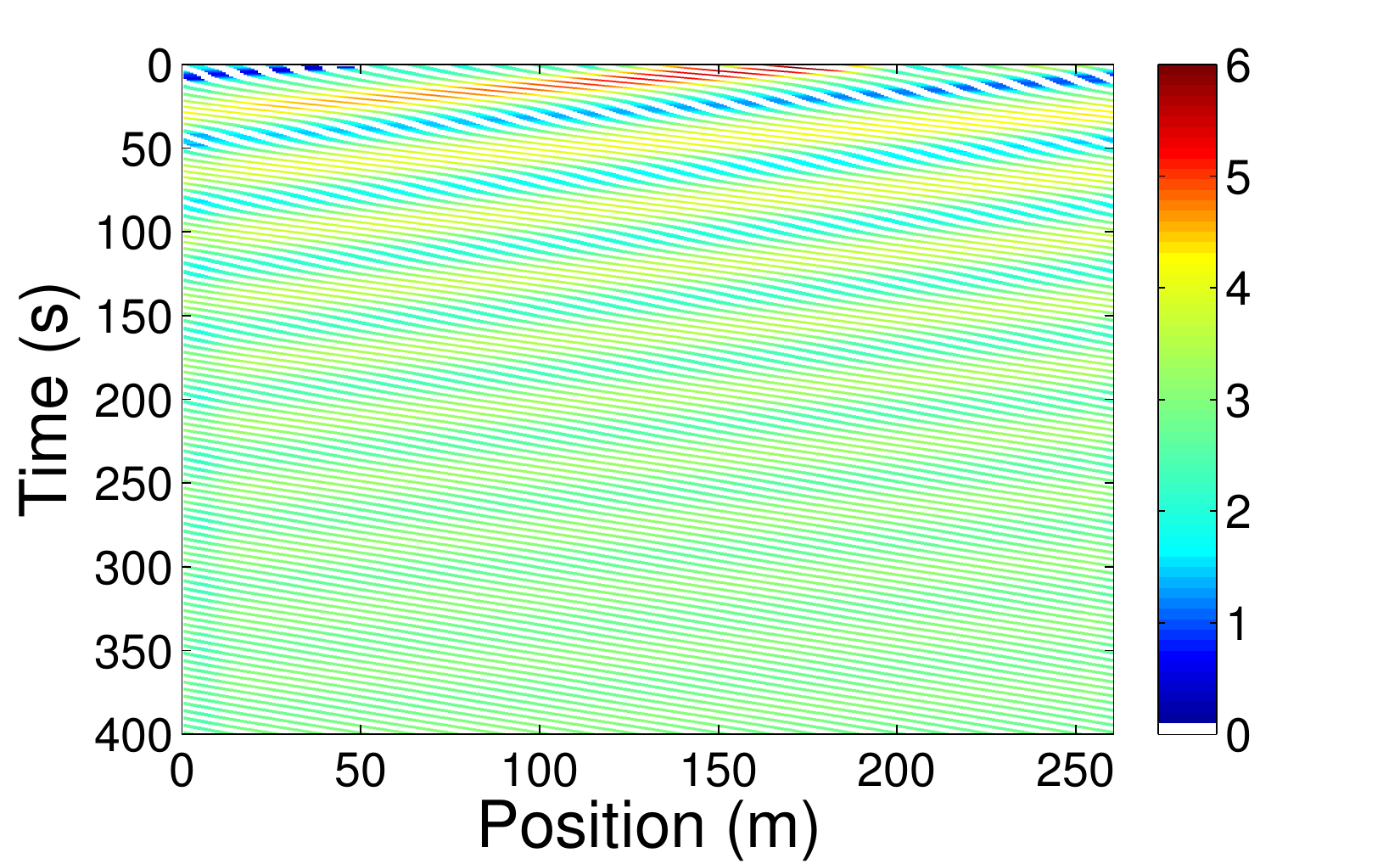}}
\subfloat[]{\includegraphics[width=.25\textwidth]{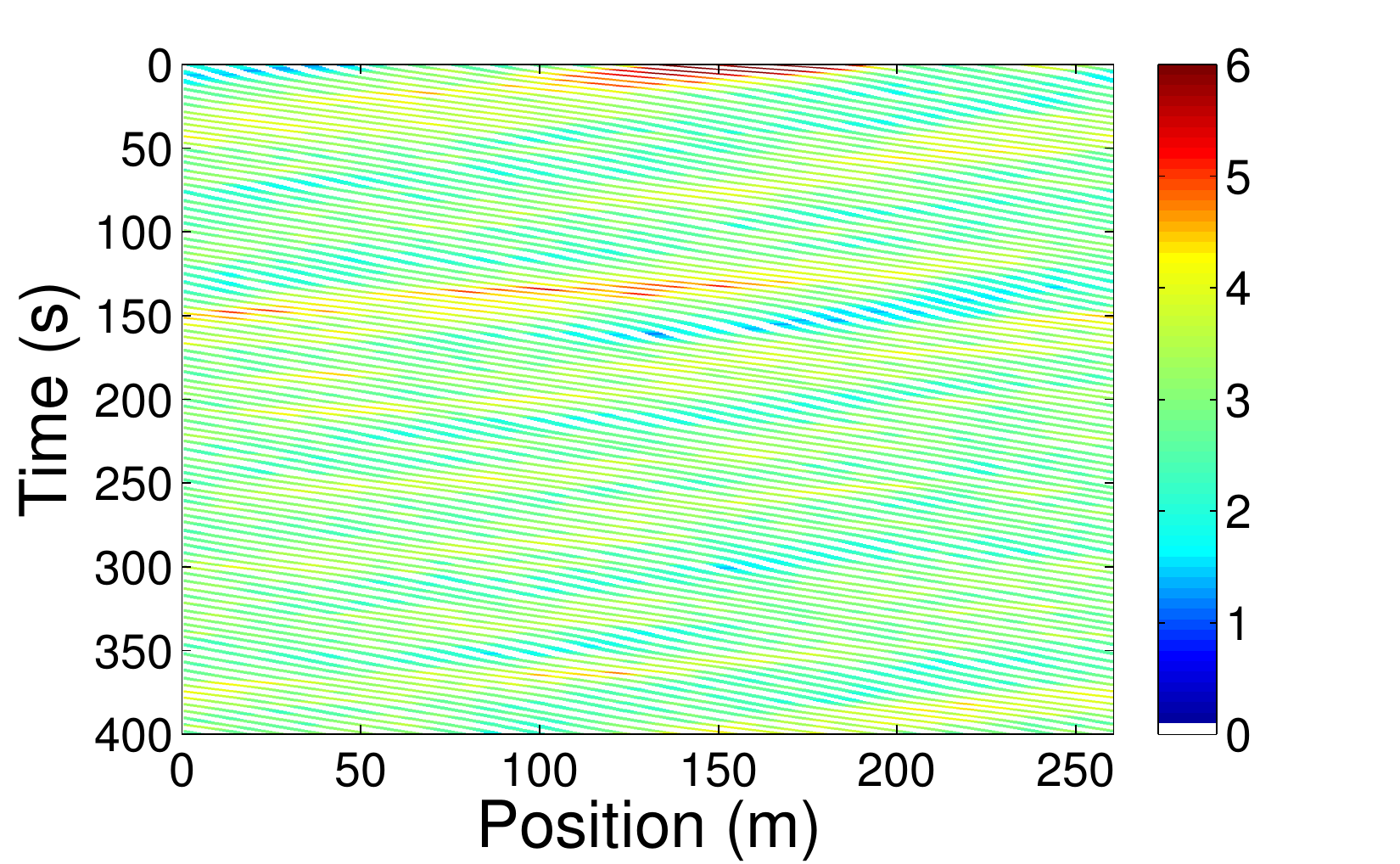}}\\
\end{tabular}
\caption{ Time-space diagrams of the results of experiments A, C, E, B, and D (left panel) and the corresponding simulations by using the IDM and the SIDM, respectively. The experiments A, C, and E are used for calibration while experiments B and D are used for validation (a-c) Experiment A (d-f) Experiment C (g-i) Experiment E (j-l) Experiment B (m-o) Experiment D. The velocity unit in the color bar is $\mathrm{(m/s)}$.}
\label{fig13}
\end{figure}

Next, we give a visual representation of the calibration (against experiments A, C, and E) and the validation (against experiments B and D) results. Figure~\ref{fig11}, Figure~\ref{fig12}, and Figure~\ref{fig13} correspond to the time-space diagrams of the OVM/SOVM, the FVDM/SFVDM, and the IDM/SIDM. From these figures, one can see that the stochastic models outperform the deterministic models in capturing the observed spontaneous appearance of traffic oscillations. The deterministic models either do not reproduce the emergence of traffic oscillations or overestimate their development. From the time-space diagrams in Figure~\ref{fig12} and Table~\ref{t3}, we can deduce that the SFVDM yields the best performance. Interestingly, the SOVM does not consider the sensitivity to the velocity difference but still performs acceptably, as shown in Figure~\ref{fig11}. Accordingly, the stochastic models have better prediction capability than the deterministic models.

To further investigate the role of stochasticity in generating traffic instability, we also assess the potential of the theoretical stability condition (equation (21)) to predict unstable traffic. For each experiment, the predicted traffic states (stable or unstable) are shown in Table~\ref{t4}. Considering the numerical errors, all the stochastic models predict unstable traffic which can indeed be observed in all the experiments. For the deterministic models, except the OVM in all experiments and the FVDM in experiment E, the stability condition predicts stable traffic. For the OVM and the FVDM in experiment E, the equation (21) predicts unstable traffic. However, the period and amplitude of oscillations significantly deviate from empirical observations (See the squared speed standard deviation in Table~\ref{t3}). Hence, our investigation suggests that stochasticity is likely the triggering factor of traffic instability in the studied experiments.


\begin{table*}[h] \footnotesize
\caption{ Calibration of the parameters of the typical stochastic and deterministic traffic models.  CRE denotes the circular road experiment, GP denotes the growth pattern and NG denotes the NGSIM data.} 

\centering

	 \begin{tabular}{ccccccc}
\textbf{Model} & \textbf{Parameter}	 &\textbf{Calibrated for CRE} &\textbf{Calibrated for GP} & \textbf{Calibrated for NG} \\

   \hline
	
    & $a \ \mathrm{(m/s^{2})}$			   & 2.77 & 1.15  & 1.18\\
    & $b \ \mathrm{(m/s^{2})}$			  & 2.61 & 3.76 & 2.24\\
   &$s_{0} \ \mathrm{(m)}$			   & 3.86 & 3.98 & 2.46\\
 IDM &    $\delta$			  & 3.76 & 3.25 & 4.02\\
   &  $T \ \mathrm{(s)}$		        & 1.23 & 1.1 & 1.72\\
 	&	$V_{max} \ \mathrm{(m/s)}$		        & 12.71 & 25 & 21.52\\

		\hline

	  & $\alpha$			  & 2.64  & 1.37 & 2.79 \\
   & $\beta \ \mathrm{(s^{-1})}$			  & 0.45 &  0.02& 0.04\\
 FVDM &  $s_{c} \ \mathrm{(m)}$			  & 6.3 & 13.15 & 9.63\\
   & $\lambda$		         	 & 0.59 & 0.03 & 0.41\\
		&$V_{max} \ \mathrm{(m/s)}$		       & 14.29 & 25 & 21.15\\

      \hline
		   & $\alpha$			 & 1.38 & 2.18 & 0.9\\
   
   & $\beta \ \mathrm{(s^{-1})}$			 &  0.7 & 0.03 & 0.17 \\
	
  OVM &  $s_{c} \ \mathrm{(m)}$			  & 13.98 & 7.37& 9.55\\
	
	&	$V_{max} \ \mathrm{(m/s)}$		        & 13 & 25 & 21\\

      \hline
			
    & $a \ \mathrm{(m/s^{2})}$			   & 1.66 & 1.25 & 1.8\\
    & $b \ \mathrm{(m/s^{2})}$			  & 2.19 & 2.39 & 3\\
   &$s_{0} \ \mathrm{(m)}$			   & 2.76 & 4.1 & 1.59\\
 SIDM &    $\delta$			  & 4.7 & 2.96 & 4.28\\
   &  $T \ \mathrm{(s)}$		        &1.63 & 1.18 & 1.39\\
 	&	$V_{max} \ \mathrm{(m/s)}$		        & 13.51 & 25 & 20\\
 	& $\sigma^2 \ \mathrm{(m/s^2)}$      & 0.14 & 0.05 & 0.28\\
		\hline
		 
   & $\alpha$			  & 2.26 & 3.32 & 2.98\\
   & $\beta \ \mathrm{(s^{-1})}$			  & 0.13 &0.24 & 0.045\\
  &  $s_{c} \ \mathrm{(m)}$			  & 7.66 &6.9 & 7.85\\
  SFVDM & $\lambda$			 & 0.6 &0.91 & 0.56\\
		&$V_{max} \ \mathrm{(m/s)}$		       &13.85 & 25 & 22\\
	& $\sigma^2 \ \mathrm{(m/s^2)}$      & 0.07 & 0.08 & 0.18\\
		\hline

   & $\alpha$			 & 1.62 & 1.55 & 1.2\\
   
   & $\beta \ \mathrm{(s^{-1})}$			 &  1.06 & 0.12 & 0.4\\
	
  SOVM &  $s_{c} \ \mathrm{(m)}$			  & 11.21 &5.76 &8.19 \\
	
	&	$V_{max} \ \mathrm{(m/s)}$		        & 13.15 & 25 &19.38\\
	
	& $\sigma^2 \ \mathrm{(m/s^2)}$       &0.09 & 0.22 & 0.31\\
		\hline
\end{tabular}  
		\label{t2}
\end{table*}

\begin{table*}[h] \footnotesize
	\caption{ Performance index (PI) and squared speed standard deviation error of each traffic model. The squared standard deviation error denotes the speed standard deviation term in $I^{2}$, See equation (51).}
\centering
	 \begin{tabular}{cccccc}
\textbf{Model} &\textbf{Experiment}	& \textbf{Density (veh/m)} & \textbf{Performance index (PI)} & \textbf{Squared standard deviation error} \\

  \hline
    & A			& 0.077   & 1.02/0.9 & 0.39/0.04 \\

   & B			 & 0.077 & 0.97/0.88 & 0.26/0.01\\
	
 IDM/SIDM   &C		  & 0.085 & 0.63/0.69 & 0.03/0.01 \\
	
		 & D		  & 0.081 & 0.99/1.02  & 0.03/0.02 \\
		
		&E		  & 0.073 &  1.06/1.04 & 0.27/0.05\\

		\hline
		
		 & A			& 0.077   & 1.02/0.86 & 0.42/0.02 \\

   & B			 & 0.077 &  0.98/0.89 & 0.23/0.01  \\
	
   FVDM/SFVDM & C		  & 0.085 & 0.65/0.64 & 0.1/0.01\\
	
		 & D		  & 0.081 & 0.9/0.86 & 0.07/0.02 \\
		
		&E		  & 0.073 &  1.08/1.01 & 0.02/0.03\\

			\hline
			
		 & A			& 0.077   & 0.86/0.86 & 0.22/0.03\\

   & B			 & 0.077 & 0.8/0.85  & 0.09/0.01 \\
	
 OVM/SOVM  & C		  & 0.085 & 0.71/0.66 & 0.01/0.01 \\
	
		 & D		  & 0.081 & 1.59/0.9 & 0.39/0.03 \\
		
		&E		  & 0.073 &  1.24/1.09 & 0.09/0.04\\
	
			\hline

\end{tabular}
	\label{t3}
\end{table*}

\begin{table*}[h] \footnotesize
\caption{ The prediction capability of each traffic model using equation (21). DM denotes the deterministic case $\sigma=0$ and SM denotes the stochastic case.}
\centering
\begin{tabular}{ccccccc}
\textbf{Experiments} & \textbf{Model} &\textbf{Stability DM} &\textbf{Stability SM} \\

     \hline
    &  SIDM & Stable & Unstable \\
  A &  SFVDM & Stable & Unstable \\
   &  SOVM & Unstable & Unstable \\
		\hline
		
		    &  SIDM & Stable & Unstable \\
  B &  SFVDM & Stable & Unstable \\
   &  SOVM & Unstable & Unstable \\  
 
		\hline
	    &  SIDM & Stable & Unstable \\
  C &  SFVDM & Stable & Unstable \\
   &  SOVM & Unstable & Unstable \\  
 
	\hline
			    &  SIDM & Stable & Unstable \\
  D &  SFVDM & Stable & Unstable \\
   &  SOVM & Unstable & Unstable \\  
 
		\hline
			    &  SIDM & Stable & Unstable \\
  E &  SFVDM & Unstable & Unstable \\
   &  SOVM & Unstable & Unstable \\  

		\hline
\end{tabular}  
		\label{t4}
\end{table*}

\subsection {The NGSIM data}

In this subsection, we calibrate the deterministic and stochastic models against NGSIM data collected on U.S. Highway 101, Los Angeles, California, on June 15, 2005. In this context, many studies have shown that lane-changing maneuvers are not responsible for the observed traffic breakdown on lane 1 \citep{chen1,chen2,lec1}. We aim to assess how the deterministic and stochastic  models can reproduce the observed stop-and-go waves on the highway segment.  

For calibration, we consider the lane 1 and vehicle trajectories leading to congestion in a specified time interval $[670\ \mathrm{s}-740\ \mathrm{s}]$. This time interval is extracted from the data sampling time interval from 07:50 a.m to 08:05 a.m on the same day. The same performance index (51) will be used to calibrate the traffic models. Next, to assess the prediction capability of each stochastic traffic model, we will validate the calibrated parameters by using the time interval $[350\ \mathrm{s}-450\ \mathrm{s}]$ corresponding to another congestion wave.  

\begin{figure}[H]
\centering
\begin{tabular}{cc} 
\centering
&\subfloat[]{\includegraphics[width=.25\textwidth]{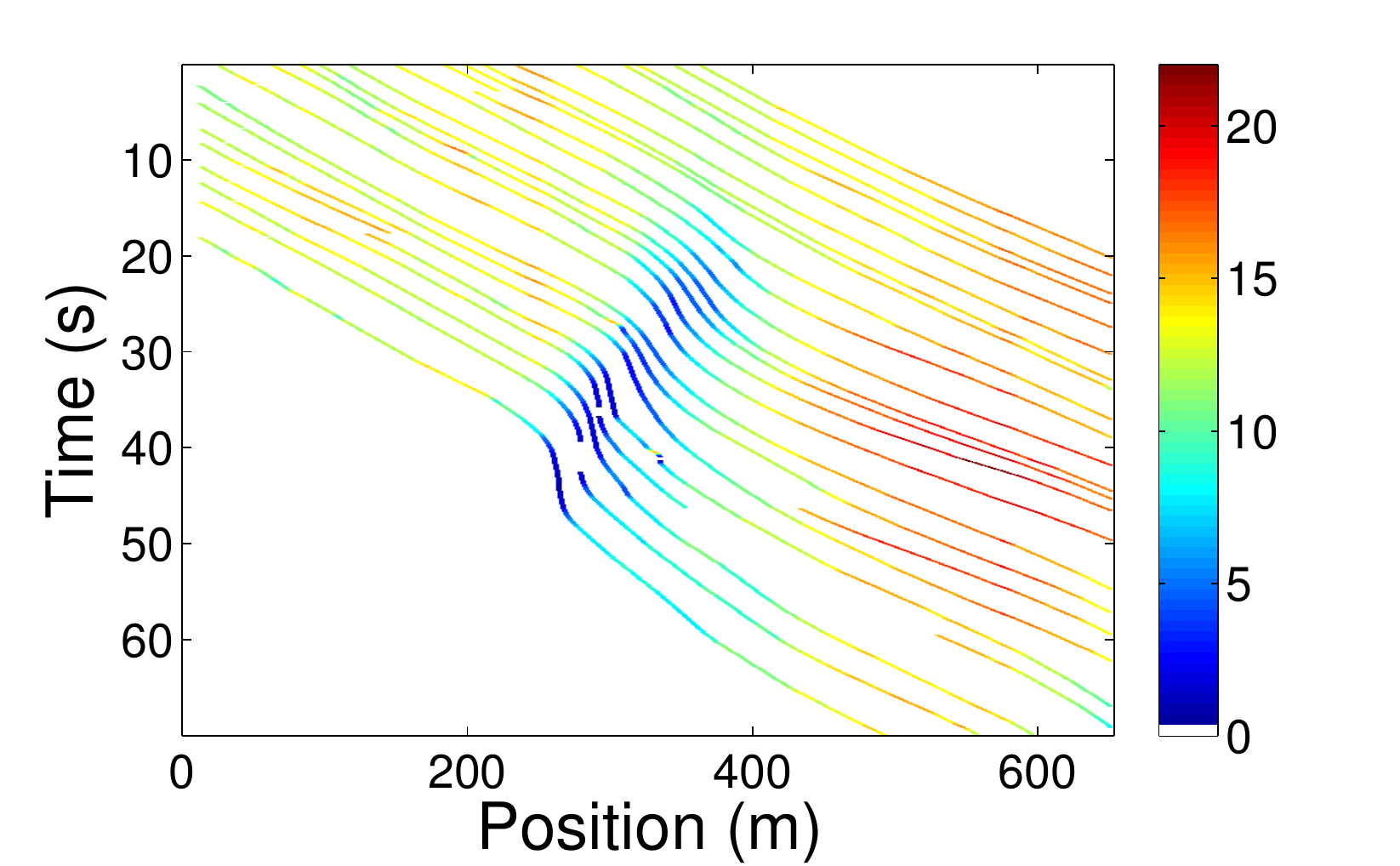}} 
\subfloat[]{\includegraphics[width=.25\textwidth]{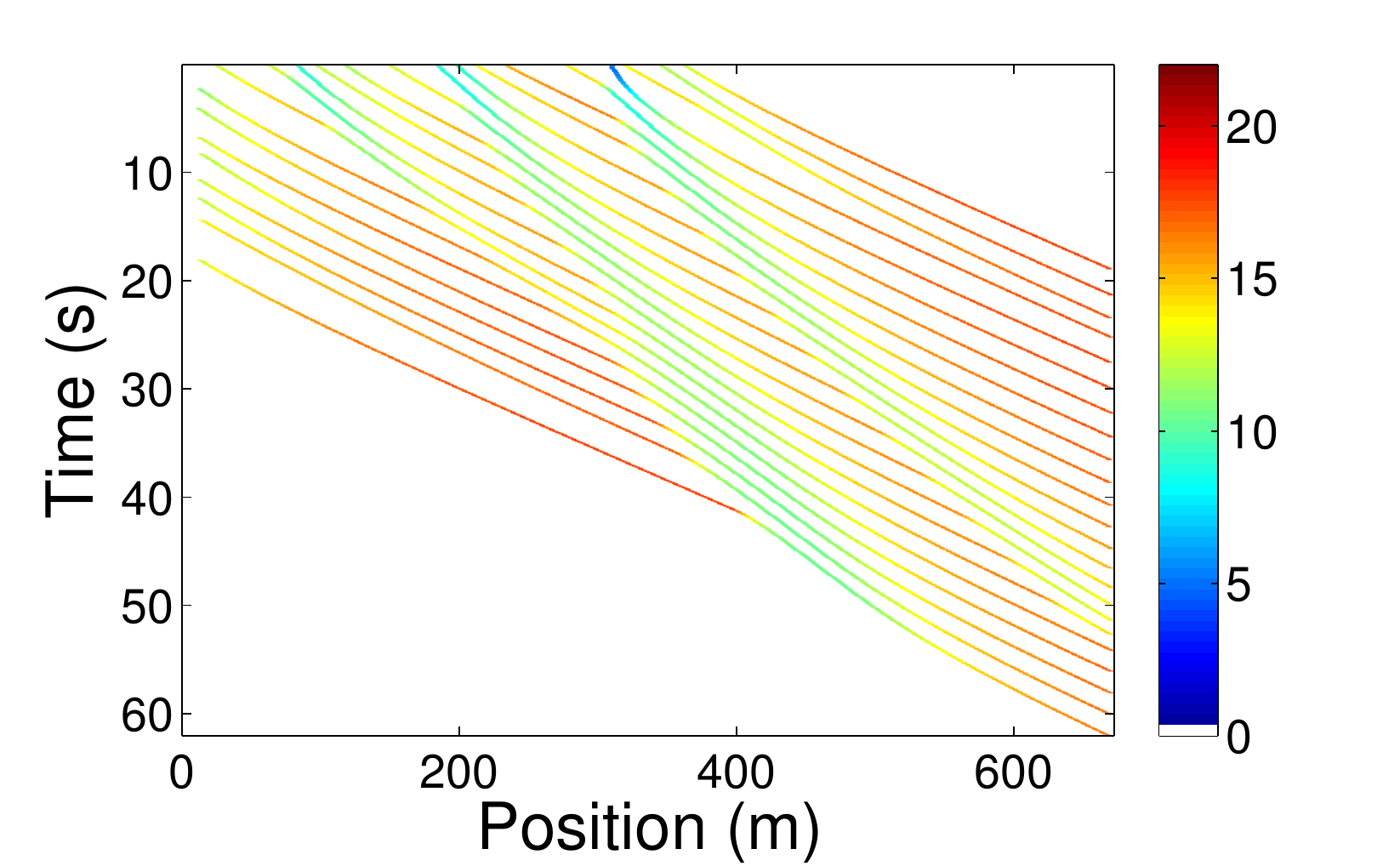}}
\subfloat[]{\includegraphics[width=.25\textwidth]{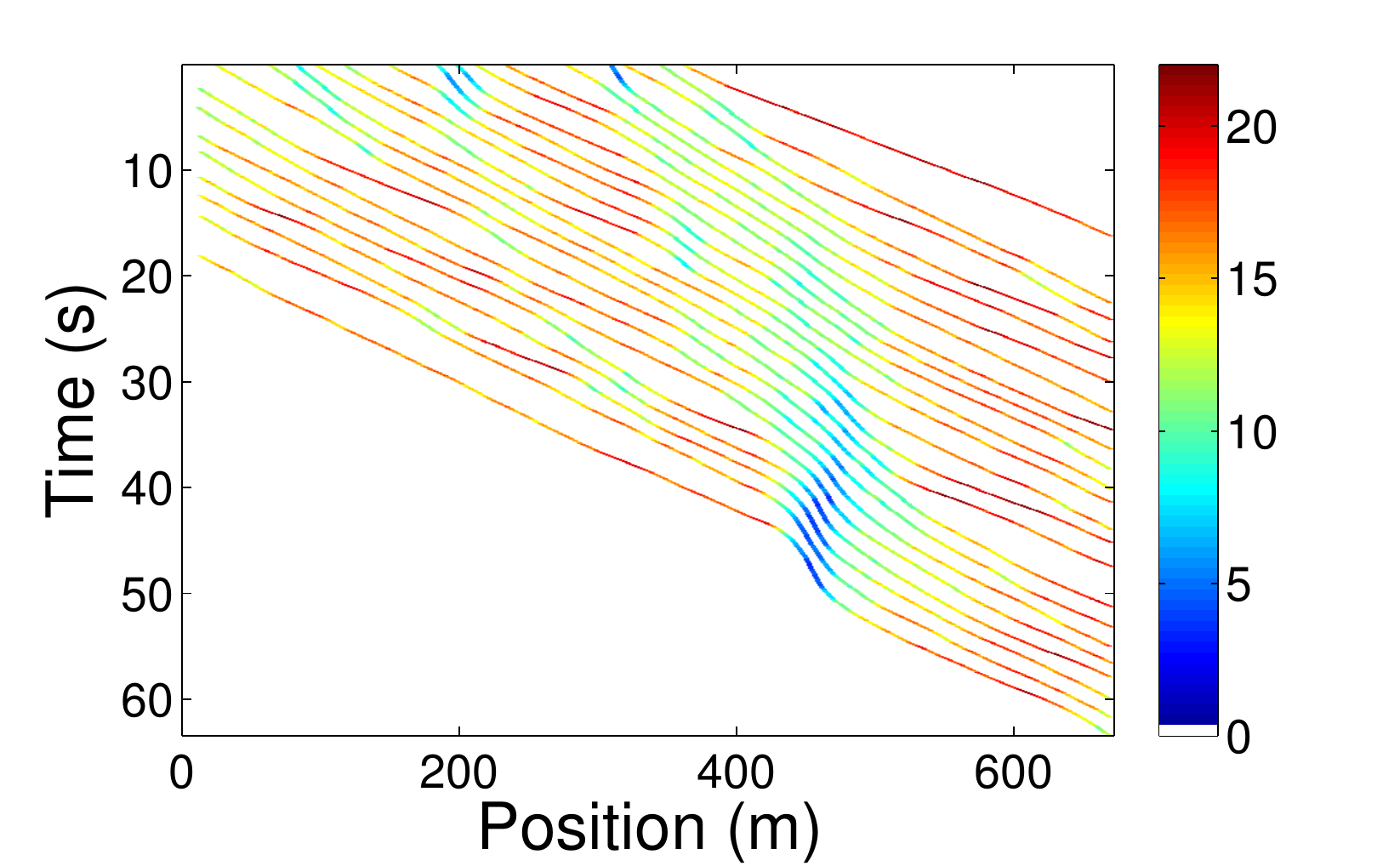}}\\
&\subfloat[]{\includegraphics[width=.25\textwidth]{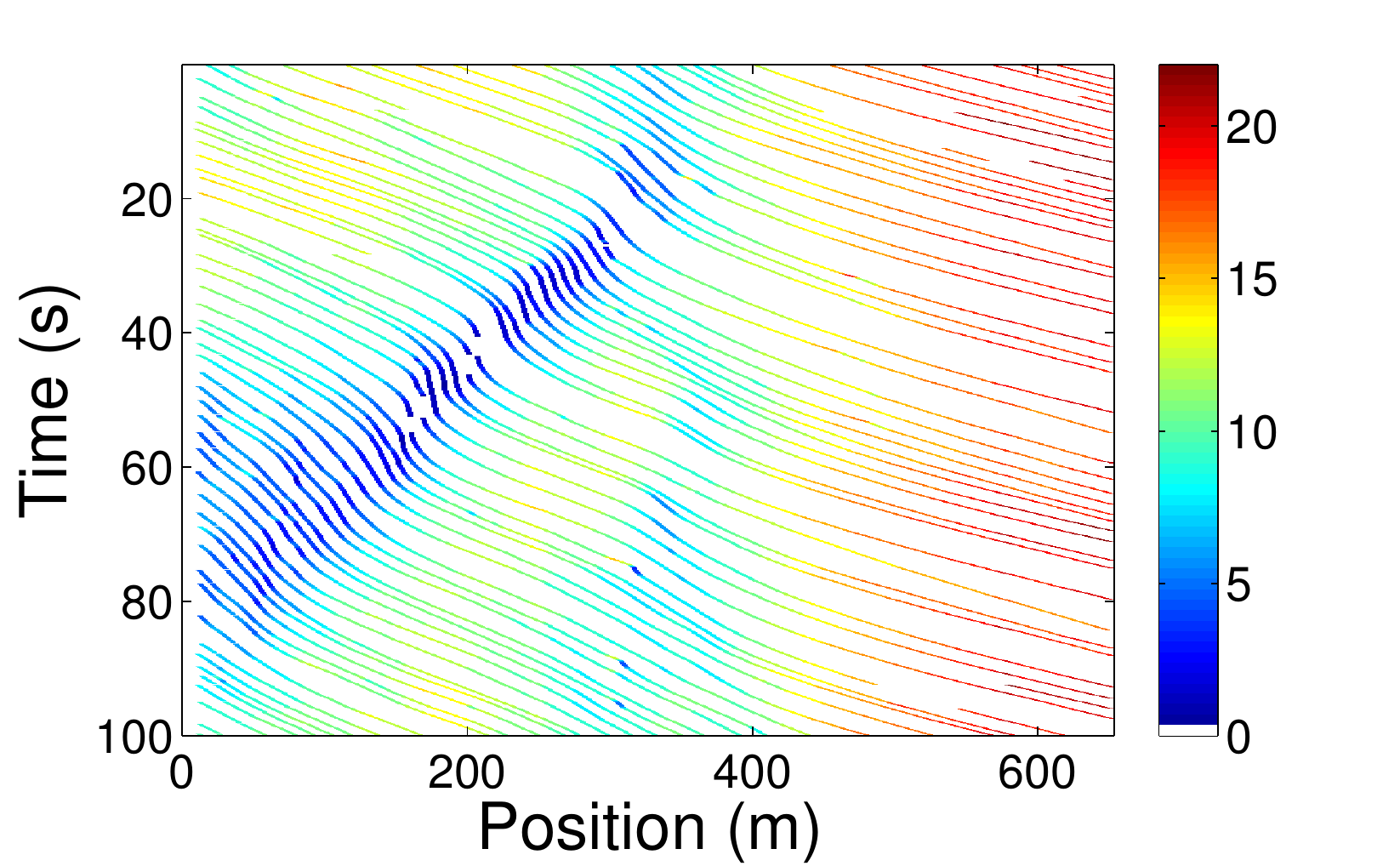}} 
\subfloat[]{\includegraphics[width=.25\textwidth]{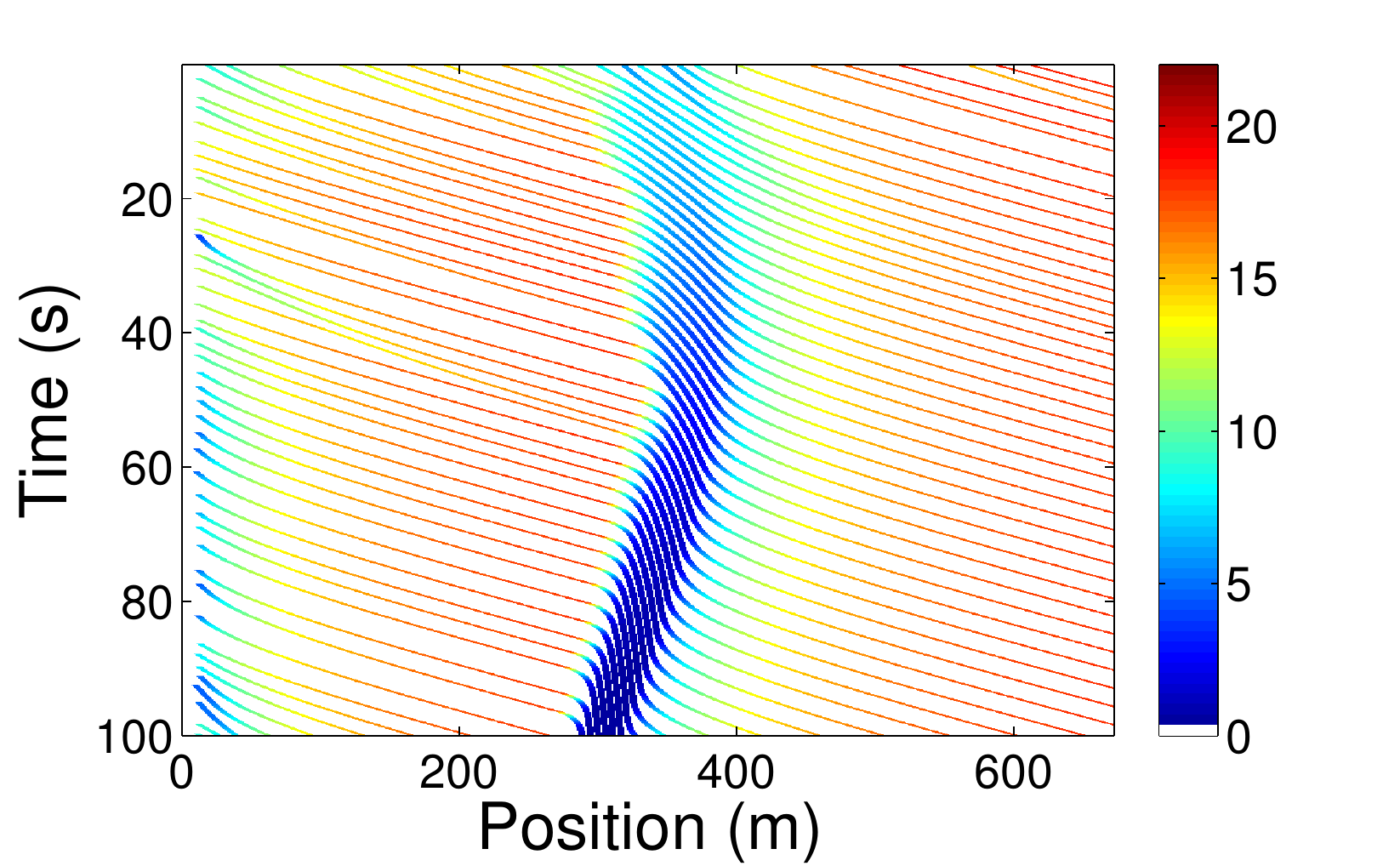}}
\subfloat[]{\includegraphics[width=.25\textwidth]{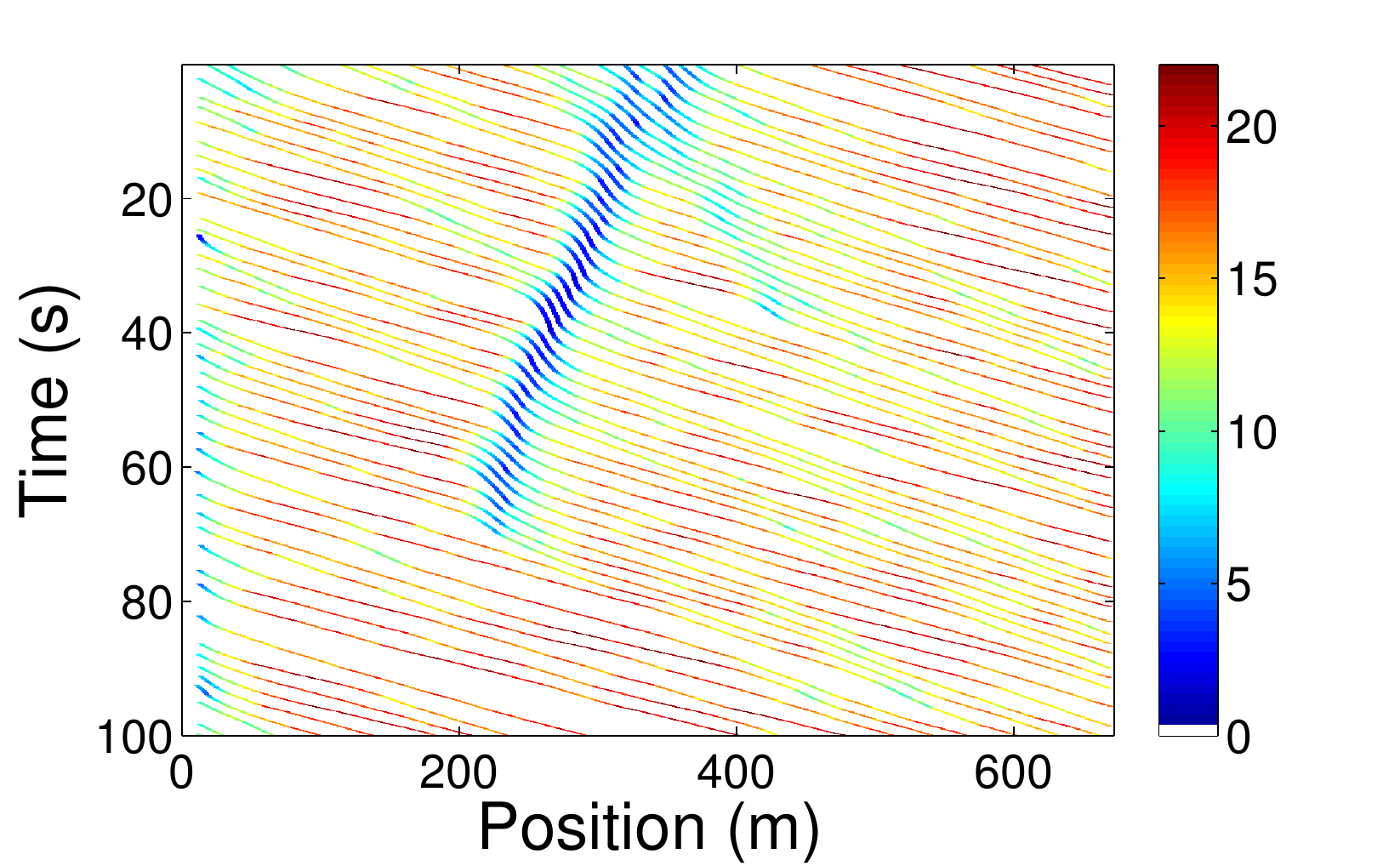}}\\
\end{tabular}
\caption{ Time-space diagrams obtained after calibration and validation of the OVM and the SOVM against NGSIM data. (a) NGSIM data in the time interval $[670s-740s]$ (b-c) Calibration by using the OVM and the SOVM, respectively (d) NGSIM data in the time interval $[350s-450s]$ (e-f) Validation by using the OVM and the SOVM, respectively. The velocity unit in the color bar is $\ \mathrm{(m/s)}$.}
\label{fig14}
\end{figure}

\begin{figure}[H]
\centering
\begin{tabular}{cc} 
\centering
&\subfloat[]{\includegraphics[width=.25\textwidth]{ngexp}} 
\subfloat[]{\includegraphics[width=.25\textwidth]{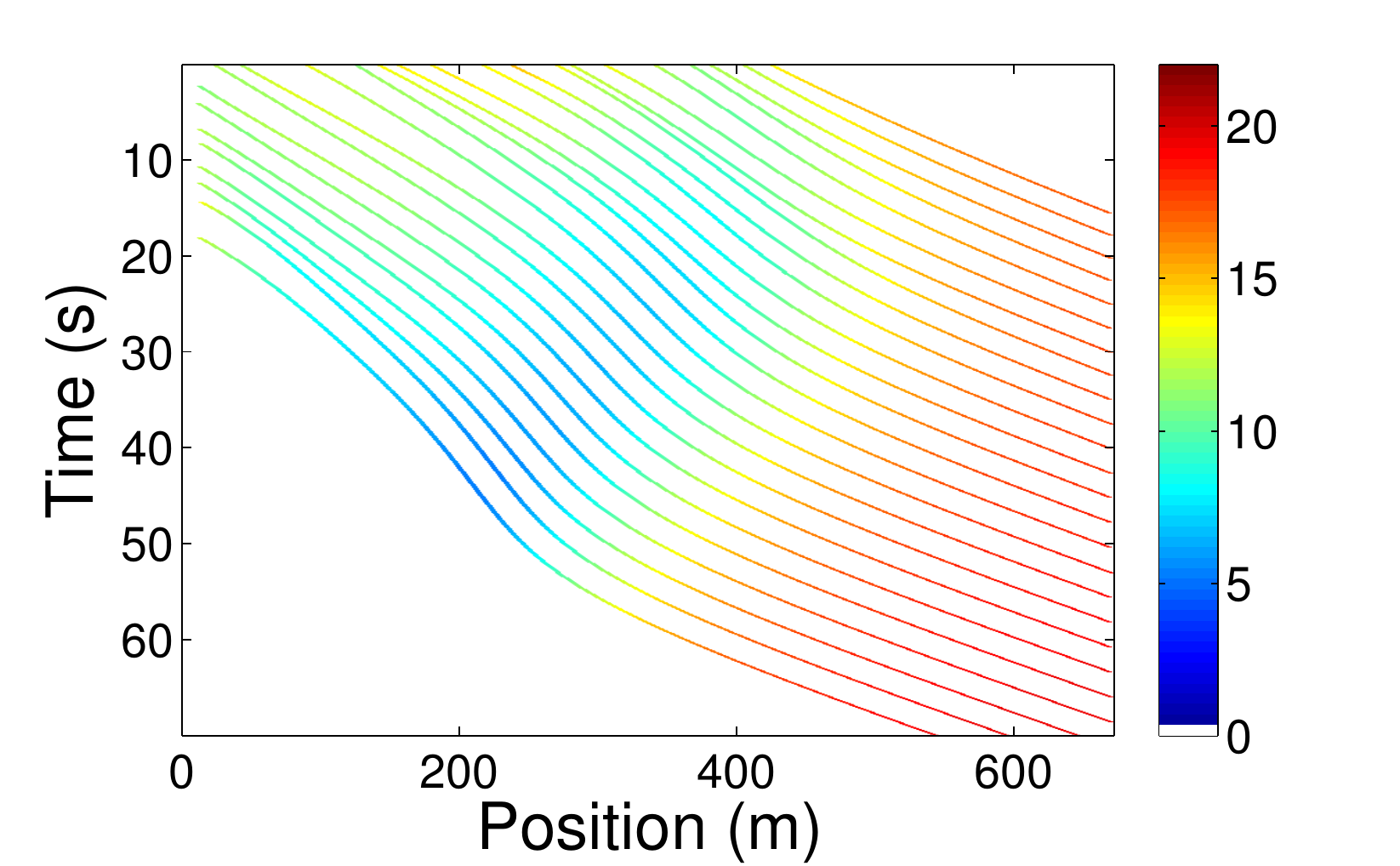}}
\subfloat[]{\includegraphics[width=.25\textwidth]{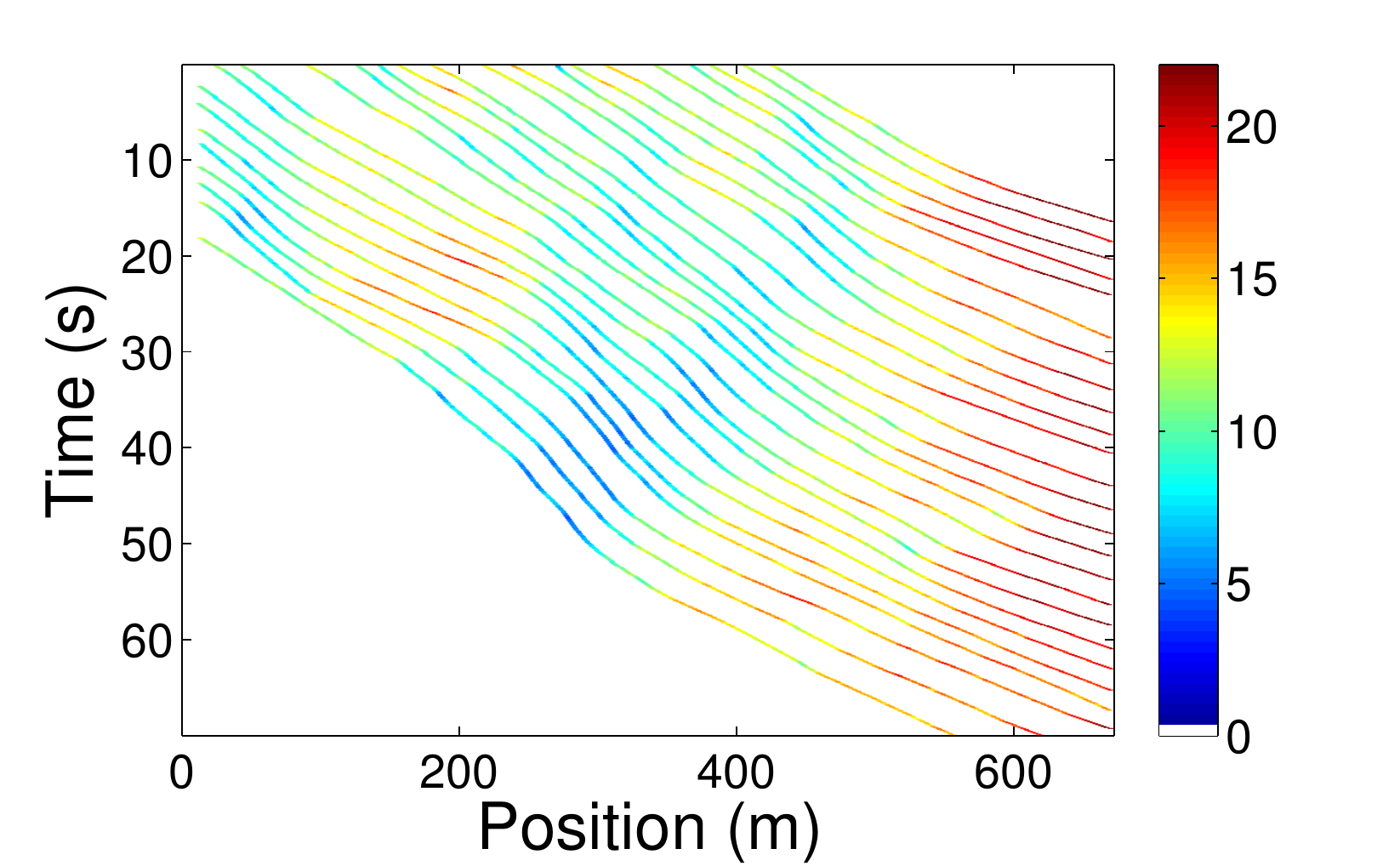}}\\
&\subfloat[]{\includegraphics[width=.25\textwidth]{ngsimval}} 
\subfloat[]{\includegraphics[width=.25\textwidth]{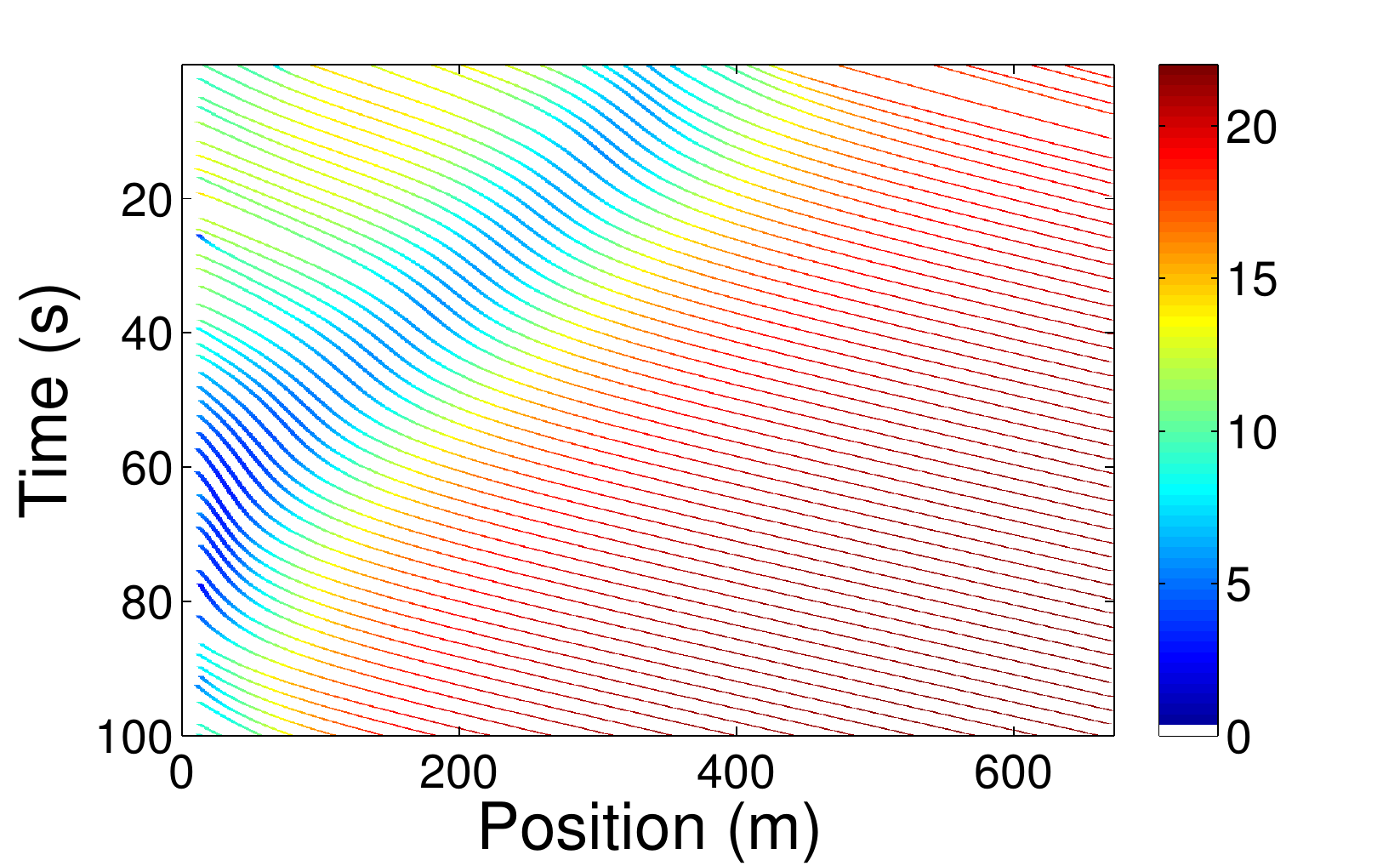}}
\subfloat[]{\includegraphics[width=.25\textwidth]{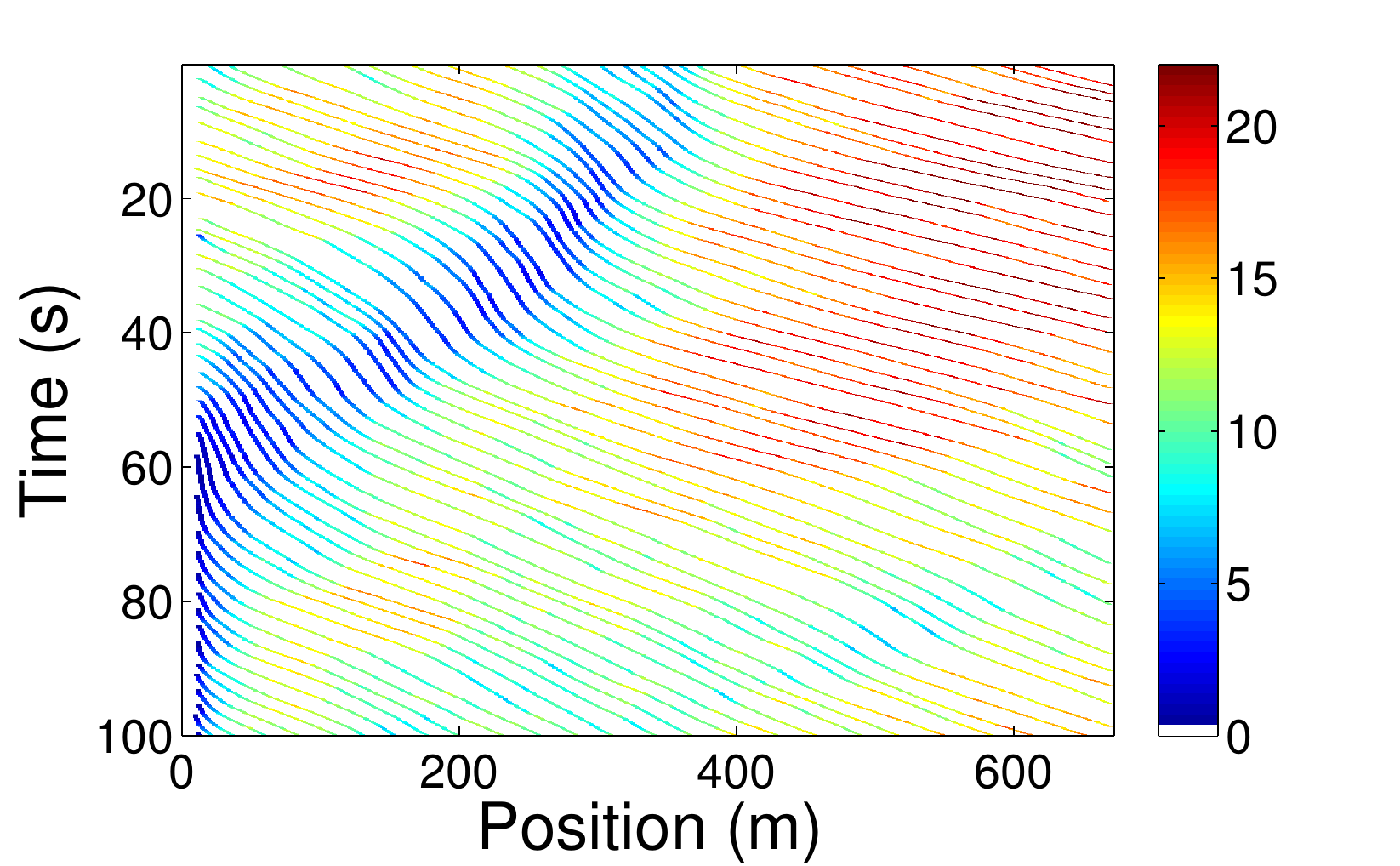}}\\
\end{tabular}
\caption{ Time-space diagrams obtained after calibration and validation of the FVDM and the SFVDM against NGSIM data. (a) NGSIM data in the time interval $[670s-740s]$ (b-c) Calibration by using the FVDM and the SFVDM, respectively (d) NGSIM data in the time interval $[350s-450s]$ (e-f) Validation by using the FVDM and the SFVDM, respectively. The velocity unit in the color bar is $\ \mathrm{(m/s)}$.}
\label{fig15}
\end{figure}

\begin{figure}[H]
\centering
\begin{tabular}{cc} 
\centering
&\subfloat[]{\includegraphics[width=.25\textwidth]{ngexp}} 
\subfloat[]{\includegraphics[width=.25\textwidth]{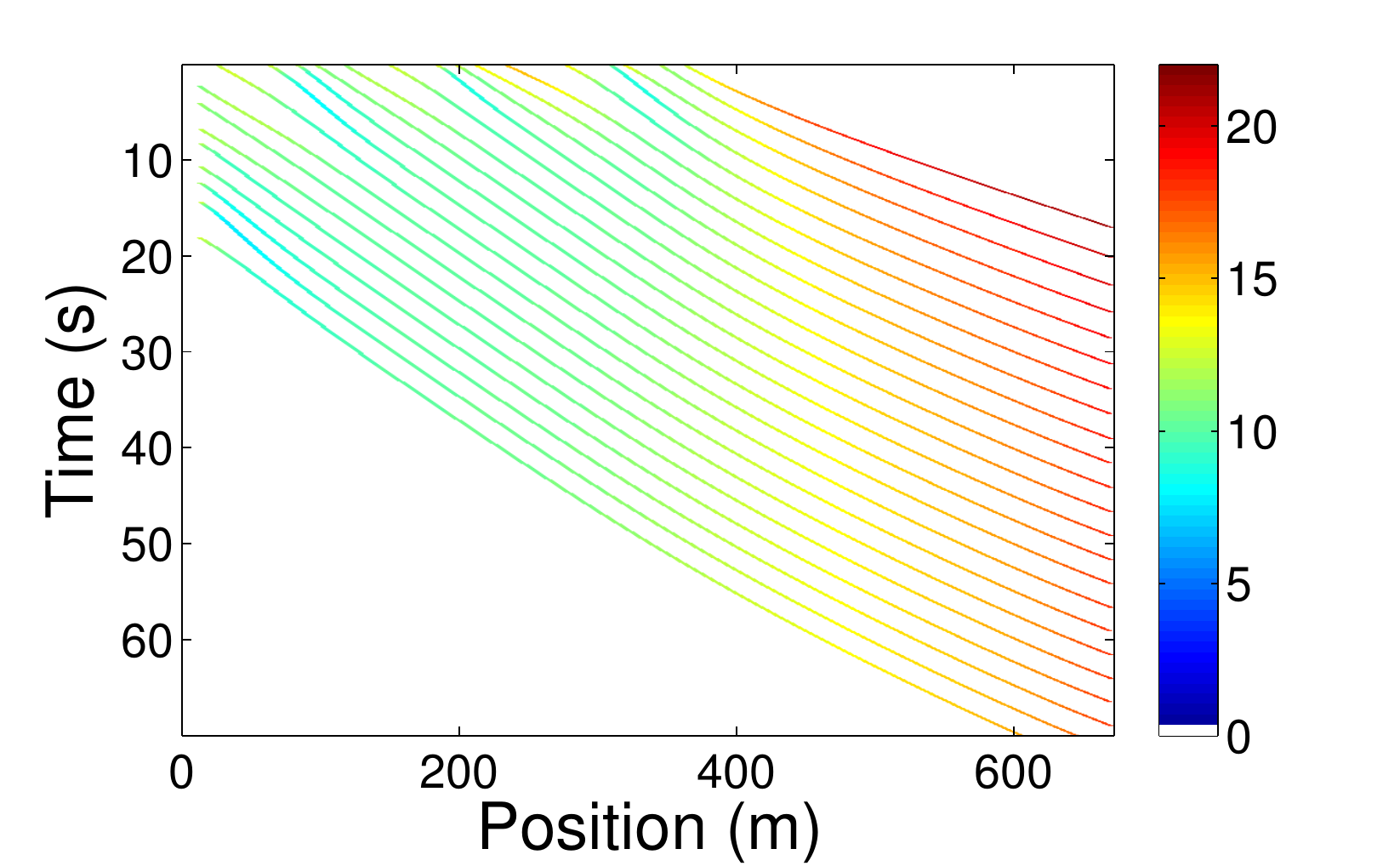}}
\subfloat[]{\includegraphics[width=.25\textwidth]{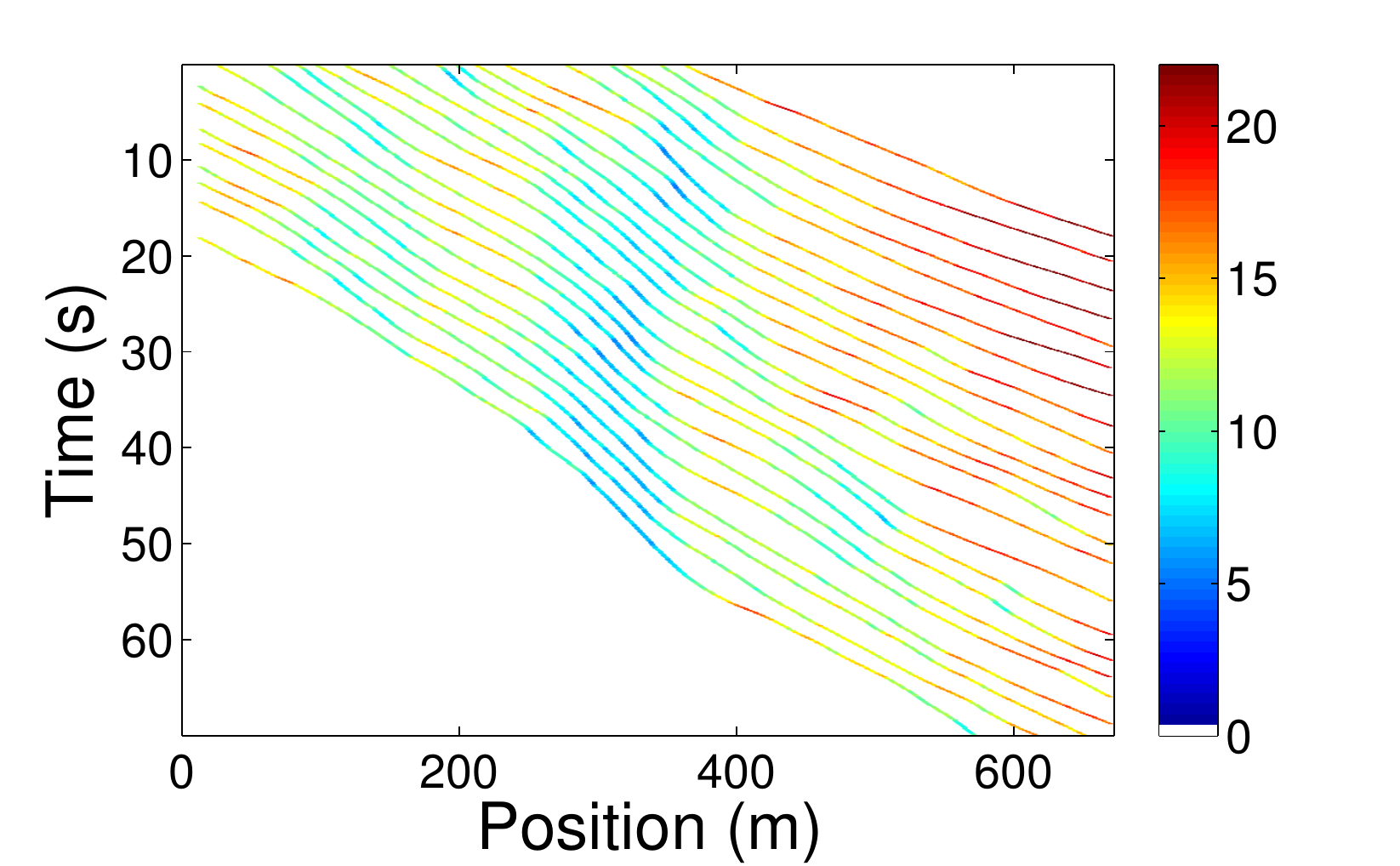}}\\
&\subfloat[]{\includegraphics[width=.25\textwidth]{ngsimval}} 
\subfloat[]{\includegraphics[width=.25\textwidth]{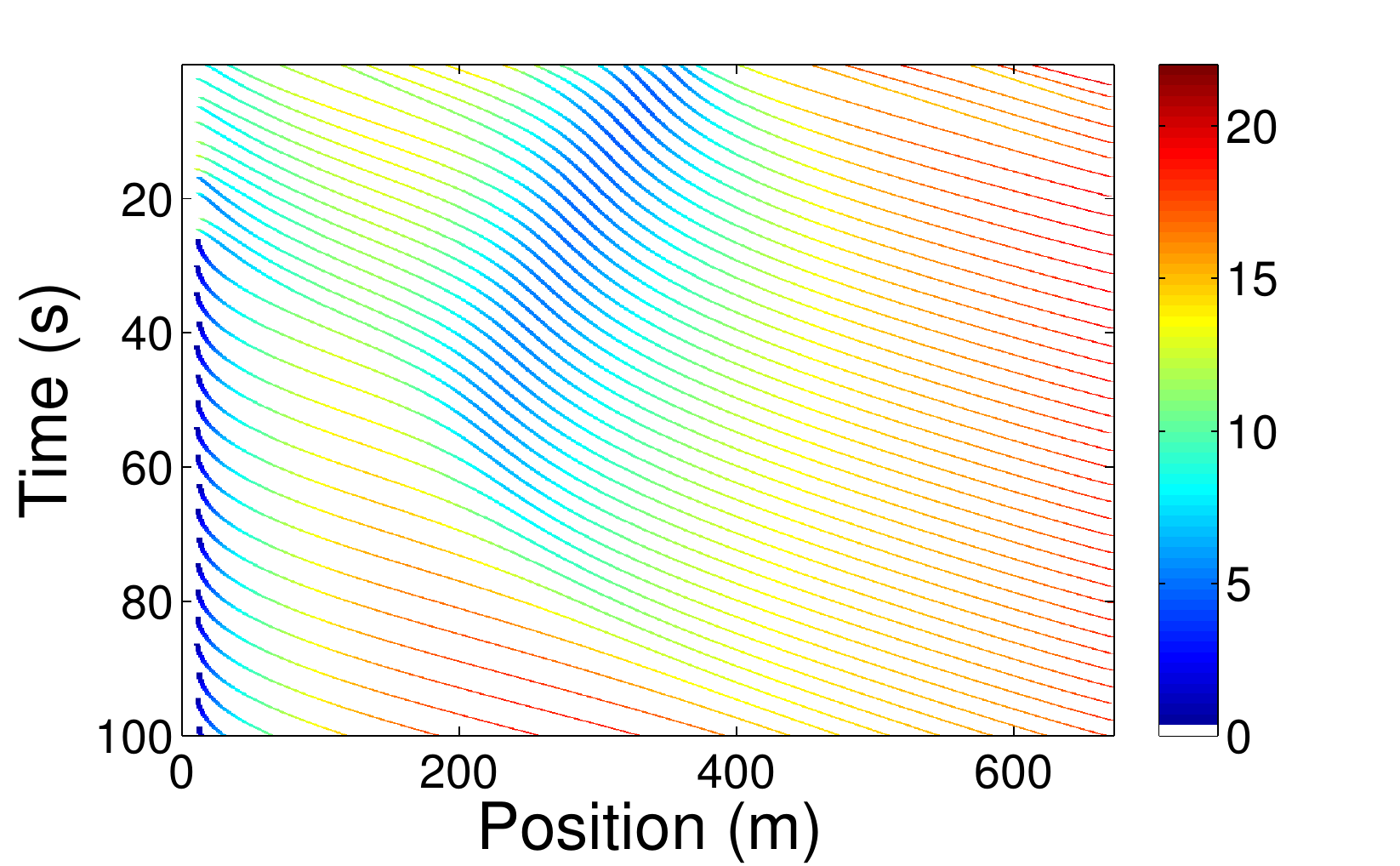}}
\subfloat[]{\includegraphics[width=.25\textwidth]{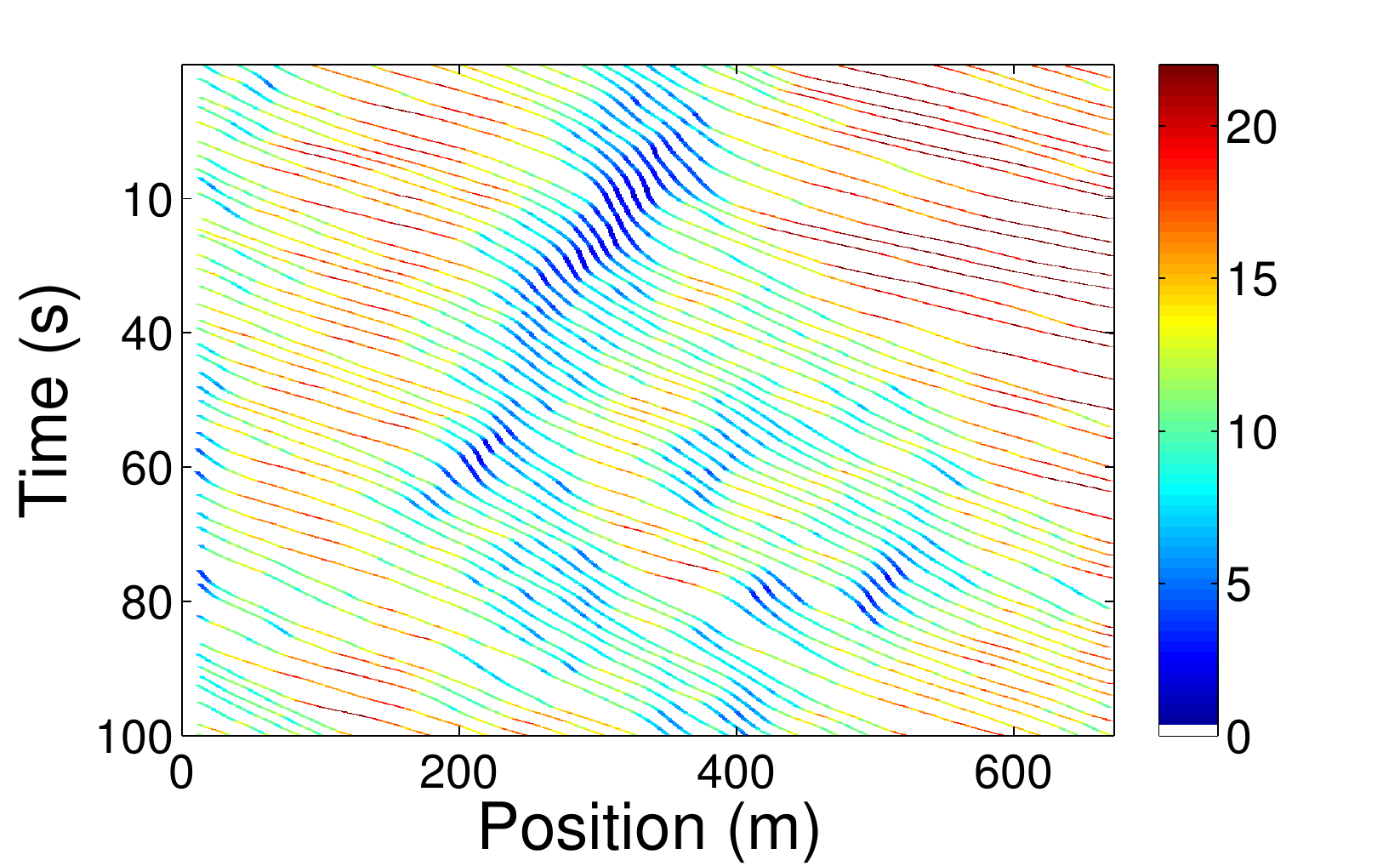}}\\
\end{tabular}
\caption{ Time-space diagrams obtained after calibration and validation of the IDM and the SIDM against NGSIM data. (a) NGSIM data in the time interval $[670s-740s]$ (b-c) Calibration by using the IDM and the SIDM, respectively (d) NGSIM data in the time interval $[350s-450s]$ (e-f) Validation by using the IDM and the SIDM, respectively. The velocity unit in the color bar is $\ \mathrm{(m/s)}$.}
\label{fig16}
\end{figure}

Table~\ref{t2} shows the calibration results of the different deterministic and stochastic traffic models. Table~\ref{t5} shows the obtained performance index after calibration and validation. To give a visual representation, we present in Figure~\ref{fig14}, Figure~\ref{fig15} and Figure~\ref{fig16} the time-space diagrams corresponding to each traffic model, i.e. the OVM/SOVM, the FVDM/SFVDM, and the IDM/SIDM after calibration and validation. For both calibration and validation, we remark that the performance of the stochastic models is quite better than the deterministic models. For the deterministic models, we remark that the FVDM yields the best result in terms of minimizing the performance index and the IDM comes in the second place. For the stochastic models, the SIDM yields the best result in terms of minimizing the performance index and the SFVDM comes in the second place. The OVM and the SOVM both yield worse performance. Finally, except the IDM, Figure~\ref{fig14}, Figure~\ref{fig15} and Figure~\ref{fig16} show that both deterministic and stochastic models can (numerically) predict unstable traffic.

\begin{table*}[h] \footnotesize
\caption{Calibration and validation of the deterministic models and stochastic models against NGSIM data. The squared standard deviation error denotes the speed standard deviation term in $I^{2}$, See equation (51). }
\centering
	 \begin{tabular}{cccccc}
\textbf{Model} & \textbf{Calibration/Validation} & \textbf{Performance index (PI)} & \textbf{Squared standard deviation error} \\

    \hline
  IDM/SIDM    	&  Calibration &  	 3.81/2.98  & 2.06/0.3 \\

   & Validation &  	3.41/4.02			 & 0.58/0.35 \\

		\hline
		
	 FVDM/SFVDM    &  Calibration	&  3.12/3.69 	   &  0.67/0.93\\

   &  Validation &   		4.39/3.78		 &  1/0.9\\

			\hline
			
		 OVM/SOVM  &  Calibration  	&  	 4.82/4.64  & 2.55/1.2 \\

   &   Validation &   5.13/5.01			 & 2.14/1 \\

			\hline

\end{tabular}
	
	\label{t5}
\end{table*}

\subsection {The growth pattern of traffic oscillations}

Previous studies have shown that the deterministic traffic models exhibit an initially convex growth pattern because of the unique dependency of the velocity on the inter-vehicular gap in the steady-state \citep{tian2}. However, as observed in real traffic, stochastic traffic models show a concave growth pattern because the relationship between the velocity and spacing covers a 2D region in the velocity-spacing plane \citep{jiang2}. The effect of stochasticity on the growth pattern of traffic oscillations has been also rigorously studied \citep{wang1,tian4}. From another perspective, \cite{mak2} have recently developed a model considering the nature of drivers and vehicle characteristics without an explicit introduction of noise, and reproduced the observed concave growth pattern of traffic oscillations.   

In this subsection, we calibrate the different stochastic and deterministic traffic models against empirical data to see how the studied stochastic models can reproduce the concave growth pattern of traffic oscillations. To this aim, we will consider the 51 car-platoon experiment conducted by \cite{jiang2}. In this experiment, 51 cars were initially lined up bumper to bumper before the leading car was instructed to adopt a given constant velocity profile, more details can be found in the paper of \cite{jiang2}. In the following, we will simulate the 51 car-platoon dynamics by adopting the same process for the different stochastic and deterministic models, i.e. the SIDM, the SFVDM, the SOVM, the IDM, the FVDM, and the OVM. The time step will be similar to the experiment which was $dt=0.1 \ \mathrm{s}$. The simulation time is $T=1500 \ \mathrm{s}$. For both experiments and simulations, we will consider three leader's velocities $v_{l}= 30 \ \mathrm{km/h} \  (8.33  \ \mathrm{m/s})$ ,  $v_{l}= 5 \ \mathrm{km/h} \ (1.38 \ \mathrm{m/s})$ and $v_{l}= 40 \ \mathrm{km/h} \ (11.11 \ \mathrm{m/s})$. The first leader's velocity will be used for calibration while the second and third ones will be used for validation. Since we focus on the growth pattern of speed standard deviation, we use the following performance index to calibrate the models:

\begin{equation}
I^{2} =\frac{1}{N} \sum_{k=1}^{k=N}{(s_{k} - \hat{s}_{k})}^{2}  
\end{equation} 

\noindent where N denotes the total number of vehicles equipped with GPS tracking device, $s_{k}$ is the simulated speed standard deviation of vehicle $k$, and $\hat{s}_{k}$ is the observed speed standard deviation of vehicle $k$.  Note that for the experimental data, 27 GPS tracking devices were used. Hence, only the trajectories of 27 vehicles were recorded. Furthermore, due to some problems related to signal or GPS tracking devices, some vehicle locations are missing.

\begin{figure}[H]
\centering
\begin{tabular}{c} 
\centering
\subfloat[]{\includegraphics[width=.5\textwidth]{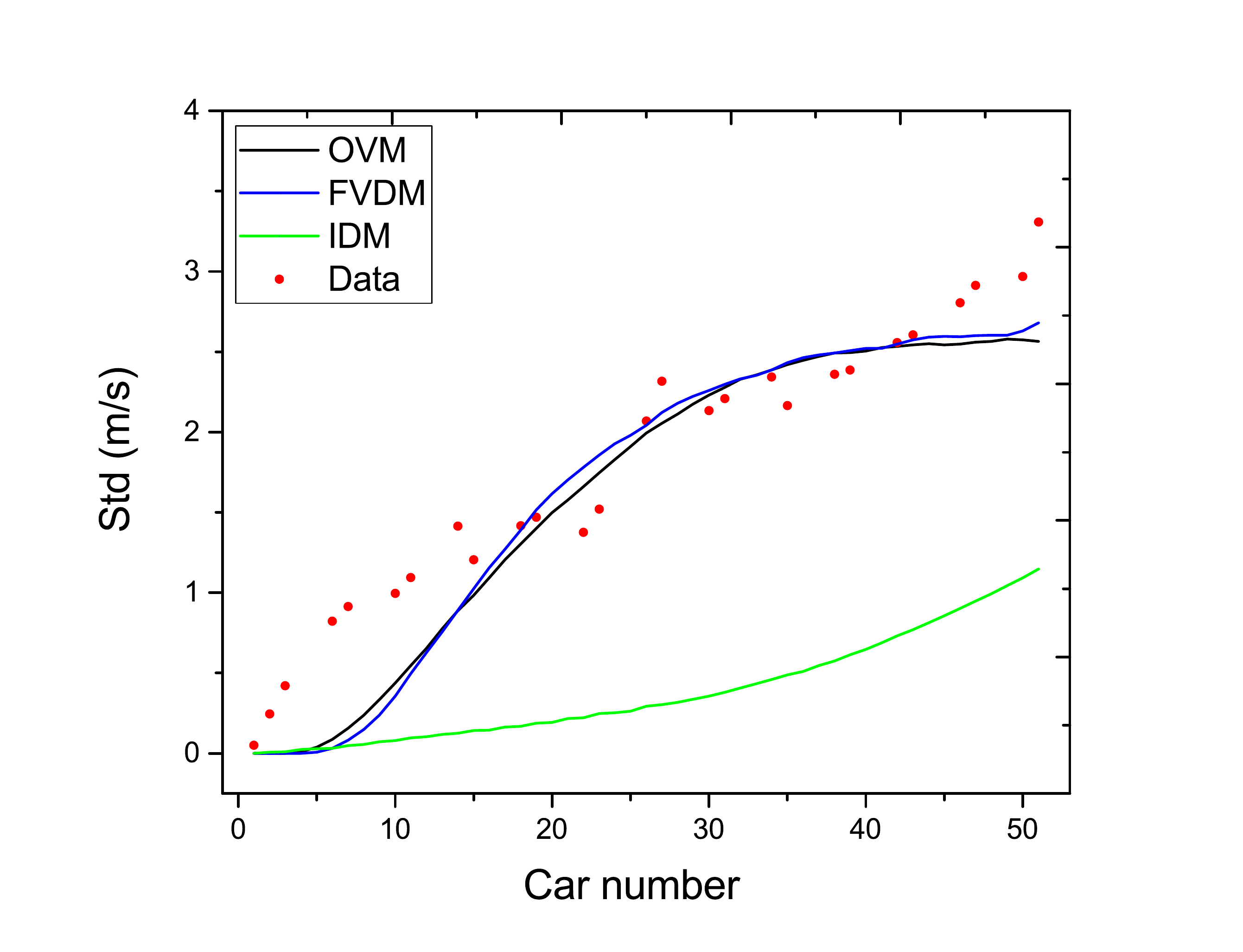}}
\subfloat[]{\includegraphics[width=.5\textwidth]{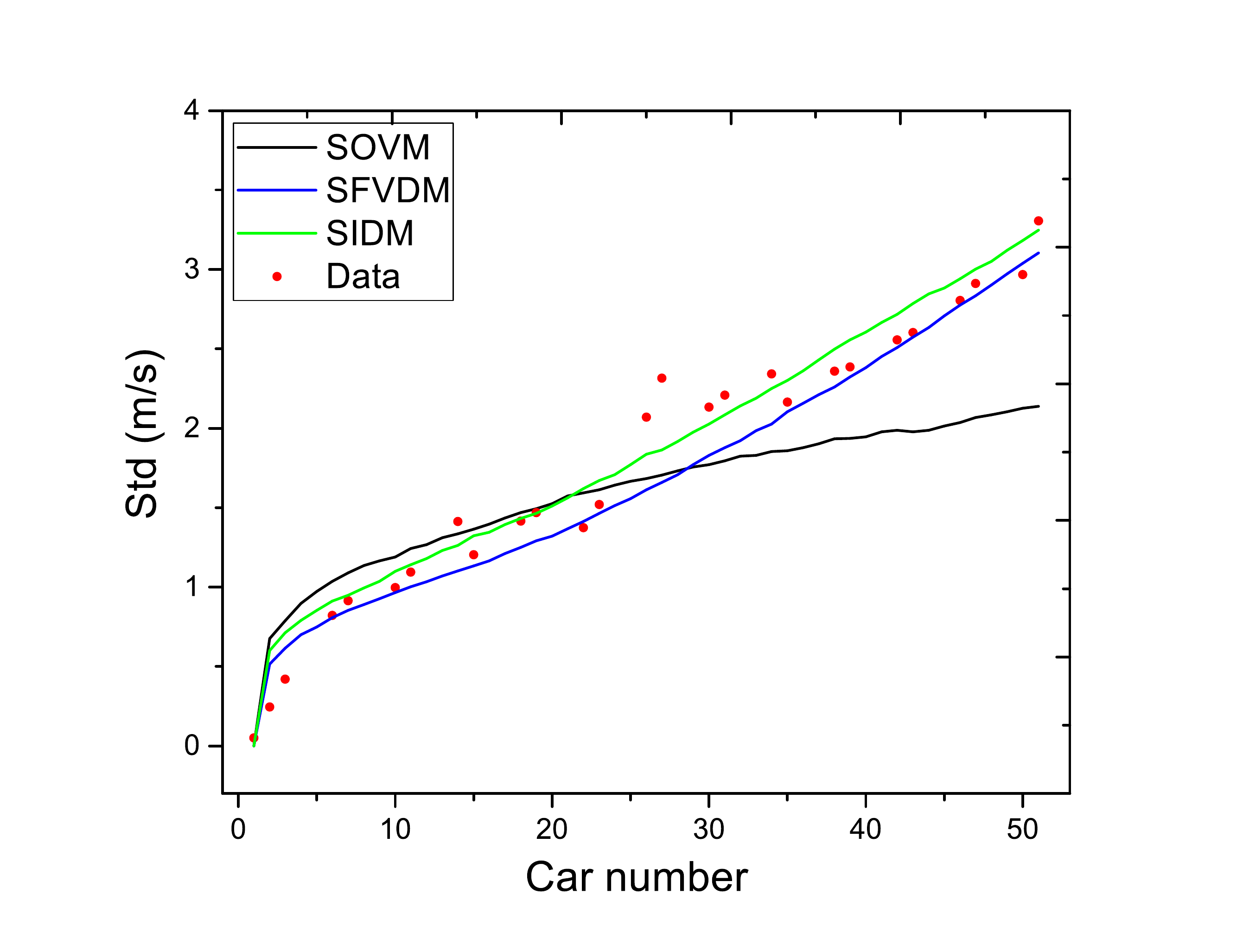}}\cr
\subfloat[]{\includegraphics[width=.5\textwidth]{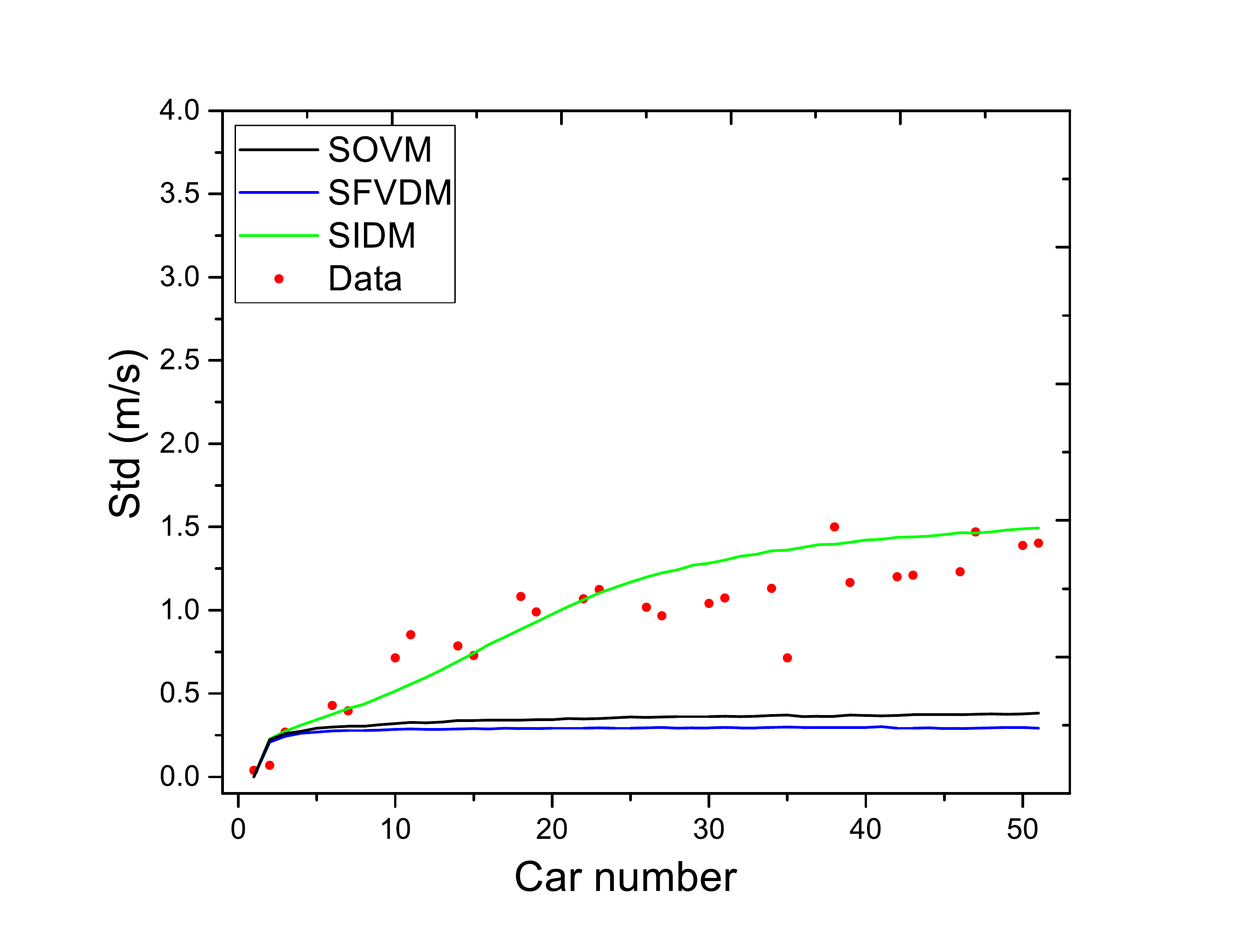}}
\subfloat[]{\includegraphics[width=.5\textwidth]{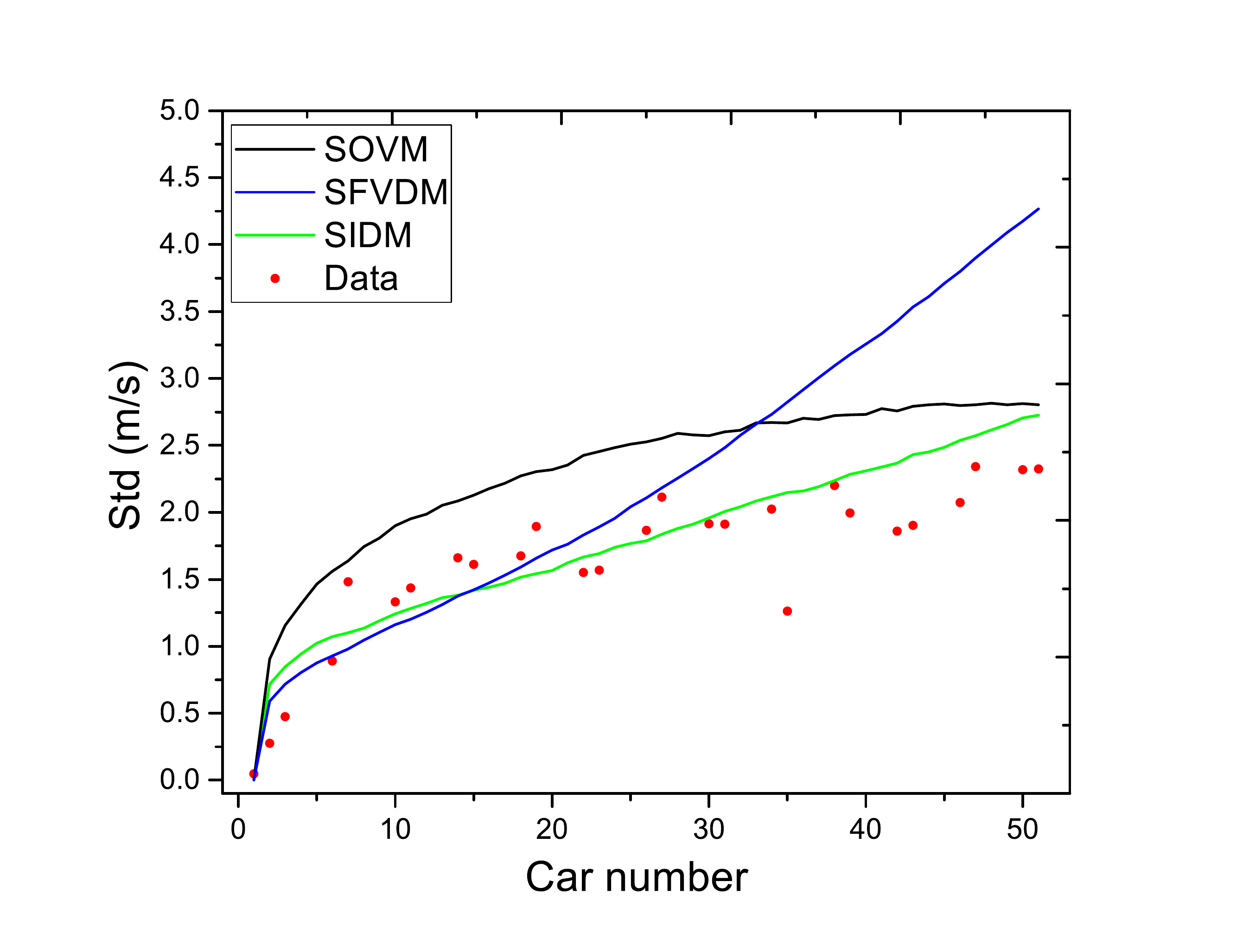}}
\end{tabular}
\caption{ Calibration and validation of the different stochastic and deterministic traffic models (a) Calibration of the deterministic models with $v_{l}=8.33 \ \mathrm{m/s}$ (b) Calibration of the stochastic models with $v_{l}=8.33 \ \mathrm{m/s}$  (c) Validation of the stochastic models with $v_{l}=1.38 \ \mathrm{m/s}$ (d) Validation of the stochastic models with $v_{l}=11.11 \ \mathrm{m/s}$.}
\label{fig17}
\end{figure}

\begin{figure}[H]
\centering
\begin{tabular}{cc} 
\centering
&\subfloat[]{\includegraphics[width=.25\textwidth]{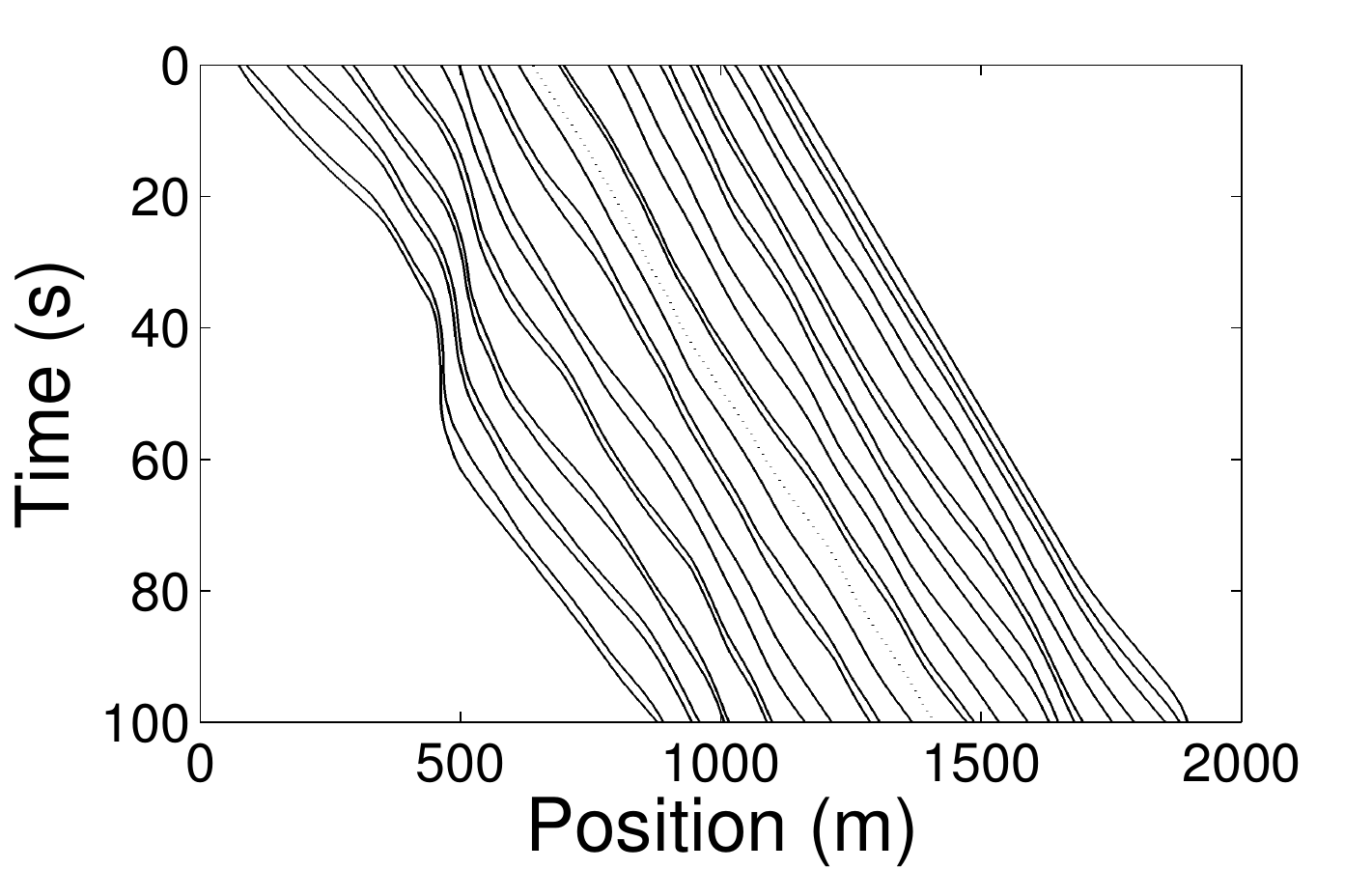}} 
\subfloat[]{\includegraphics[width=.25\textwidth]{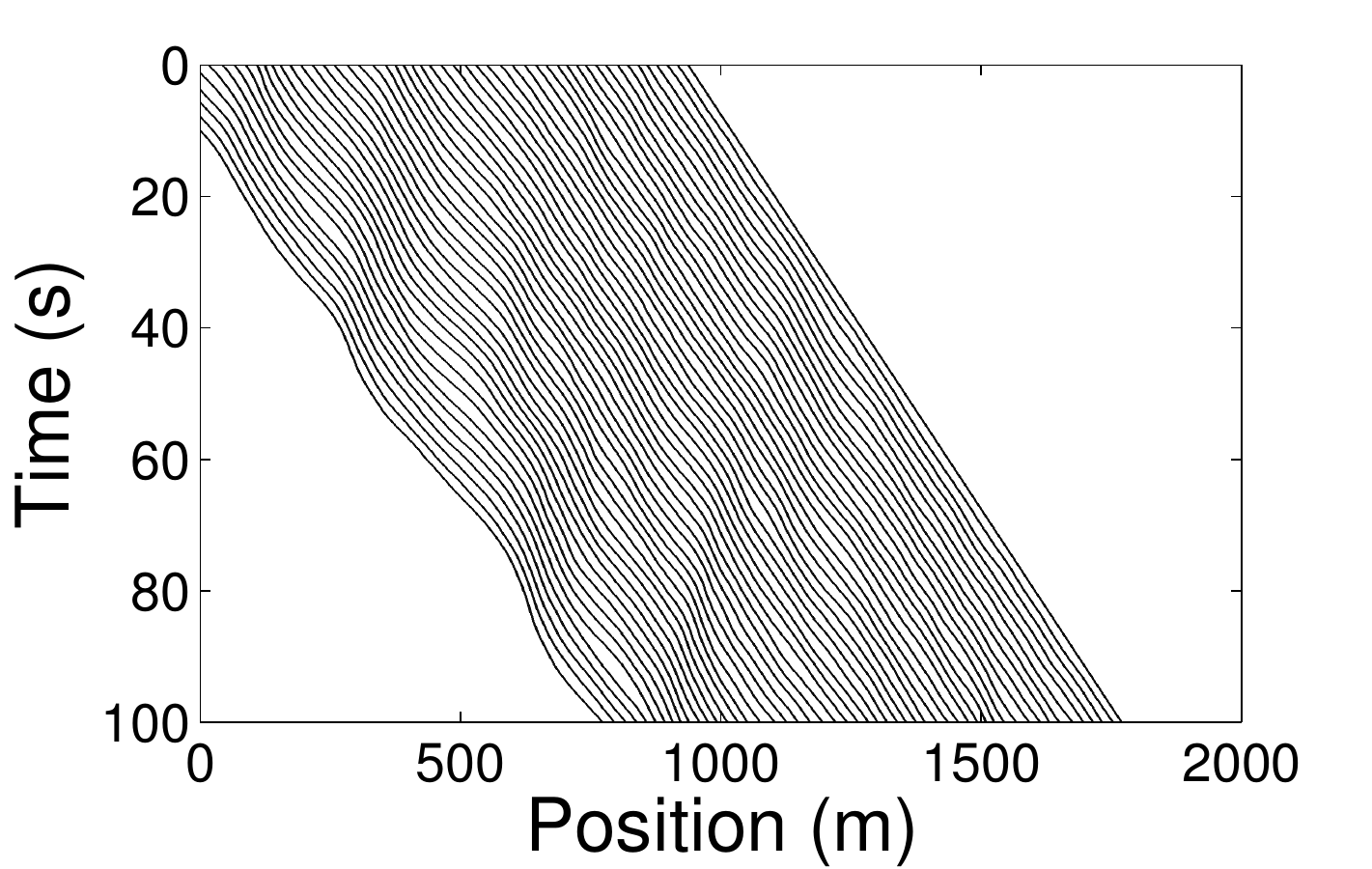}}
\subfloat[]{\includegraphics[width=.25\textwidth]{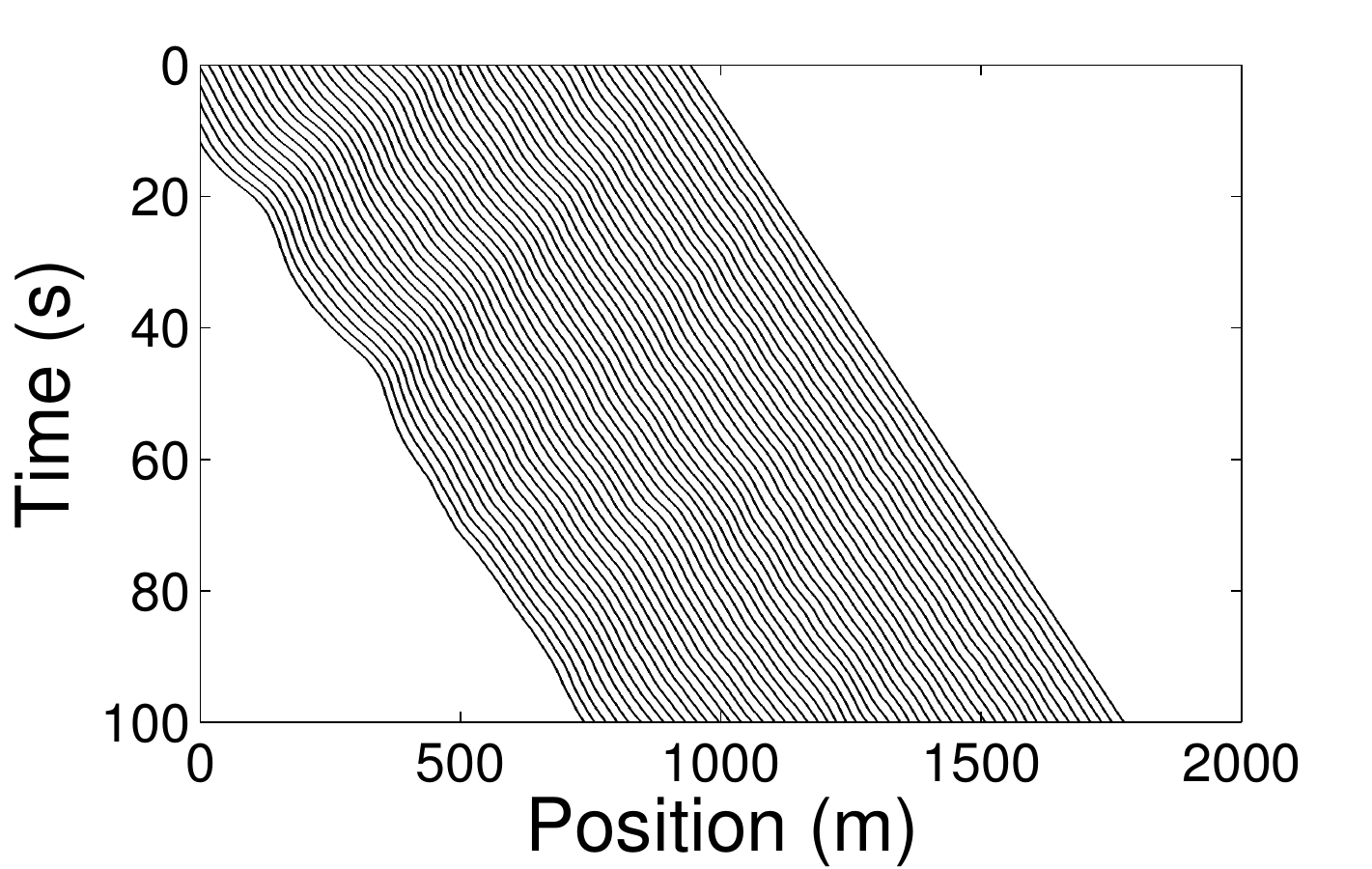}}
\subfloat[]{\includegraphics[width=.25\textwidth]{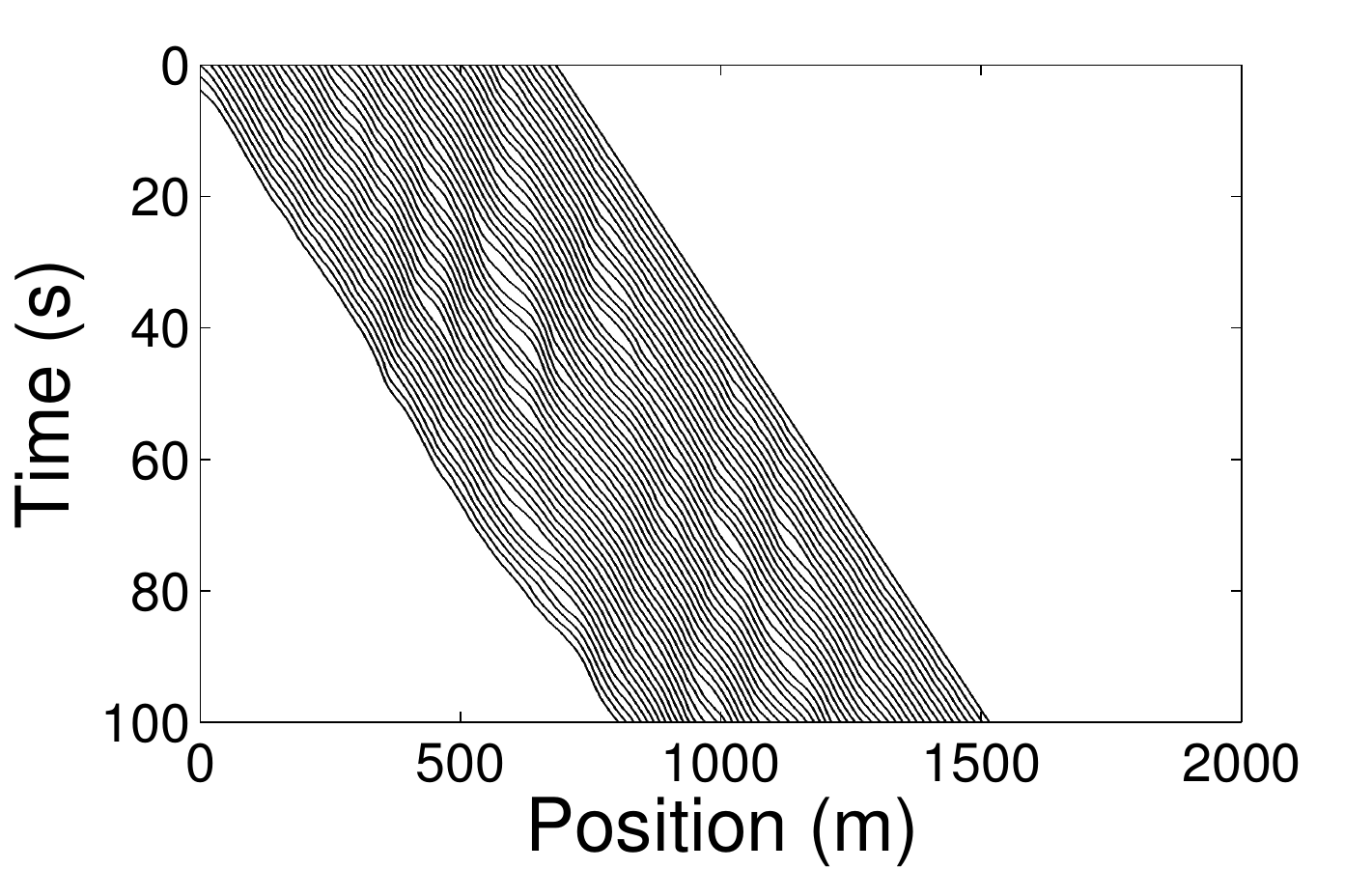}}\\
&\subfloat[]{\includegraphics[width=.25\textwidth]{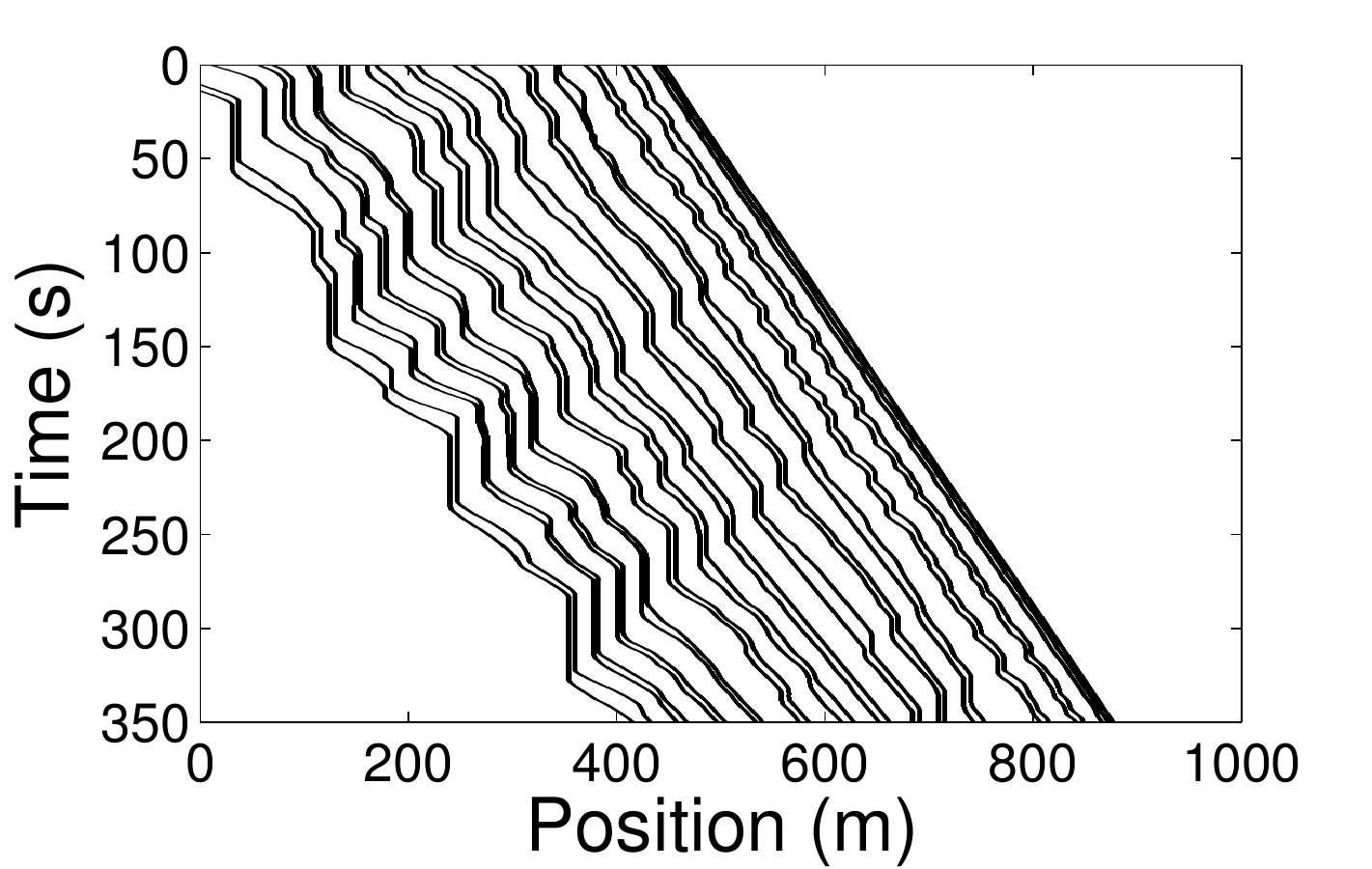}} 
\subfloat[]{\includegraphics[width=.25\textwidth]{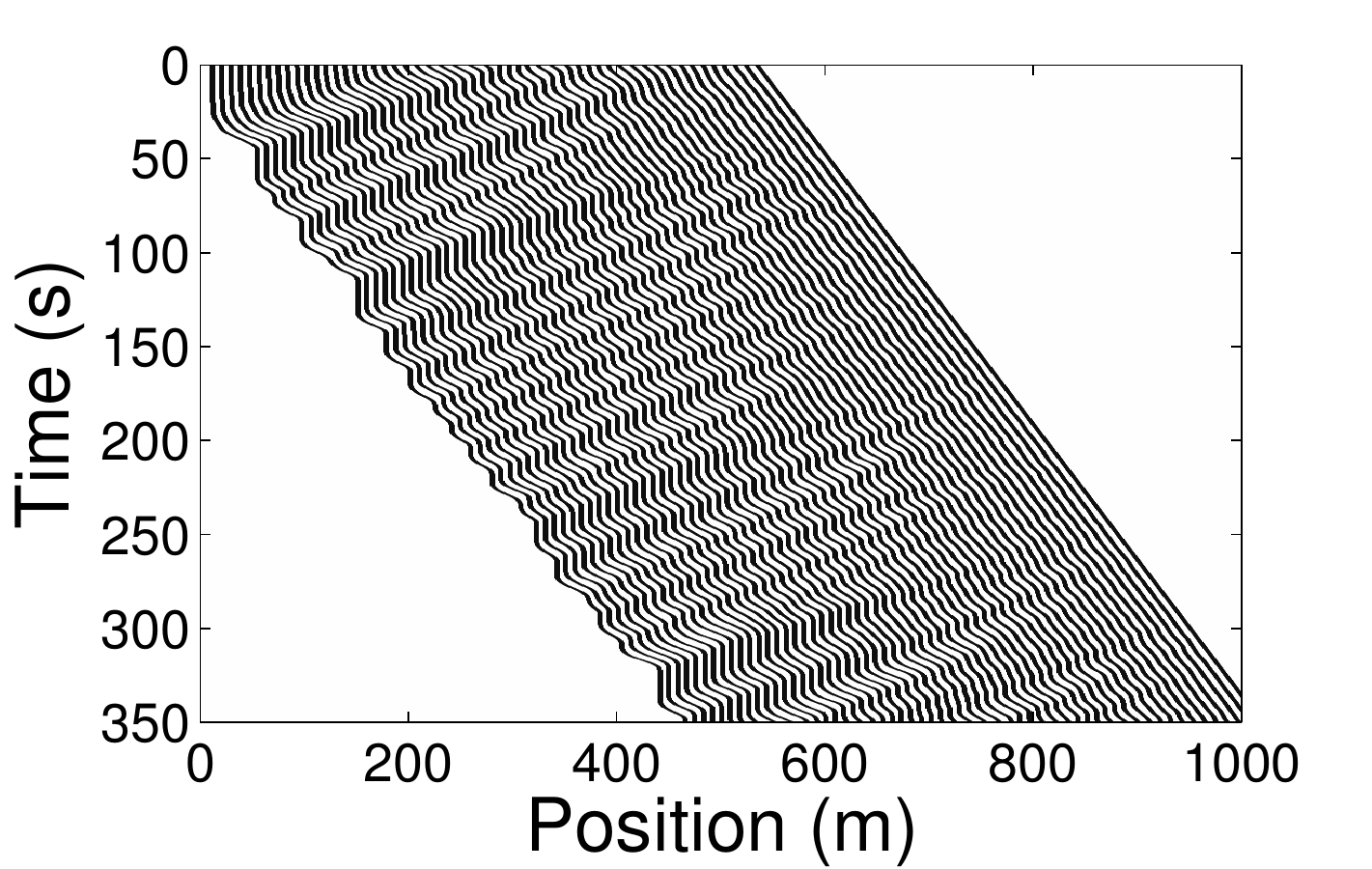}}
\subfloat[]{\includegraphics[width=.25\textwidth]{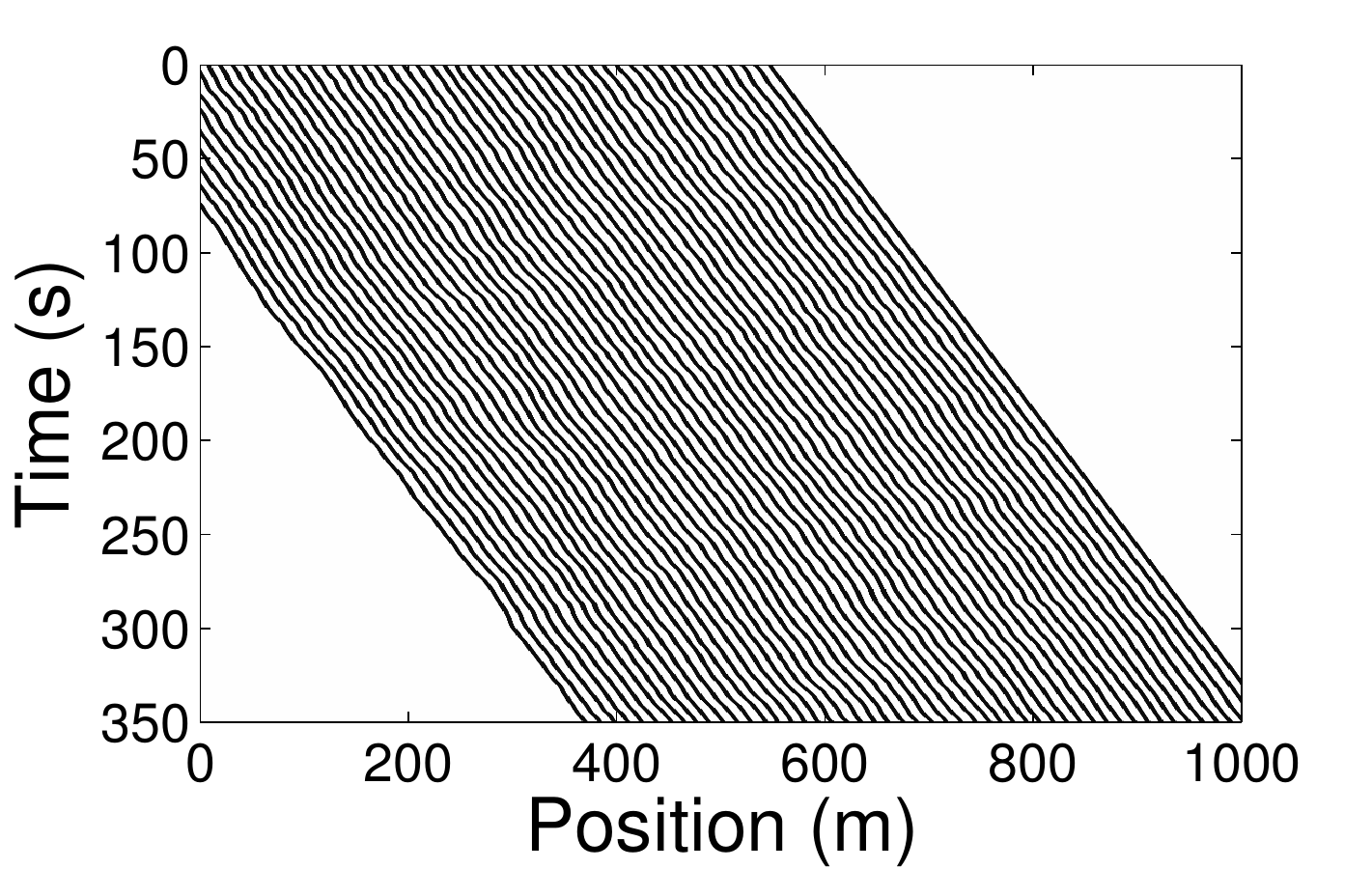}}
\subfloat[]{\includegraphics[width=.25\textwidth]{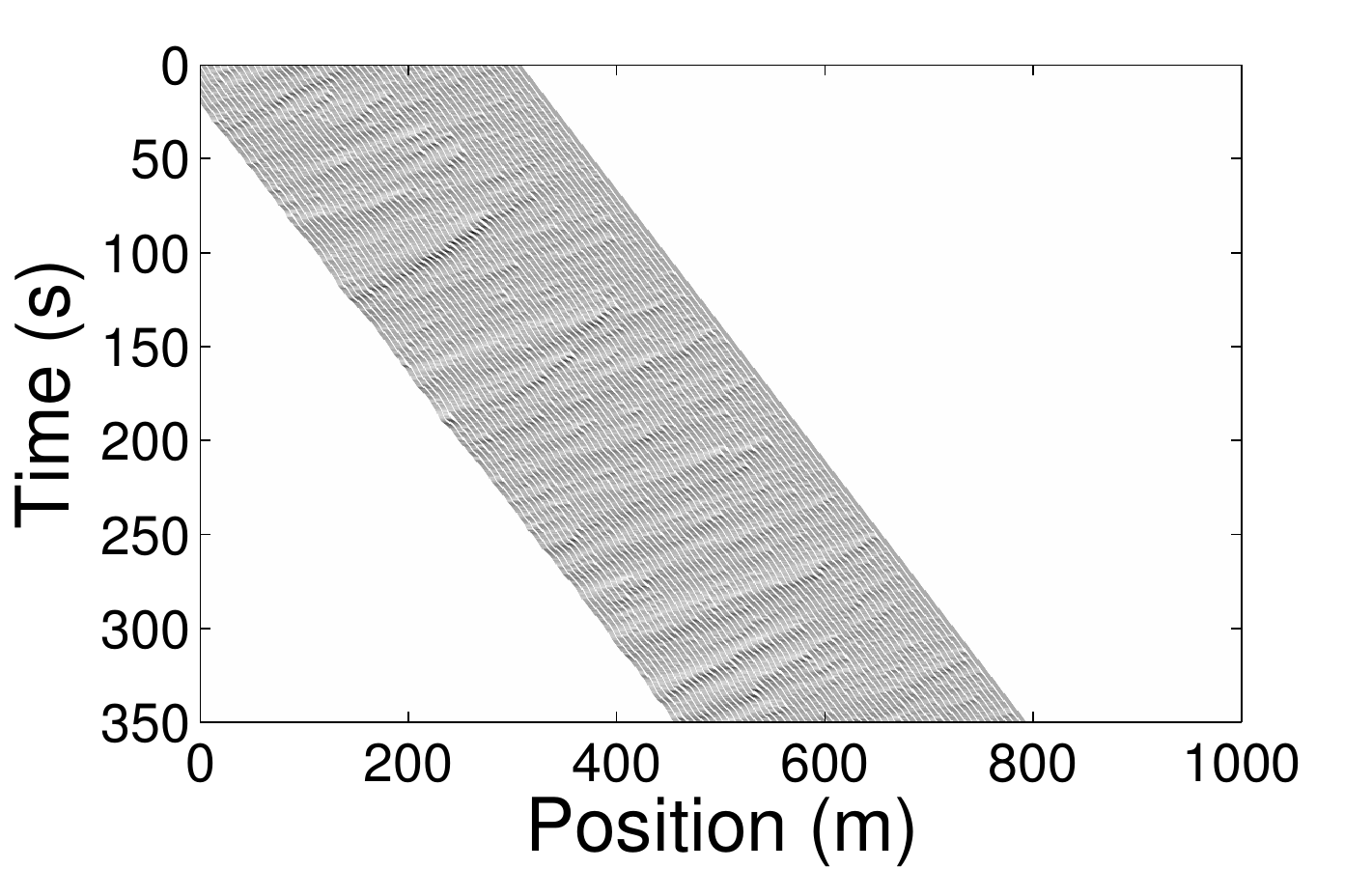}}\\
&\subfloat[]{\includegraphics[width=.25\textwidth]{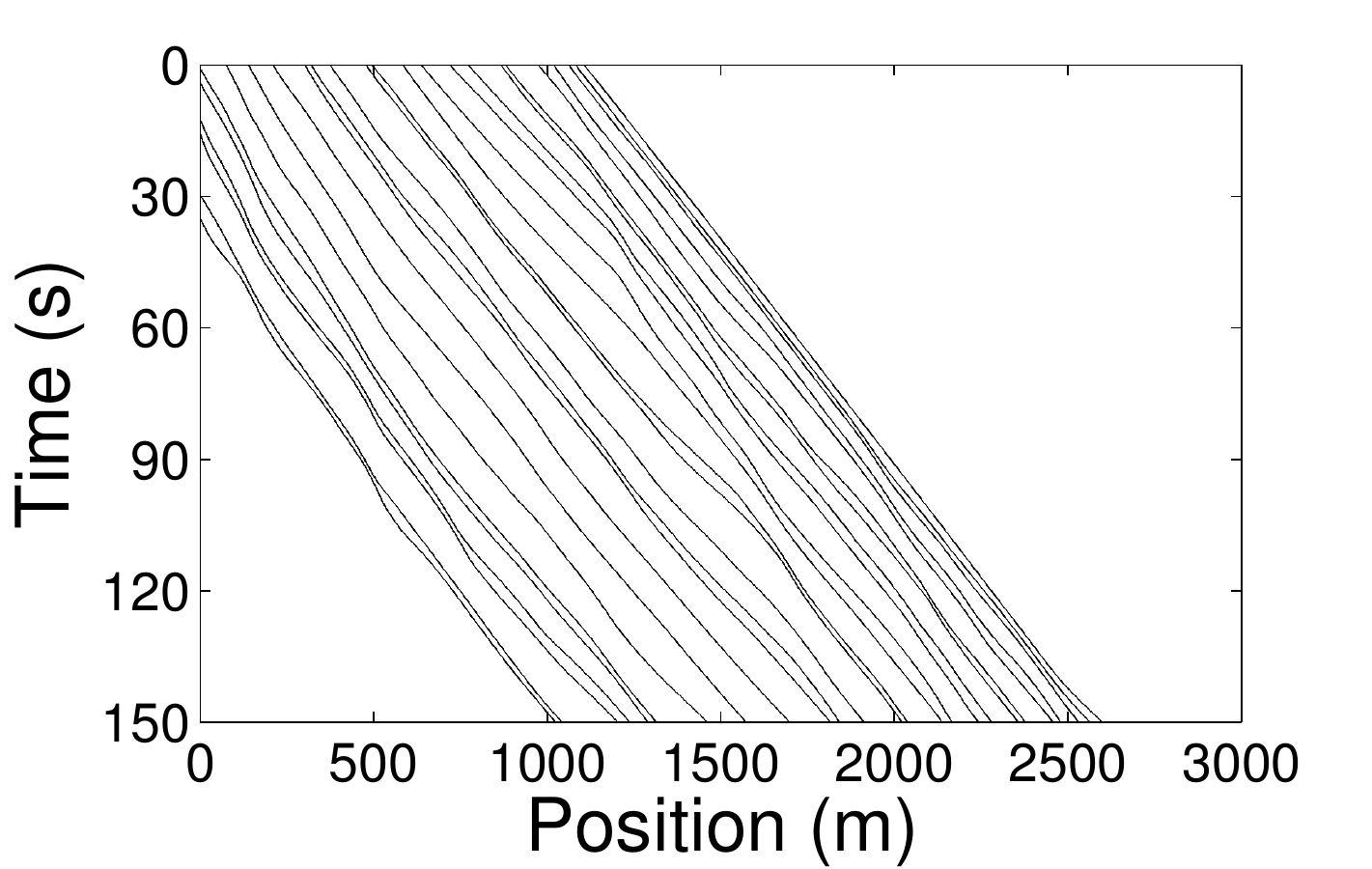}} 
\subfloat[]{\includegraphics[width=.25\textwidth]{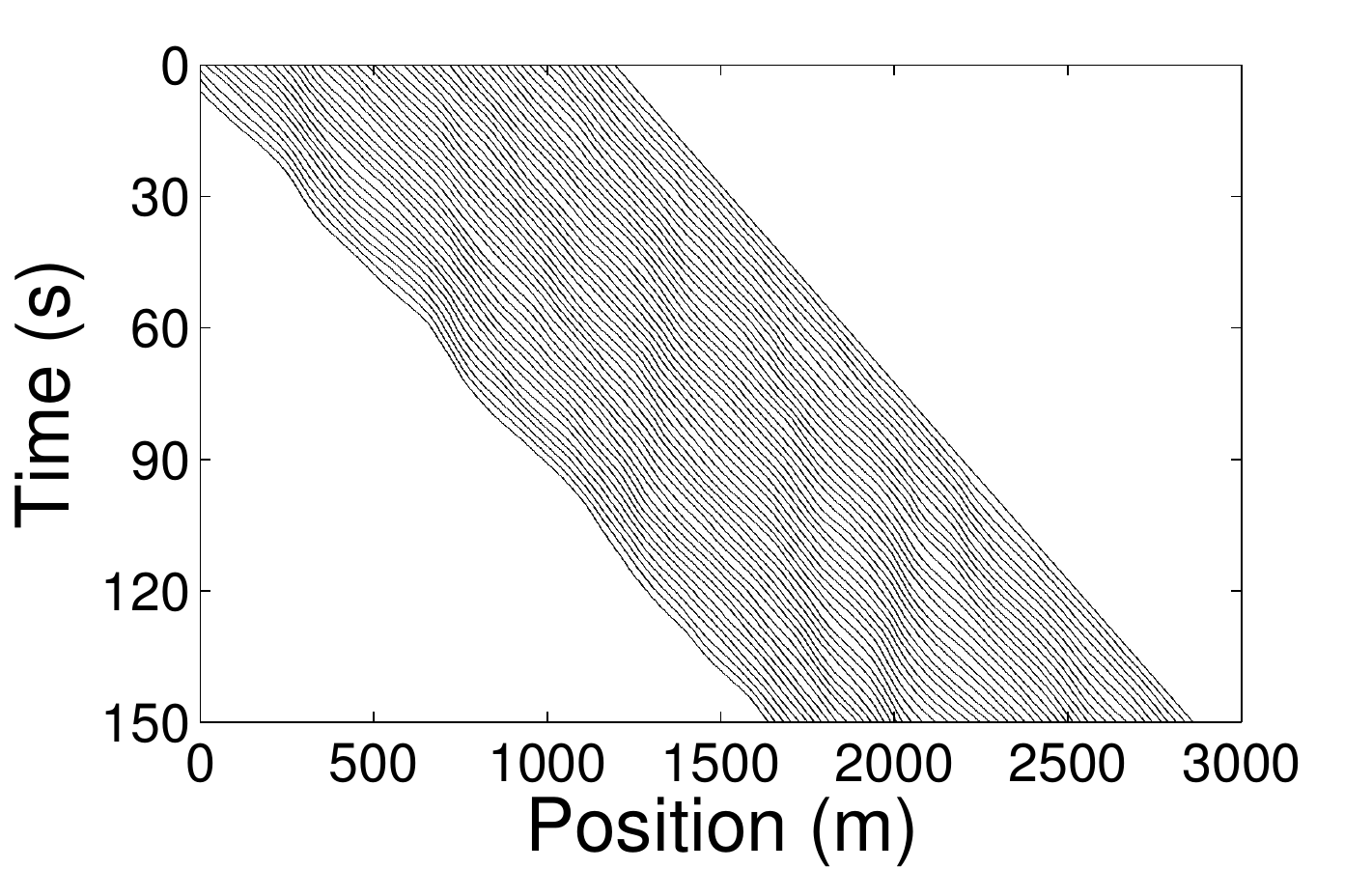}}
\subfloat[]{\includegraphics[width=.25\textwidth]{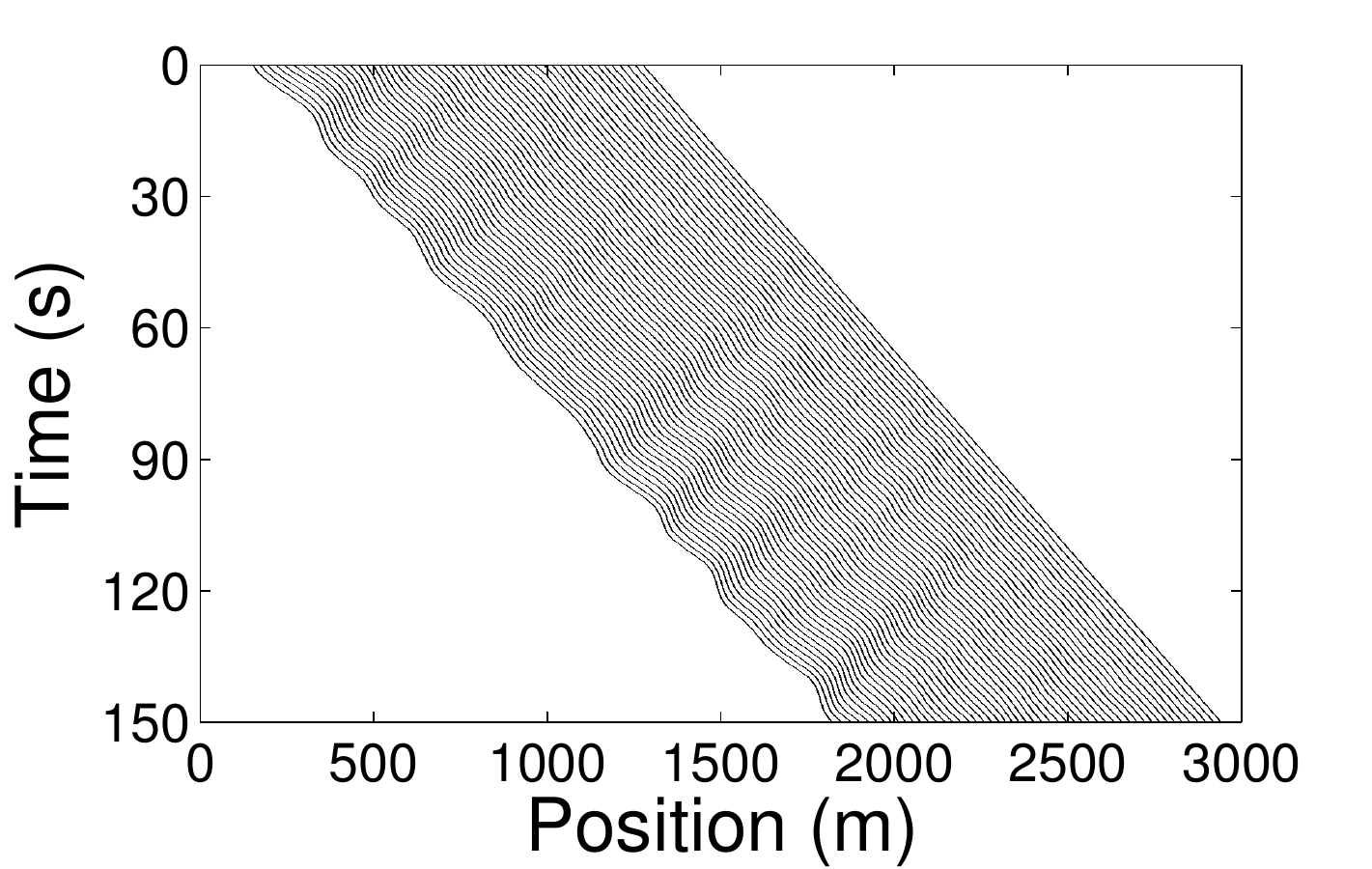}}
\subfloat[]{\includegraphics[width=.25\textwidth]{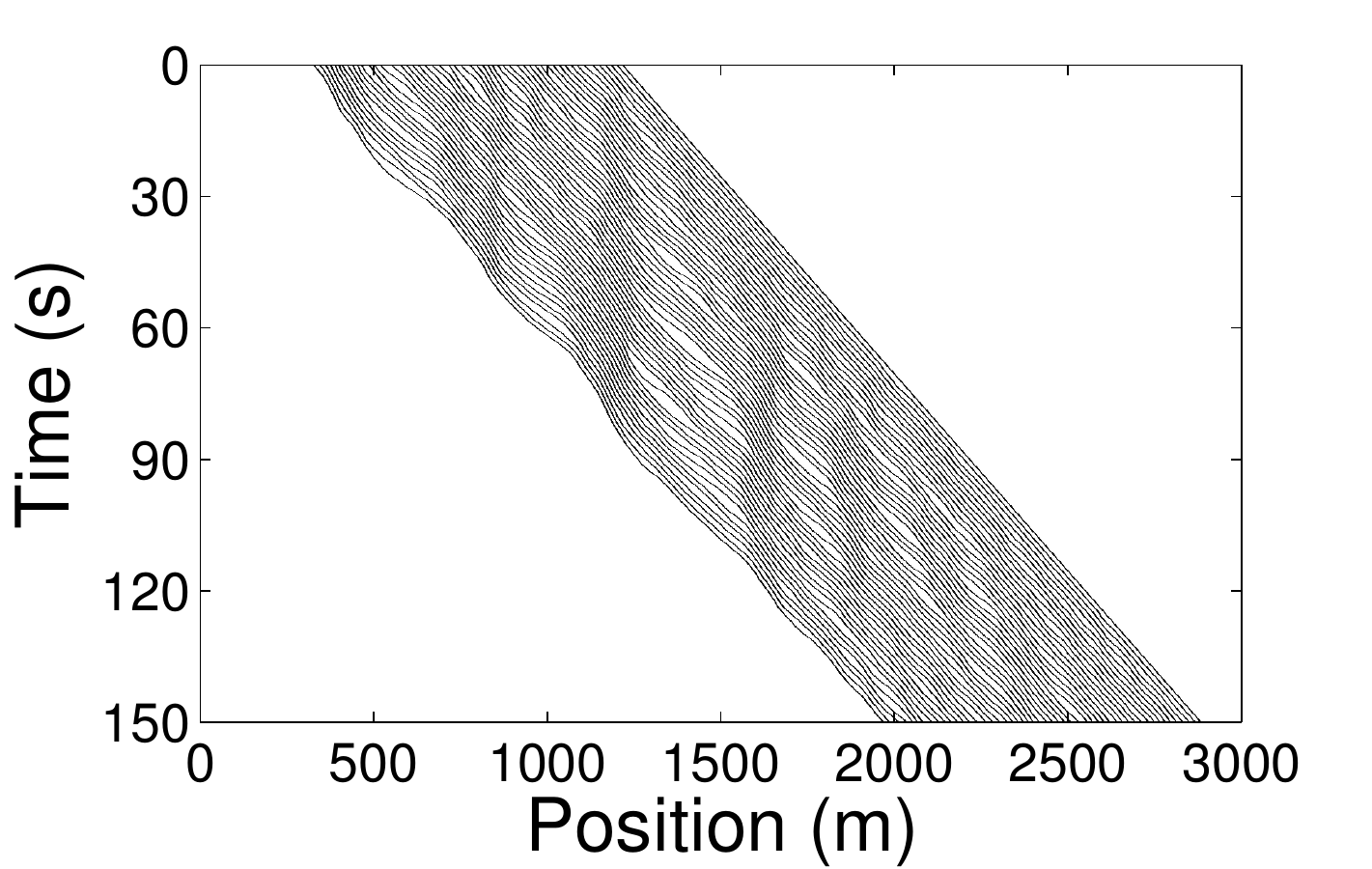}}\\
\end{tabular}
\caption{ Time-space diagrams of the 51 car-platoon experiment (left panel) and the corresponding simulations by using the calibrated SIDM, SFVDM, and SOVM and different leader's velocity (a-d) $v_{l}=8.33 \ \mathrm{(m/s)}$ (Calibration) (e-h) $v_{l}=1.38 \ \mathrm{(m/s)}$ (Validation) (i-l) $v_{l}=11.11 \ \mathrm{(m/s)} $ (Validation). }
\label{fig18}
\end{figure}

Figure~\ref{fig17}(a,b) display a comparison between the experimental observations and the calibrated traffic models. Table~\ref{t2} shows the calibrated parameters of the deterministic and stochastic traffic models. Table~\ref{t6} shows the corresponding performance index. Figure~\ref{fig18}(a-d) show the time-space diagrams corresponding to the experiment with $v_{l}= 30 \ \mathrm{km/h} \  (8.33  \ \mathrm{m/s})$ and the simulation results. Figure~\ref{fig17}(a) shows that the initial growth pattern of the calibrated deterministic traffic models is convex and deviates both qualitatively and quantitatively from the experimental data. On the other hand, as shown in Figure~\ref{fig17}(b), the stochastic traffic models exhibit a concave growth pattern of traffic oscillations which is in qualitative agreement with the experimental data. While Figure~\ref{fig17}(b) and the corresponding trajectories in Figure~\ref{fig18}(d) show that the SOVM is not in a quantitative agreement with the experimental observations, the SIDM and the SFVDM can reproduce the observed concave growth pattern of traffic oscillations not only qualitatively but also quantitatively, see also the trajectories in Figure~\ref{fig18}(a-c). 

Next, we assess the prediction capability of the stochastic models. Figure~\ref{fig17}(c,d) show the validation result of the different stochastic traffic models for both $v_{l}= 5 \ \mathrm{km/h} \ (1.38 \ \mathrm{m/s})$ and $v_{l}= 40 \ \mathrm{km/h} \ (11.11 \ \mathrm{m/s})$. Figure~\ref{fig18} (e-l) show the corresponding time-space diagrams. Note that the total number of vehicles in simulations is 51 vehicles while in the experiment (due to the GPS problem), only the trajectories of a maximum of 27 vehicles can be shown.

Figure~\ref{fig17}(c) and the vehicle trajectories in Figure~\ref{fig18}(f) show that the SIDM is in good agreement with data for a leader's velocity $v_{l}= 5 \ \mathrm{km/h} \ (1.38 \ \mathrm{m/s})$. Figure~\ref{fig17}(d) and Figure~\ref{fig18}(j) show that numerical simulation overestimates the development of traffic oscillations for $v_{l}= 40 \ \mathrm{km/h} \ (11.11 \ \mathrm{m/s})$. On the other hand, the SFVDM has shown a good calibration performance (Figure~\ref{fig18}(c)) but its prediction potential is low as the SOVM. The difference between simulation and experimental observations is more likely related to the similarity of the dynamics of the two models in a low-velocity environment, i.e. the SFVDM and the SOVM. In such a low velocity and low spacing conditions, the desired velocity is almost constant for both the SFVDM and the SOVM which prevents a significant development of traffic oscillations (stop and go waves) as experimentally observed (See Figure~\ref{fig18}(e,g,h) for a leader's velocity of $v_{l}= 5 \ \mathrm{km/h} \ (1.38 \ \mathrm{m/s})$. Thus, the speed standard deviation values shown in Figure~\ref{fig17}(c) are smaller than the experimental results. In Figure~\ref{fig18}(k) ($v_{l}= 40 \ \mathrm{km/h} \ (11.11 \ \mathrm{m/s})$), the frequency of oscillations generated by the SFVDM significantly exceeds the observations in Figure~\ref{fig18}(i). The high frequency of oscillations leads to a sharp increase of the speed standard deviation values in Figure~\ref{fig17}(d). Finally, Table~\ref{t6} and Figure~\ref{fig18}(d,h,l) show that the SOVM has the worse performance in both calibration and validation. These results suggest that the SIDM has the best overall prediction capability.


\begin{table*}[h] \footnotesize
\caption{ Performance index (PI) corresponding to each leader's velocity, $v_{l}=8.33 \ \mathrm{m/s}$	is used for calibration while $v_{l}=1.38  \ \mathrm{m/s}$ and $v_{l}=11.11 \ \mathrm{m/s}$ are used for validation. }

\centering
	 \begin{tabular}{cccccc}
\textbf{Model} &\textbf{Leader's velocity $v_{l}$}	& \textbf{Performance index (PI)} \\

  \hline
  IDM    	&  $8.33 \ \mathrm{m/s}$	   & 1.45 \\

		\hline
		
	 FVDM    	&  $8.33 \ m/s$	   & 0.44 \\

			\hline
			
		 OVM    	&  $8.33 \ m/s$	   & 0.42 \\

     \hline
  SIDM    	&  $8.33 \ \mathrm{m/s}$	   & 0.33 \\

   & $1.38  \ \mathrm{m/s}$				 & 0.31\\
	
	 & $11.11 \ \mathrm{m/s}$				 & 0.39\\
		
		\hline
		
	 SFVDM    	&  $8.33 \ m/s$	   & 0.3 \\

   & $1.38  \ \mathrm{m/s}$				 & 1.02\\
	 & $11.11 \ \mathrm{m/s}$				 & 1.79\\
	
			\hline
			
		 SOVM    	&  $8.33 \ m/s$	   & 0.59 \\

   & $1.38 \ \mathrm{m/s}$				 & 0.91\\
	
	 & $11.11 \ \mathrm{m/s}$				 & 1.04\\
	
			\hline 
	
\end{tabular}
		\label{t6}
\end{table*}

\section{Conclusion}

The first objective of this work is to carry out an analytical investigation of a general stochastic continuous car-following model considering both the inter-vehicular gap and the velocity difference. Based on the direct generalized Lyapunov method, we have extracted a string stability condition of the model and demonstrated that the presence of stochastic factors has a non-negligible destabilizing effect. 

Next, we have carried out numerical simulations to validate our theoretical analysis where we have proven that our methodology yields more accurate results than \cite{ngod}. The Lyapunov method is hence more suited to study the stability of the present class of stochastic traffic models. 

Finally, we have shown that the stochastic continuous car-following models outperform the existing deterministic traffic models in reproducing the observed traffic oscillations and the concave growth pattern of traffic oscillations.  

Our study can be extended in a few directions in future work. 
(i) The general model does not take reaction time and anticipation effect into account, which might increase the accuracy of the stochastic traffic models in reproducing the empirically observed traffic oscillations; (ii) The formulation of stochastic factors is rather simple; (iii) The stability analysis is performed only for homogeneous vehicles. In particular, in the coming era of connected and automated driving vehicles, the traffic composition will be a mixture of human-driven vehicles and automated-driven vehicles.

In the present study, we have shown that considering stochasticity improves the existing traffic models by achieving more credible simulations. Nevertheless, our approach is still not complete in capturing the complexity of traffic dynamics. Indeed, the stochastic factors reflect our limited knowledge of the complex car-following process and do not constitute the unique direction to improve traffic modeling. Subsequently, efforts are still needed to develop traffic models that can better capture the vehicle/driver characteristics for a more accurate prediction of traffic dynamics.





\hspace{10pt}

\begin{flushleft}
\normalsize{\bfseries{\large{Acknowledgments}}}
\end{flushleft}

\hspace{10pt}
This work was supported by the National Natural Science Foundation of China (No. 71971015, 71621001, and 71931002).

\bibliographystyle{APA}
\bibliography{Article-SCF}

\begin{appendices}
\counterwithin{figure}{section}
\counterwithin{table}{section}
\counterwithin{equation}{section}

\section{String instability of stochastic car-following models}

\begin{figure}[H]
\centering
\begin{tabular}{c} 
\centering
\subfloat[]{\includegraphics[width=.5\textwidth]{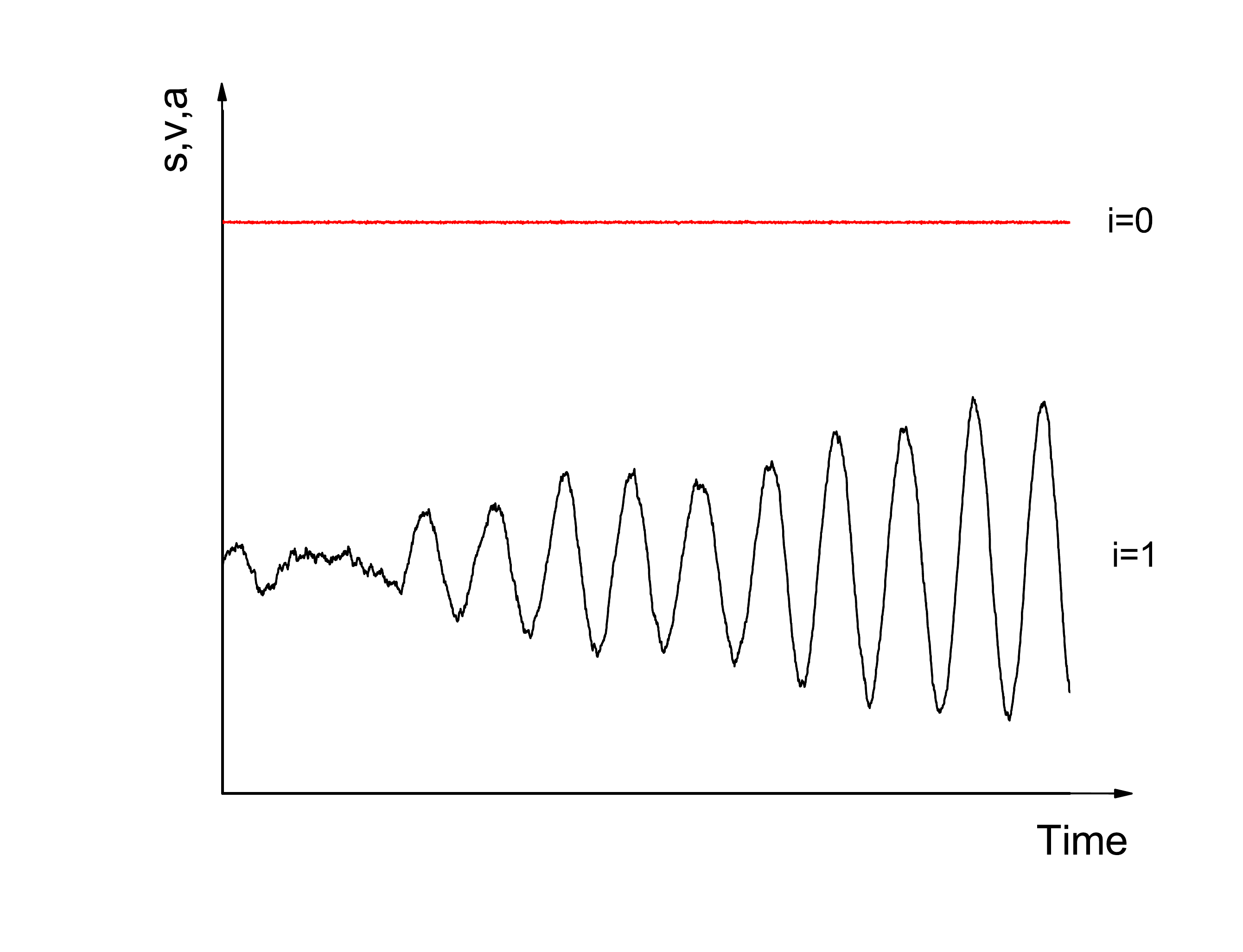}}
\subfloat[]{\includegraphics[width=.5\textwidth]{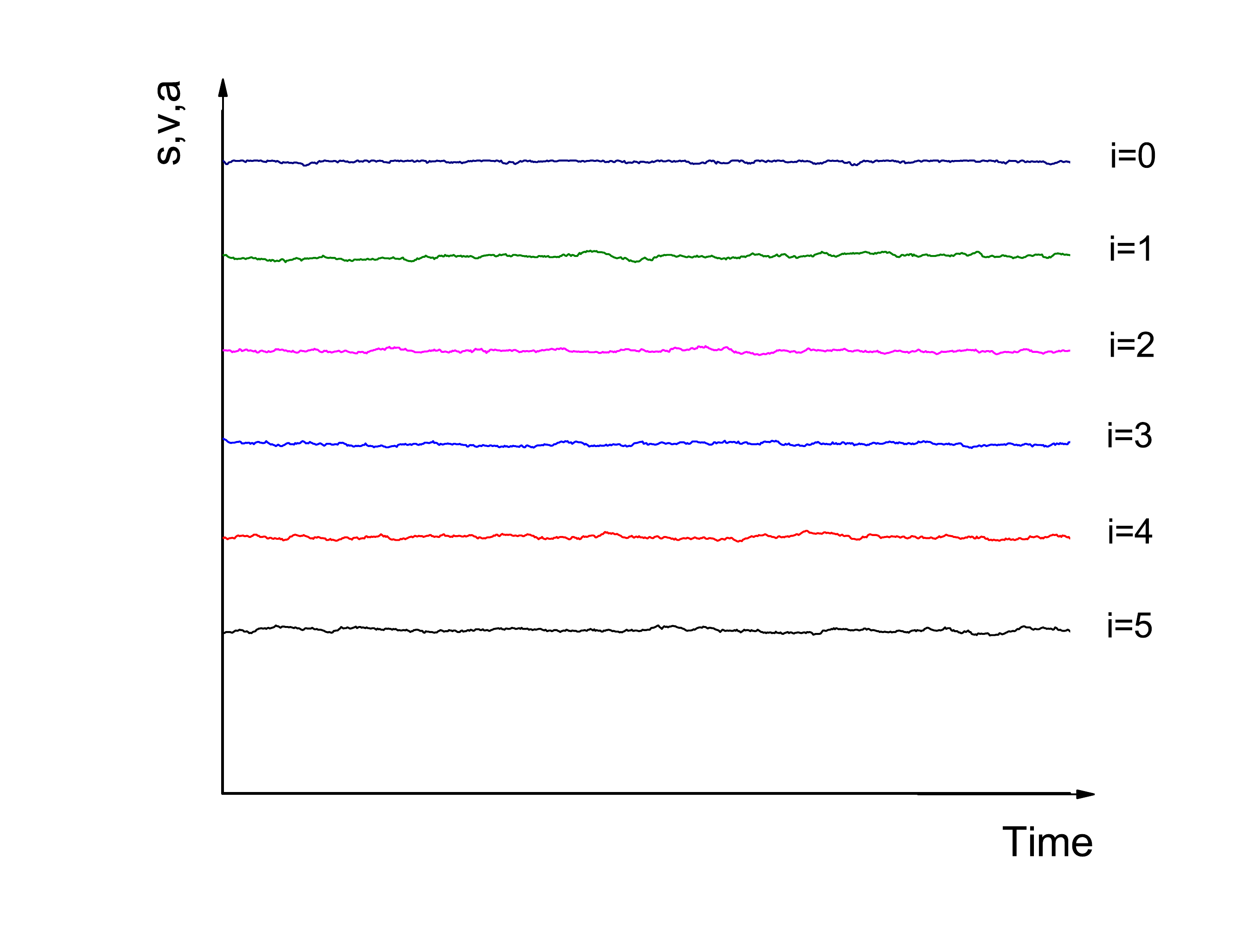}}\cr
\subfloat[]{\includegraphics[width=.5\textwidth]{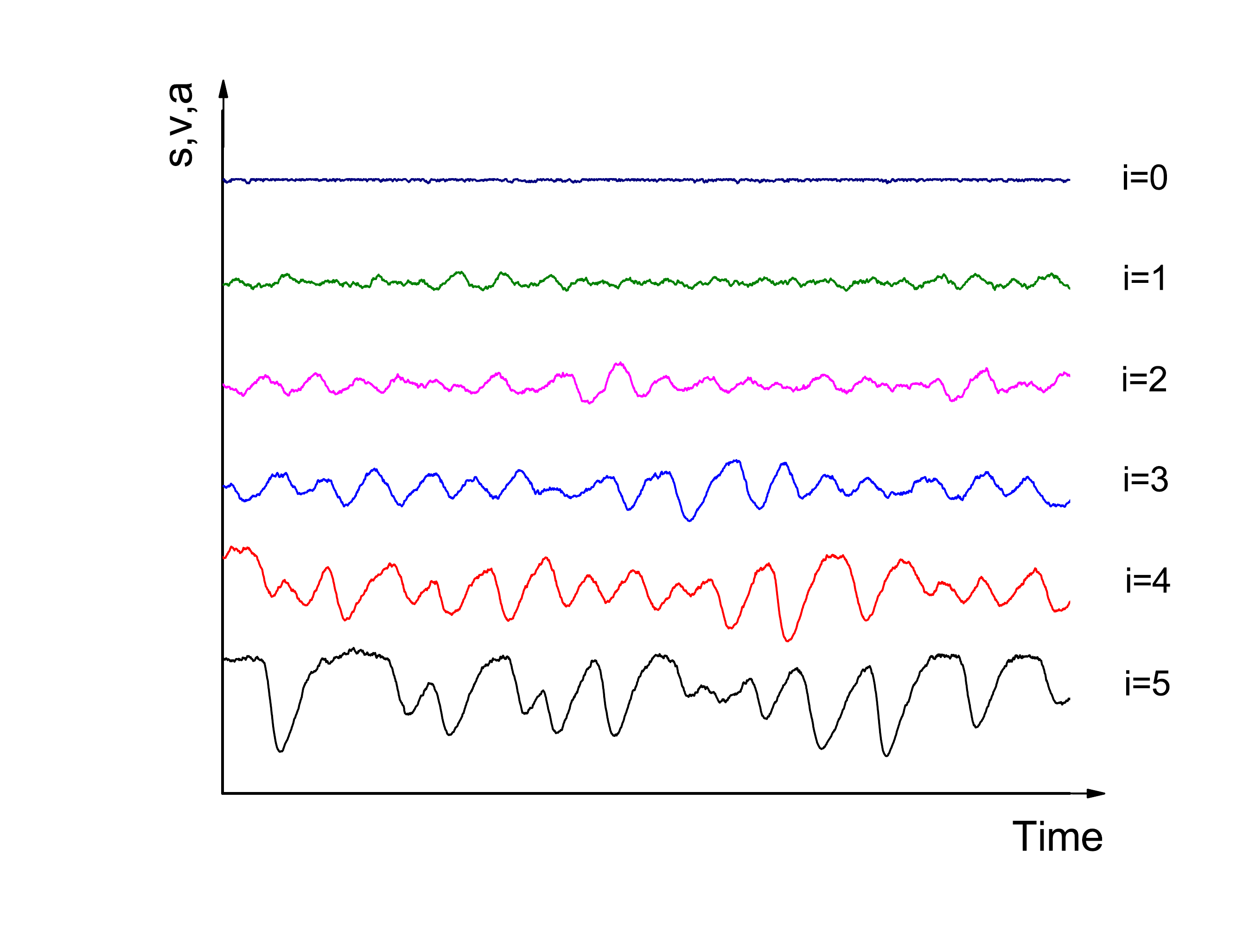}}
\end{tabular}
\caption{Sketch of the string stability and the local stability of a stochastic car-following process. (a) local instability, (b) string stability, (c) string instability. (s,v,a) denote the spacing, the velocity or the acceleration of the car i. i=0 is the leading vehicle.}
\label{figy}
\end{figure}

To give a visual illustration, we show a sketch of the local unstable traffic situation in Figure~\ref{figy}(a). In this case, the follower’s velocity oscillation amplifies with time. Figure~\ref{figy}(b) shows a string stable car-following process while Figure ~\ref{figy}(c) shows a string unstable traffic situation in a stochastic environment.  

Analytically, the string stability property has been encoded in the matrix (9) through the relation $d\delta s_{n}= (\delta v_{n-1} - \delta v_{n})dt$ before inserting the previously mentioned wave forms. Hence, equation (7) quantifies how speed fluctuations evolve along a platoon.


\section{The impact of the speed standard deviation term on improving the calibration quality of stochastic traffic models}


To demonstrate the role of the speed standard deviation term in the performance index, we also performed a simulation, using the performance index without the speed standard deviation term as below:

\begin{equation}
I^{2} =\frac{1}{NT_{m}} \sum_{i=1}^{i=T_{m}}\sum_{k=1}^{k=N} {(v_{ik} - \hat{v}_{ik})}^{2}
\end{equation} 

We present the results corresponding to experiment A only. The conclusion also applies to the other experiments. As shown in Figure~\ref{fig19} below, it is obvious that the performance index with the speed standard deviation term reproduces the time-space diagram better than the one without the speed standard deviation term.

\begin{figure}[H]
\centering
\begin{tabular}{c} 
\centering
\subfloat[]{\includegraphics[width=.25\textwidth]{expA}}
\subfloat[]{\includegraphics[width=.25\textwidth]{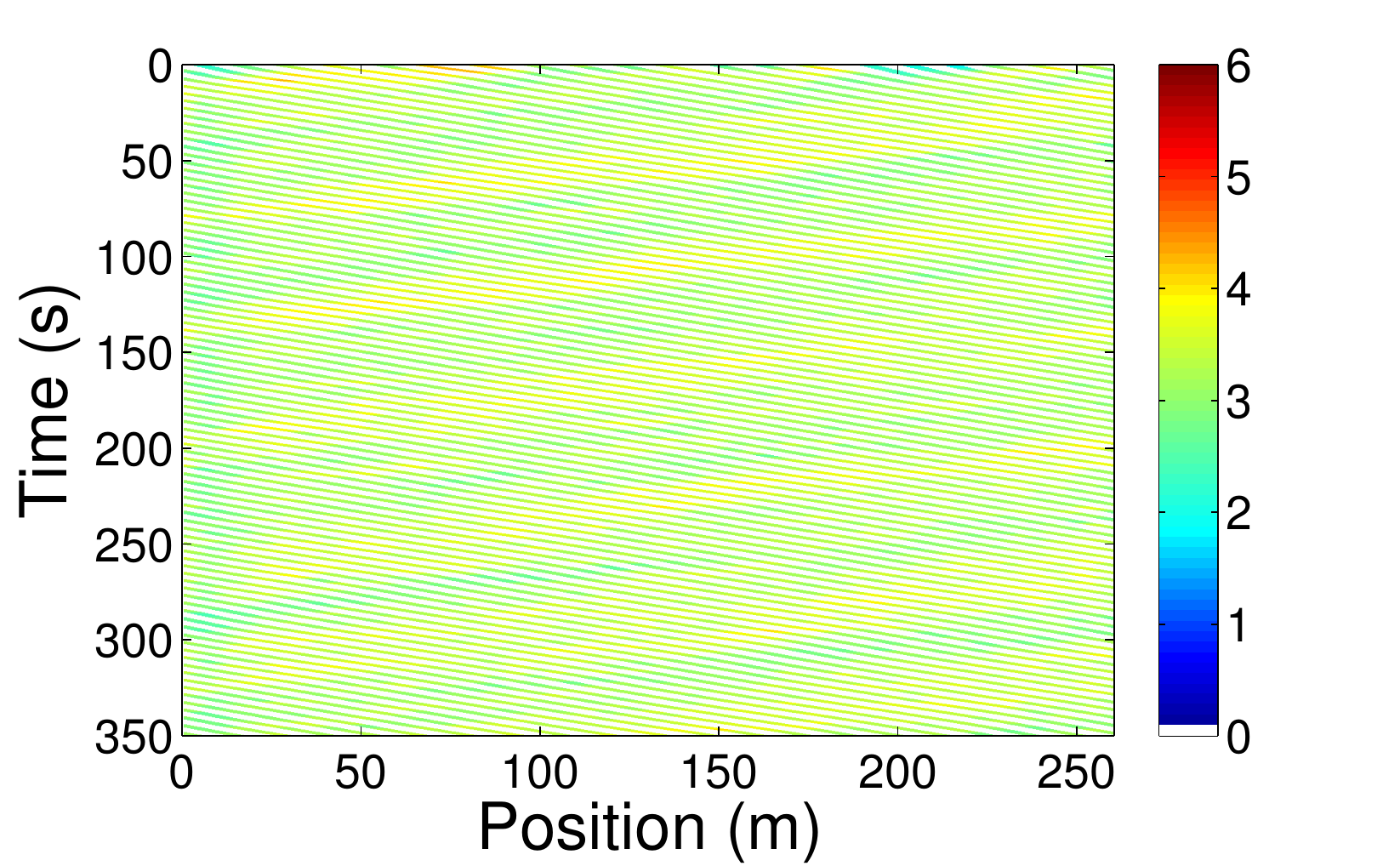}}
\subfloat[]{\includegraphics[width=.25\textwidth]{expAsidm}}

\end{tabular}

\caption{ Time space diagrams corresponding to the experiment A and the calibration using the objective functions in equation (51) and (B.1). (a) Experiment A, (b) calibration by using equation (B.1), (c) calibration by using equation (51). The velocity unit in the color bar is $\ \mathrm{(m/s)}$.}
\label{fig19}
\end{figure}

\end{appendices}

\end{document}